 \documentclass[reprint,
superscriptaddress,
preprintnumbers,
nofootinbib,
 amsmath,amssymb,
 aps, preprintnumbers,floatfix
]{revtex4-2}  
\usepackage[normalem]{ulem}
\usepackage{amsmath}
\usepackage{physics}
\usepackage{amssymb}
\usepackage{epsfig}
\usepackage{graphicx}
\usepackage{hyperref}
\usepackage{tabu}
\usepackage{boldline}
\usepackage{xspace}
\usepackage{slashed}
\usepackage{multirow}
\usepackage{diagbox}
\usepackage{tabularx}
\usepackage{comment}
\usepackage{array}
\usepackage{makecell}
\usepackage{cleveref}
\usepackage{floatrow}

\newfloatcommand{capbtabbox}{table}[][\FBwidth]

\usepackage{color}
\usepackage[normal]{subfigure}
\usepackage[table]{xcolor}
\usepackage{enumitem}
\usepackage[utf8]{inputenc}
\usepackage{colortbl}
\definecolor{nicered}{rgb}{0.7,0.1,0.1}
\definecolor{nicegreen}{rgb}{0.1,0.5,0.1}
\definecolor{red}{rgb}{1.0, 0, 0}
\definecolor{pink}{RGB}{255, 0, 145}

\usepackage{lineno}

\begin{document}

\title{
Constraints on Strongly-Interacting Dark Matter \\ from the James Webb Space Telescope
}

\author{Peizhi Du}
\email{peizhi.du@rutgers.edu}
\affiliation{New High Energy Theory Center, Department of Physics and Astronomy,
Rutgers University, Piscataway, NJ 08854}

\author{Rouven Essig}
\email{rouven.essig@stonybrook.edu}
\affiliation{C.N. Yang Institute for Theoretical Physics, Stony Brook University, Stony Brook, NY, 11794, USA}

\author{Bernard J.~Rauscher}
\email{Bernard.J.Rauscher@nasa.gov}
\affiliation{NASA Goddard Space Flight Center, Observational Cosmology Laboratory, Greenbelt, MD, 20771, USA}

\author{Hailin Xu}
\email{hailin.xu@stonybrook.edu}
\affiliation{C.N. Yang Institute for Theoretical Physics, Stony Brook University, Stony Brook, NY, 11794, USA}

\preprint{YITP-SB-2024-31}

\begin{abstract}
    Direct-detection searches for dark matter are insensitive to dark matter particles that have large interactions with ordinary matter, which are stopped in the atmosphere or the Earth's crust before reaching terrestrial detectors.  We use ``dark'' calibration images taken with the HgCdTe detectors in the Near-Infrared Spectrograph (NIRSpec) on the James Webb Space Telescope (JWST) to derive novel constraints on sub-GeV dark matter candidates that scatter off electrons. 
    We supplement the JWST analysis pipeline with additional masks to remove pixels with high-energy background events. 
    For a 0.4\% subcomponent of dark matter that interacts with an ultralight dark photon, we disfavor all previously allowed parameter space at high cross sections, and constrain some parameter regions for subcomponent fractions as low as $\sim$0.01\%. 
\end{abstract}

\maketitle

\section{Introduction}\label{sec:introduction} 
Although there is extensive evidence for dark matter (DM) from astrophysics and cosmology, its particle nature remains unknown. 
Direct-detection experiments provide an important probe of the interactions between DM and ordinary matter. Tremendous progress has been made in searching for ever smaller interactions, in particular also for sub-GeV DM masses~\cite{Essig:2011nj,Tiffenberg:2017aac,Crisler:2018gci,Agnese:2018col,Abramoff:2019dfb,Aguilar-Arevalo:2019wdi,SENSEI:2020dpa,Arnaud:2020svb,Amaral:2020ryn,DAMIC_2022,CDEX:2022kcd,Essig:2022dfa,DAMIC-M:2023gxo,DAMIC-M:2023hgj,SENSEI:2023zdf,SuperCDMS:2024yiv}. In order to achieve sensitivity to small interactions, the experiments are typically placed underground to avoid cosmic rays and well-shielded to avoid radiogenic backgrounds.  This means, however, that they are insensitive to DM that has large interactions with ordinary matter, which would be scattered or significantly slowed by the Earth's atmosphere, crust, and/or the shielding before reaching the detector.  
This effect is reflected as an upper limit on current direct-detection constraints~\cite{Starkman:1990nj,Zaharijas:2004jv,Emken:2019tni}.

Collider, accelerator, astrophysical, and cosmological data place stringent constraints on strongly-interacting DM,\footnote{``Strong'' refers in this paper to the size of the interaction between DM and ordinary matter, not to the strong force described by Quantum Chromodynamics.} generally requiring interactions below a specific threshold~\cite{Chen:2002yh,Dvorkin:2013cea,Gluscevic:2017ywp,McDermott:2010pa,Slatyer:2015jla,Dolgov:2013una,Dubovsky:2003yn,Xu:2018efh,Ali-Haimoud:2015pwa,Boddy:2018wzy,Vogel:2013raa,Davidson:2000hf,Creque-Sarbinowski:2019mcm,Boehm:2013jpa,Bhoonah:2018gjb,Chang:2018rso,Davidson:2000hf,Davidson:2000hf,Prinz:1998ua,Magill:2018tbb,ArgoNeuT:2019ckq,Harnik:2020ugb,Plestid:2020kdm,PhysRevD.102.032002,SENSEI:2023gie,Munoz:2018pzp,Liu:2019knx,Rich:1987st,Mahdawi:2018euy,Prabhu:2022dtm}. 
The DM bounds depend sensitively on the DM mass and specific model and, for concreteness, we focus here on 
DM coupled to an ultralight dark-photon mediator, which is subject to many of the bounds applicable to millicharged particles. 
In particular, if this constitutes all of DM, constraints from the Cosmic Microwave Background (CMB) already rule out parameter space above the direct-detection upper limit~\cite{Buen-Abad:2021mvc,Nguyen:2021cnb}. However, 
the CMB constraint vanishes if this constitutes only a sub-component of DM with fractional abundance $f_{\chi}\equiv \rho_{\chi}/\rho_{\rm DM} \lesssim 0.4\%$~\cite{Boddy:2018wzy}, where $\rho_{\rm DM}$ is the total DM density. An allowed region above the ground-based direct-detection bounds then exists, see Fig.~\ref{fig:DM-rates}. 

The unconstrained region can be probed by proposed accelerator- and collider-based experiments~\cite{Kelly:2018brz,Gninenko:2018ter,Foroughi-Abari:2020qar,Oscura:2023qch,deMontigny:2023qft,Kalliokoski:2023cgw,Tsai:2024wdh} or by leveraging the fact that the density can be significantly enhanced near the Earth's surface~\cite{Pospelov:2020ktu,Budker:2021quh,Berlin:2021zbv,Kim:2021eix,Kling:2022ykt,McKeen:2022poo,Berlin:2023zpn,McKeen:2023ztq,Pospelov:2023mlz,Ema:2024oce}. 
In addition, low-threshold balloon- and space-borne detectors can have reach beyond ground-based experiments due to the reduced atmospheric shielding~\cite{Rich:1987st,snowden1990search,1990ApJ...364L..25S,Wandelt:2000ad,Erickcek:2007jv,Emken:2019tni,Adams:2019nbz,Saffold:2024lsj}, provided they only have minimal shielding surrounding the detectors. For example, the DarkNESS mission proposes to use a Skipper-CCD on a satellite to probe the open region to very small fractional abundances $f_{\chi}$~\cite{Saffold:2024lsj}.  

In this letter, we show that \textit{existing} space-based sensors can already probe the allowed region.  In particular, we analyze ``dark'' images obtained with 
the Near Infrared Spectrograph (NIRSpec) on the James Webb Space Telescope (JWST) to place novel constraints on strongly-interacting sub-GeV DM.  We simulate the expected DM signal assuming DM-electron interactions with an ultralight dark-photon mediator,  and include the effects of the detector shielding. Alongside the JWST image-processing pipeline, we develop a masking process to reduce high-energy events and their secondary impacts, while largely preserving the DM signal.  Assuming the backgrounds after masking consist solely of Poisson-distributed dark current,  we place constraints on the DM-electron cross-section for each DM mass based on the shape of the pixel charge distribution. Our results show that current JWST data can constrain a wide range of open parameter space between $1$~MeV and $10$~GeV DM masses for $f_\chi$ as low as $\sim$0.01\%.
The rest of this paper is outlined as follows. In \S\ref{sec:DM-Signal}, we describe the expected DM signal. In \S\ref{sec:Pipeline-and-masks}, we discuss the JWST NIRSpec data and the data analysis masks. \S\ref{sec:DM-Constraints} gives the constraints. The Supplemental Material (SM) contains additional details. 

\section{Dark Matter Signals at JWST NIRSpec}
\label{sec:DM-Signal}

The JWST NIRSpec instrument features two detectors (NRS1 and NRS2) consisting of $\text{Hg}_{\text{1-x}}\text{Cd}_\text{x}\text{Te}$  with $\text{x}=0.3$, which has a bandgap $E_\text{gap}=0.234\,\text{eV}$. Each detector has $2048 \times 2048$~pixels, including a 4-pixel wide frame of ``reference pixels'' on all sides that electronically mimic regular pixels, although they do not respond to light. The reference pixels are used within the JWST pipeline to remove correlated noise. Each pixel has an area of $18\,\text{$\mu$m}\times18\,\text{$\mu$m}$ and a thickness of about $5\,\text{$\mu$m}$, corresponding to a detector mass of 0.05037~gram.
We focus on data taken with NRS2. 

The analysis of the JWST data will depend on the properties of the DM candidate.  For concreteness, we consider a fermionic DM particle $\chi$  
coupled to an ultralight dark-photon mediator $A'$ that is kinetically mixed with the ordinary photon~\cite{Holdom:1985ag,Galison:1983pa,Essig:2011nj,Emken:2019tni} (we leave other models to future work). 
The DM signal rate in the HgCdTe target is given by~\cite{Essig:2011nj,Essig:2015cda,Hochberg:2021pkt,Knapen:2021run,Knapen:2021bwg,Griffin:2021znd,Kahn:2021ttr,Trickle:2022fwt,Dreyer:2023ovn} 
\begin{eqnarray}\label{eq: DM signal rate}
        \frac{\dd R}{\dd\ln E_e}=f_\chi \frac{\rho_\chi}{m_\chi}N_{\text{cell}}\alpha\frac{\overline\sigma_e m_e^2}{\mu_{\chi e}^2}\int\frac{\dd q}{q^2}E_e\abs{F_{\text{DM}}(q)}^2\nonumber\\
        \times\abs{\frac{f_{\text{crys}}(E_e,q)}{\epsilon(E_e,q)}}^2 \int_{v_\chi>v_{\text{min}}}\dd v_\chi\frac{1}{v_\chi}f_{\rm det}(v_\chi)\,,
\end{eqnarray}
 where $m_\chi$ is the DM mass, $m_e$ is the electron mass, $\mu_{\chi e}$ is the reduced DM-electron mass, $N_{\text{cell}}$ is the number of unit cells in the crystal target, $f_\text{crys}(E_e,q)$ is the crystal form factor, $v_{\text{min}}=\frac{E_e}{q}+\frac{q}{2m_\chi}$,
which is the minimum DM speed $v_\chi$ for the incoming DM particle to transfer energy $E_e$ and momentum $q$ to the target, $\rho_\chi=0.3\;\text{GeV}/\text{cm}^3$ is the local DM density, and $f_{\rm det}(v)$ is the DM velocity distribution at the detector (see below).  For the ultralight dark photon, we define the DM-electron reference cross section as $\overline\sigma_e\equiv16\pi\alpha\alpha_D\epsilon^2\mu_{\chi e}^2/(\alpha m_e)^4$,
where $\alpha$ ($\alpha_D$) is the electromagnetic (dark) fine structure constant, and $\epsilon$ is the kinetic-mixing parameter. The DM form factor is $F_{\text{DM}}(q)\equiv (\alpha m_e/q)^2$.
We use \texttt{QEDark}~\cite{Essig:2015cda,QEdark} to calculate the crystal form factor $f_\text{crys}$ for both HgTe and CdTe.  A more accurate determination of the scattering rate is in progress~\cite{Dreyer:in-progress}. The factor $1/|\epsilon(E_e,q)|^2$ encodes screening effects, and we use the Lindhard formula to approximate the dielectric function $\epsilon(E_e,q)$~\cite{Dressel_Gruner_2002}. To convert $E_e$ to the number of electron-hole pairs (denoted $e^-$) that are produced, $Q$, we use 
$Q(E_e) =1+\lfloor (E_e -E_\text{gap})/\epsilon_\text{eh} \rfloor$, for $E_e\geq E_\text{gap}$,
where the half-brackets represent the floor function, and the average energy needed to create an electron-hole pair is $\epsilon_\text{eh}\sim 3 E_{\rm gap}$. Examples of the event rate are given in the SM.  
For each $Q$, we conservatively choose the lower rate between HgCd and CdTe. 

\begin{figure}[t]
    \centering
    \includegraphics[width=\linewidth]{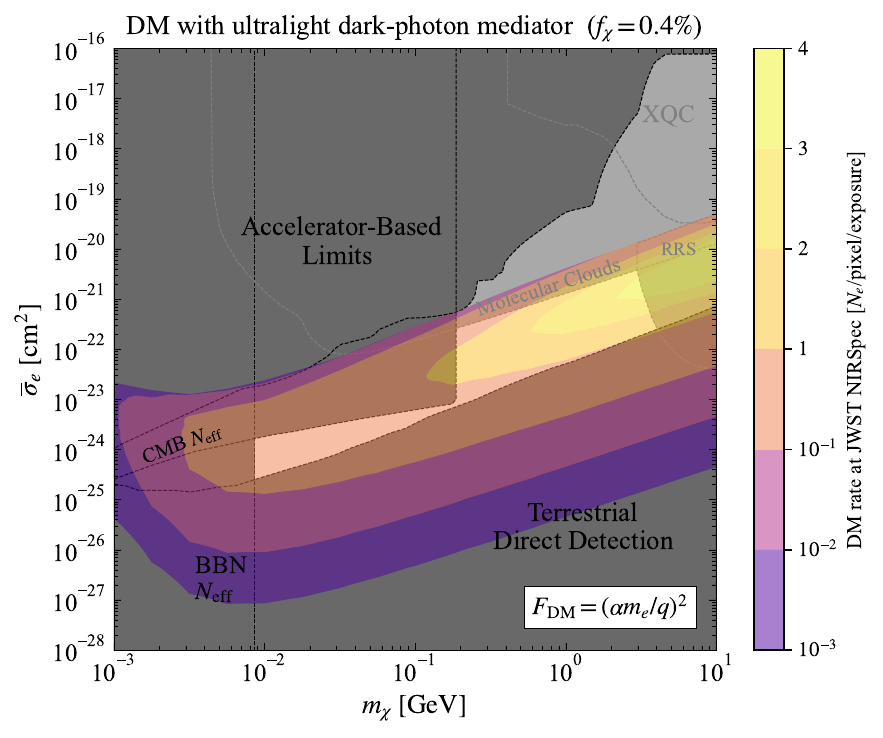}
    \caption{The allowed parameter space for DM interacting with an ultralight dark photon for a fractional DM abundance of $f_\chi = 0.4\%$.  
    The colored regions show the number of electrons created by DM-electron scattering in a NIRSpec HgCdTe detector in one full effective exposure (3559.7~seconds), divided by the number of pixels.  The effects from the detector shielding are included. Dark gray regions are bounds that are either independent or only scale mildly with $f_\chi$~\cite{Boehm:2013jpa,Munoz:2018pzp,Emken:2019tni,Chen:2002yh,Dvorkin:2013cea,Gluscevic:2017ywp,McDermott:2010pa,Slatyer:2015jla,Dolgov:2013una,Dubovsky:2003yn,Xu:2018efh,Ali-Haimoud:2015pwa,Boddy:2018wzy,Vogel:2013raa,Davidson:2000hf,Creque-Sarbinowski:2019mcm,Boehm:2013jpa,Bhoonah:2018gjb,Chang:2018rso,Davidson:2000hf,Davidson:2000hf,Prinz:1998ua,Magill:2018tbb,ArgoNeuT:2019ckq,Harnik:2020ugb,Plestid:2020kdm,PhysRevD.102.032002,SENSEI:2023gie}, while light-gray regions are bounds that scale with $f_\chi$~\cite{Rich:1987st,Mahdawi:2018euy,Prabhu:2022dtm}.}
    \label{fig:DM-rates}
\end{figure}

JWST is located at Lagrange point L2, and 
we take its velocity the same as the Earth velocity in the galactic rest frame. 
We assume the standard halo model for the DM distribution in the galactic halo~\cite{Baxter:2021pqo}, where DM follows a Maxwell-Boltzmann velocity distribution truncated at galactic escape velocity. 
However, the velocity distribution of strongly-interacting DM at the detector $f_{\rm det}(v)$ can be drastically different due to the shielding surrounding the detector, as the DM can scatter and even be stopped by the shield for very large $\overline\sigma_e$. For the dark-photon mediator, the scattering off nuclei in the shield dominates the scattering, and we neglect the interactions with electrons in the shield~\cite{Emken:2019tni}. 
The shielding consists of a camera housing, various mirrors, an optical bench, and other structures, and thus varies in different directions and is complicated to model precisely. Nevertheless, we can model it sufficiently well by assuming that the detector is surrounded on the front, top, left, and right by $\sim$20~mm of SiC, on the back  by 12~mm Mo, and on the bottom (which contains the optical bench) by $\lesssim$25~mm of SiC~\cite{Jakobsen_2022,Birkmann2022}. The 12~mm Mo-shield is roughly equivalent to 120~mm of SiC; we conservatively assume it cannot be penetrated by DM and reduce the incoming DM flux by a factor of two~\cite{JWST_communication}.
To determine $f_{\rm det}(v)$, we use \texttt{DaMaSCUS-CRUST}~\cite{Emken2018a,Emken:2018run} to perform Monte-Carlo simulations of DM passing through 20~mm of SiC. 

Fig.~\ref{fig:DM-rates} shows the number of electrons created by DM-electron scattering in a HgCdTe detector in one full exposure (3559.7~seconds), divided by the number of pixels. 
The impact of the shielding is clearly visible towards larger $\overline{\sigma}_e$.

\section{JWST Data and Masks}\label{sec:Pipeline-and-masks}
In this study, we consider the ``full-frame-dark'' images taken with the NRS2 NIRSpec detector and an opaque filter that have the longest exposure (3574.3 seconds) (we leave analyzing data from the NRS1 detector to future work). The data are obtained from the Barbara A.~Mikulski Archive for Space Telescopes~\cite{MAST-website}. We first study 20~datasets to design a preliminary analysis procedure. Ten datasets are the first integration of an observation, which we found to have significantly less noise than the other ten datasets obtained from the second integration. 
We then downloaded an additional 20 first-integration datasets, further refined our analysis procedure, and then applied it to all 30 datasets, which we list in the SM.

During a single integration of the dark image, the charges in each pixel are accumulating for $\sim$3574.3~seconds.  However, during integration, the entire image is read out non-destructively 245~times each $\sim$14.6~seconds.  Each readout will produce an image that contains a snapshot of the accumulated charges in each pixel at the time of readout; this image is called a ``frame.'' 
After the full integration, the charges are discarded. 
The signal in each pixel is measured in digital numbers (DN), which can be converted to an inferred number of charges ($N_e$) after applying a gain factor correction~\cite{JWST_gain}.

The JWST image processing pipeline corrects pixel values and flags bad pixels~\cite{Bushouse2023}. 
We apply to our datasets the subset of the pipeline  that does not affect a DM signal, which typically consists of $\lesssim$15~$e^-$ (see \S\ref{sec:DM-Signal}).  In addition, we develop custom masks to remove the impact of ``high-energy'' background events.  

\begin{figure}
    \centering
    \includegraphics[width=\linewidth]{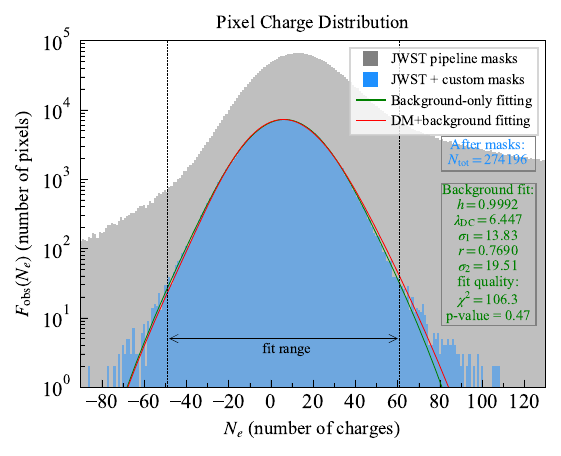}
    \caption{Pixel charge distribution $F_\text{obs}(N_e)$ for the dataset \texttt{obs\_id: jw01121008001\_02102\_00001\_nrs2} after applying a subset of the steps in the JWST image-processing pipeline (gray) and after applying our custom masks (blue) (see text for details). The green curve is the fit to the data within the fit range (gray dashed lines) of the background model, corresponding to a Poisson distributed dark current with expectation value $\lambda_{\rm DC} = 6.447~e^-$ and smeared by two Gaussian distributions ($\sigma_1$ and $\sigma_2$) to mimic the readout noise, with $r$ being their relative contributions and $h$ a normalization factor.  The red curve shows the contribution of a DM signal together with a dark-current background, for $m_\chi = 1$~GeV and $\overline\sigma_e = 3.09\times 10^{-22}$~cm$^2$, which is excluded at 95\%~CL.}
    \label{fig: pvd pipeline mask}
\end{figure}

\noindent\textbf{The JWST Image-Processing Pipeline.} 
Starting with the uncalibrated files, we process each dataset through the JWST pipeline, which corrects the pixel values and flags their properties.  From the pipeline, we use the following corrections: 
\texttt{superbias subtraction}, \texttt{reference pixel correction}, and \texttt{linearity correction}.  We use the Improved Reference Sampling and Subtraction (\texttt{IRS$^2$}) readout scheme, which produces the lowest readout noise~\cite{Rauscher_2017}.  
In addition, we remove from the subsequent analysis all pixels that have been flagged by the \texttt{saturation} step ($\sim$$1.4\%$ of pixels), as these have DN values above the saturation threshold. 
We do not implement the \texttt{dark current subtraction} step, since this could remove part of a potential DM signal. In addition, we do not implement the \texttt{jump detection} step, since it could potentially remove some DM events; instead, we develop a similar but more conservative mask ourselves (see below). 

To remove charges that accumulated before the start of the integration, we subtract from the DN value $y_i$ for each pixel in each frame $i$ ($i=1, 2, \ldots, 245$), the DN value of the corresponding pixel in the first frame, $y_1$ (we suppress a pixel index).  The total integrated DN for the full integration is then $y_\text{int}\equiv y_{245} - y_1$, which corresponds to an effective exposure of $\sim$3559.7~seconds. The noise in $y_\text{int}$ is then the superposition of the readout noise of two single frames, but we simply use ``readout noise'' to refer to the noise in $y_\text{int}$.
In Fig.~\ref{fig: pvd pipeline mask}, we show in gray the charge distribution for one dataset after applying the above JWST image-processing pipeline steps and converting DN to $N_e$ by the gain factor.  The charge distribution peaks near $\sim$15~$e^-$, with a long tail extending to very large charge values.  Although the ``dark'' images are taken with an opaque filter, each image contains many cosmic-ray and radiogenic background events, in addition to a sizable dark current.  We next apply custom masks to mitigate the high-energy backgrounds, while keeping potential DM events.

\noindent\textbf{Custom DM Masks.} 
If a pixel has a sudden increase in DN value from one frame to the next frame that is larger than expected from a DM event plus a factor of few times the readout noise, then it very likely disturbed by a ``high-energy'' background event.  We develop a \texttt{jump mask} to identify such pixels. In addition, high-energy events from cosmic rays or radiogenic backgrounds will deposit large amounts of charge in a set of connected pixels, which we call ``clusters''. We find that these clusters are also surrounded by a ``halo'' of charges, which can extend in radius to $O(100)$ pixels (such combined clusters and halos are called ``snowballs'' in the JWST user community~\cite{JWST_snowball}). We will not study these here, but based on similar halos seen around high-energy events in SENSEI data~\cite{sensei2020}, we hypothesize that the halos arise from Cherenkov radiation of high-velocity charged particles within the HgCdTe detectors~\cite{Du_2022,Du:2023soy}. The jump mask is not efficient at removing the pixels within the halo, as they do not often have a large jump. We thus design a \texttt{cluster \& halo mask} to reject pixels that are spatially correlated with a hot cluster. We use an \texttt{edge mask} to mask the edges of the image, since we find them to be less stable. Moreover, residual correlated noise remains after the \texttt{IRS$^2$} correction, appearing as ``hot'' columns in the image. We develop and implement a \texttt{hot-column mask} to account for this issue. Finally, we employ a \texttt{brightness mask} to remove pixels whose DN values are much larger than those expected by DM. Then we implement the gain correction to convert the DN of each pixel to $N_e$. 

\begin{table}[t]
\begin{tabular}{| >{\centering\arraybackslash}m{0.3\linewidth} | >{\centering\arraybackslash}m{0.3\linewidth} | >{\centering\arraybackslash}m{0.3\linewidth} |}
\hline \makecell[c]{\textbf{Mask}}& \makecell[c]{\textbf{Efficiency}}&\textbf{Pixels left}\\
\hline \makecell[c]{\texttt{jump mask}} & \makecell[c]{29.9($\pm$0.7)\%} & \makecell[c]{69.0($\pm$0.8)\%} \\
\hline \makecell[c]{\texttt{cluster \&}\\\texttt{halo mask}} & \makecell[c]{40.3($\pm$5.8)\%} & \makecell[c]{41.3($\pm$4.4)\%} \\
\hline \makecell[c]{\texttt{edge mask}} & \makecell[c]{16.5($\pm$1.3)\%} & \makecell[c]{34.5($\pm$4.1)\%} \\
\hline \makecell[c]{\texttt{hot-column mask}} & \makecell[c]{50.5($\pm$0.1)\%} & \makecell[c]{17.1($\pm$2.2)\%} \\
\hline \makecell[c]{\texttt{brightness mask}} & \makecell[c]{68.2($\pm$1.0)\%} & \makecell[c]{5.4($\pm$0.8)\%} \\
\hline
\end{tabular}
\caption{The list of masks for mitigating backgrounds, together with the fraction of pixels removed (``efficiency'') and the number of pixels left after each mask.  The errors in parentheses show the variation across the 30 datasets.}
\label{table: pipeline and mask}
\end{table}

We next list the masks in detail. 
For each mask, we tried a range of cut values with the goal of removing much of the high-$N_e$ tails of the pixel charge distribution shown in gray in Fig.~\ref{fig: pvd pipeline mask}.  
\begin{itemize}[leftmargin=*]\addtolength{\itemsep}{-0.7\baselineskip}
\vspace{-1mm}
    \item \texttt{jump mask:} This mask is inspired by the \texttt{jump detection} step in the JWST pipeline.  We mask pixels whose value increases by $>60~\rm DN$ between any two frames, i.e., which have $\left|y_{i}-y_{i-1}\right|>60~\rm DN$ for at least one frame $i=2, \ldots, 245$. 
    \item \texttt{cluster \& halo mask:} We divide the 2048$\times$2048 pixels into 128$\times$128 ``mosaic pixels,'' each of which consists of 16$\times$16 regular pixels. We then count the number of pixels flagged by the \texttt{jump mask} in each mosaic. If more than 70\% (40\%) of pixels in a given mosaic are flagged, we discard this mosaic along with the surrounding mosaics that are completely contained within a radius 3 (1) mosaics.
    \item \texttt{edge mask:} We mask 64~pixels (4~mosaic pixels) all around the edges of the image. 
    \item \texttt{hot-column mask:} The image has 128 ``mosaic columns''. Each of them is 16 regular-pixels wide. The edge mask removes a total of 8 mosaic columns. For each of the remaining 120 mosaic columns, we sum the charge in the pixels that are not flagged by the pipeline or removed by the \texttt{jump mask} and that have $20~\text{DN} < y_\text{int}\leq 400~\text{DN}$. The upper bound of 400~DN is to ignore the few pixels with very large DN, which would contribute significantly to the summation but which are unlikely due to correlated noise. We then discard the 60 mosaic columns that have the largest summed charges. 
    \item \texttt{brightness mask:} We mask any pixel with $\left|y_\text{int}\right|>150~\rm DN$ and a circular patch around it with a radius of 2 pixels. 
\end{itemize}

\begin{figure*}[t!]
    \centering
        \includegraphics[width=0.48\textwidth]{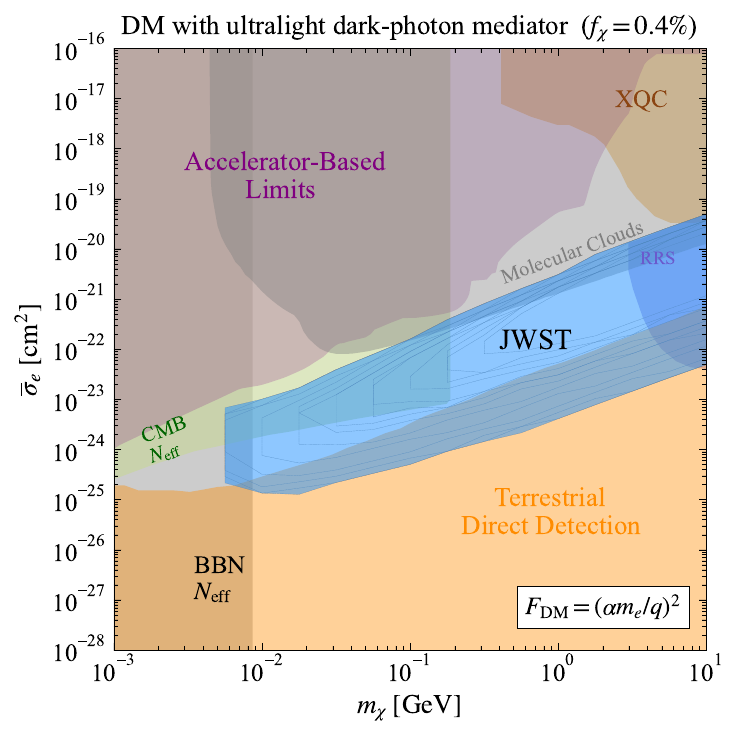}
    \hfill
        \includegraphics[width=0.48\textwidth]{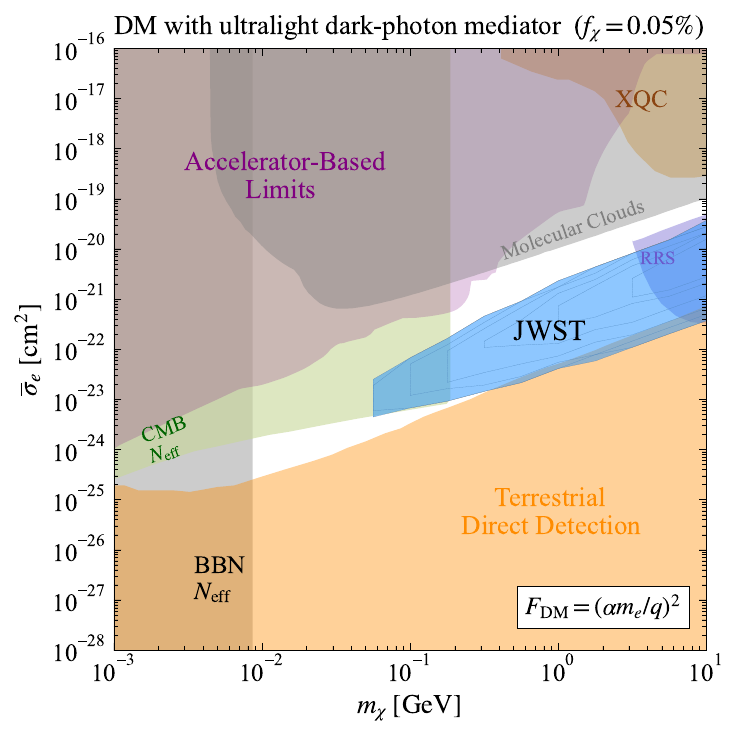}
    \caption{\textbf{Blue regions} show the 95\% CL constraints from 14 JWST NIRSpec dark images on the DM-electron scattering cross section $\overline{\sigma}_e$ versus the DM mass $m_\chi$ for DM interacting with ultralight dark-photon mediator, for a sub-component of DM with fractional abundance $f_\chi=0.4\%$ (\textbf{left}) and $f_\chi=0.05\%$ (\textbf{right}). The strongest bound is %from dataset \texttt{obs\_id: jw01121008001\_02102\_00001\_nrs2} and 
    shaded blue, while the blue lines within show the excluded regions from the other datasets (in the right plot, not all 14 JWST datasets provide a constraint at 95\%~CL).  Other limits are from~\cite{Creque-Sarbinowski:2019mcm, Munoz:2018pzp, Prinz:1998ua,Ball:2020dnx,Magill:2018tbb,ArgoNeuT:2019ckq,Plestid:2020kdm,Mahdawi:2018euy,Rich:1987st}.}
    \label{fig:JWST-constraint}
\end{figure*}

After applying the above masks, we apply the gain correction to convert each pixels value in DN to the number of electrons, $N_e$. The gain factor of ``good'' pixels in NRS2 is on average %0.996 $e^-/\rm DN$ for NRS1 and 
1.137 $e^-/\rm DN$; we remove pixels whose gain is $<0.6~e^-/{\rm DN}$. An example of the resulting charge distribution after the custom DM masks is shown in blue in Fig.~\ref{fig: pvd pipeline mask}, which looks approximately Gaussian. In particular, we see that much of the high-energy tail is removed.  In addition, the \texttt{jump mask} also reduces the low-$N_e$ tail significantly.  
We list the average mask efficiencies across all 30~datasets in Table~\ref{table: pipeline and mask}.

\section{Dark Matter Constraints}\label{sec:DM-Constraints}
The observed charge distribution of the surviving pixels, $F_{\rm obs}(N_e)$, resembles the distribution expected from a Poisson distributed dark current (DC) convolved with readout noise. The DC values of the different datasets varies significantly (see SM), which is consistent with the variation in the peak of the charge distribution after the JWST pipeline steps (gray histogram in Fig.~\ref{fig: pvd pipeline mask}), and likely due to instrument instabilities. 
Nevertheless, a DM signal is expected to slightly skew the pixel charge distribution towards higher values of $N_e$, since DM will in some pixels deposit charges well above the average DC. 
To derive a constraint, we compare the measured pixel charge distribution within  $|N_e-N_e^{\rm peak}|\leq55$ ($N_e^{\rm peak}$ is the peak of the charge distribution after masks) with the one expected from a DM signal plus a DC signal. For a given $m_\chi$ and $f_\chi$, the latter is given by 
\begin{eqnarray}\label{eq:F_model}
    F(N_e)&=&N_{\rm tot}\,h\sum_n f(n;\overline\sigma_e,\lambda_{\rm DC})g_{\rm read}(N_e,n),\\
   g_{\rm read}(N_e,n)&\equiv&  r\, \textrm{Gauss}(N_e;n,\sigma_{1})+(1-r)\textrm{Gauss}(N_e;n,\sigma_{2}) \ .\nonumber 
\end{eqnarray}
Here, $f(n;\overline\sigma_e,\lambda_{\rm DC})$
is the probability distribution function of the pixel charge distribution from the convolution of a DM signal (calculated using \S\ref{sec:DM-Signal}) and a Poisson distributed DC background with mean $\lambda_{\rm DC}$; $N_{\rm tot}$ is the number of surviving pixels; $h$ accounts for the normalization factor of the distribution not necessarily being equal to $N_{\rm tot}$ (we find $0.998<h\leq1.000$ in the background-only fit of all datasets); $g_{\rm read}$ is the readout noise, modeled with two Gaussian distributions (since the readout noise differs between different pixels) indicated by $\textrm{Gauss}(N_e;n,\sigma)$ with mean $n$ and variance $\sigma$; and the fractions $r$ and $(1-r)$ ensure the probability density function $g_{\rm read}$ integrates to unity. 
%{The normalization factor $h$ accounts for our model $F(N_e)$ fits the observed $F_{\rm obs}(N_e)$ very well in the fit range but goes slightly lower than the observation in both tails due to inadequate modeling of the tails, resulting in lower overall normalization than $N_{\rm tot}$. Nonetheless, we find $0.998<h\leq1.000$ in the background-only fit of all datasets.}
%\xhl{We introduce an additional normalization factor $h$ to account for the possibility that our model $F(N_e)$ fits the observed $F_{\rm obs}(N_e)$ very well in the fit range but goes slightly lower than the observation in both tails due to inadequate modeling of the tails, resulting in lower overall normalization than $N_{\rm tot}$. Nonetheless, we find $0.998<h\leq1.000$ in the background-only fit of all datasets.}

The background-only model is given by setting $\overline{\sigma}_e=0$ in Eq.~\ref{eq:F_model}: $f(n;0,\lambda_{\rm DC})$=Poisson$(n;\lambda_{\rm DC})$. 
We constrain our fit range to avoid the tails of the charge distribution, which likely contain non-DC backgrounds and pixels with poor noise performance, which we cannot adequately model. Fig.~\ref{fig: pvd pipeline mask} shows the background fit curve in green for one example dataset, together with the fit parameters (the fit range includes $\sim99.8\%$ of surviving pixels).  
We note that the measured DC ($\lambda_{\rm DC}=6.447~e^-$) is comparable to the rates expected from DM-electron scattering via an ultralight mediator in the open parameter space for $f_\chi$ near 0.4\%, see Fig.~\ref{fig:DM-rates}. 

We calculate $\chi^2_\text{fit}$ and $p$-value for the background model in each of the 30~datasets. We find that not all are fit well by the background model, likely due to an inadequate modeling of all noise sources. We thus select the 14 datasets with $\chi^2_\text{fit}\leq121$ ($p$-value$>0.15$), for which the background model does provide a good fit. We show the charge distributions and the background fit for the other 13 datasets in the SM.

We next use the profile likelihood ratio test to derive constraint on $\overline\sigma_e$ for different $m_\chi$ and $f_\chi$~\cite{Cowan:2010js}. For a given $m_\chi$ and $f_\chi$, we first obtain best-fit values for $(\overline\sigma_e,h,\lambda_{\rm DC},\sigma_1,\sigma_2,r)$ that maximize the likelihood. We then scan over different values of $\overline\sigma_e$ while treating $(h,\lambda_{\rm DC},\sigma_1,\sigma_2,r)$ as nuisance parameters to get the likelihood ratio with respect to the best-fit ones. 
We consider the one-sided test statistics $q_\mu$ to set the upper limit on $\overline{\sigma}_e$~\cite{Cowan:2010js}. In Fig.~\ref{fig:JWST-constraint}, we show the $95\%$ confidence level (CL) constraints for $f_\chi=0.4\%$ (left) and $f_\chi=0.05\%$ (right), together with bounds from terrestrial direct-detection, accelerator-based searches, and cosmological bounds. 
The SM contains additional details about the statistics and bounds for other values of $f_\chi$. 

We see that the JWST NIRSpec data provide novel and important bounds on strongly-interacting DM, disfavoring the allowed parameter region for $f_\chi=0.4\%$ for DM interacting with an ultralight dark photon.  However, much viable parameter space remains for smaller $f_\chi$, which can be probed by the proposed DarkNESS mission~\cite{Saffold:2024lsj}.  DarkNESS will use Skipper-CCDs, which have significantly lower and more stable noise than the NIRSpec detectors, the ability to count individual charges, and will have very little shielding, allowing for even larger cross sections to be probed. 

\begin{acknowledgments}
We especially thank Megan Hott and Aman Singal for their help in calculating the DM-electron scattering form factor for HgCdTe. 
We thank Jialu Li, Weizhe Liu, Giacinto Piacquadio, Oren Slone, and Javier Tiffenberg for useful discussions.  
We thank Maurice te Plate for information on the NIRSpec detector shielding.  
PD is supported by DOE Grant DOE-SC0010008.
RE acknowledges support from DOE Grant DE-SC0025309, Simons Investigator in Physics Award~623940, Heising-Simons Foundation Grant No.~79921, and Binational Science Foundation Grant No.~2020220.  
BR is supported by NASA as part of the James Webb Space Telescope Project.
HX is supported by 
DOE Grant DE-SC0009854, Simons Investigator in Physics Award~623940, and Binational Science Foundation Grant No.~2020220. 
\end{acknowledgments}

\begin{center}
    \large{\textbf{Supplemental Material}}
\end{center}

\setcounter{section}{0}

In this Supplemental Material, we provide additional information.  In \S\ref{sec:shielding}, we present the dark matter (DM) flux incident on the NIRSPec sensor before and after passing through the detector shielding. In \S\ref{sec:DMspectrum}, we show the expected electron recoil spectrum in the JWST NIRSpec HgCdTe detectors for various DM masses.  In \S\ref{sec:statistical}, we provide additional details about the statistical analysis of the pixel charge distribution. 
\S\ref{sec:other-fchi} presents the JWST NIRSpec DM bounds for $f_\chi=0.01\%$, 0.1\%, 10\%, and 100\%. In \S\ref{sec:sample-image}, we show sample images before and after masking.  In \S\ref{sec:data-list}, we list the JWST NIRSpec datasets on which we developed the selected JWST pipeline and custom DM masks. Finally, in \S\ref{sec: all background fittings}, we show the pixel charge distributions together with the background fits for the 14 images that pass our selection criteria and that we use to generate the constraints.

\begin{figure}[b]
    \centering
    \includegraphics[width=\linewidth]{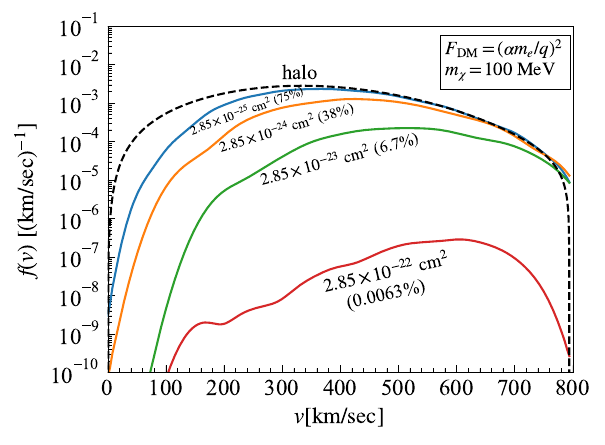}
    \caption{The DM speed distribution after passing through a 20~mm SiC shield, simulated with \texttt{DaMaSCUS-CRUST}~\cite{Emken2018a,Emken:2018run}, for a DM particle interacting with an ultralight dark-photon mediator, with a DM mass $m_\chi=100$ MeV, and for various reference cross sections, $\overline\sigma_e$. The black dashed curve is the halo DM's speed distribution in the Earth's frame. The normalization indicates the ratio of the number of particles passing the shield over the number of initially-sampled halo DM particles.}
    \label{fig: shielded speed distribution}
\end{figure}

\section{Dark matter flux before and after detector shielding}\label{sec:shielding}
For strongly-interacting DM, the DM flux incident on the detector is severely attenuated by any surrounding shielding. The shielding is described in the main paper. We use \texttt{DaMaSCUS-CRUST}~\cite{Emken2018a,Emken:2018run} to do Monte-Carlo simulations of DM passing through the shield and to obtain the DM flux that reaches the sensor. An example of the simulated results is given in~Fig.~\ref{fig: shielded speed distribution}, where the DM speed distribution, $f(v)$, is shown after the shielding, for a DM particle with mass $m_\chi=100$~MeV and reference cross sections, $\overline\sigma_e$, spanning three orders of magnitude. The black dashed curve is the normalized speed distribution in the Earth's rest frame of the DM in the halo. Indicated along with the cross sections are the ratios of the number of particles that pass through the shield to the number of initially sampled halo DM particles. 
The flux drops sharply when the cross section is larger than a specific value (here, $\sim10^{-23}\;\text{cm}^2$), which is when the shield becomes opaque to DM particles. The flux at low speeds is attenuated more than at higher speeds due to the interaction form factor $F_{\text{DM}}(q)=(\alpha m_e/q)^2$, which heavily favors soft scattering.

\begin{figure}[t]
    \centering
    \includegraphics[width=0.9\linewidth]{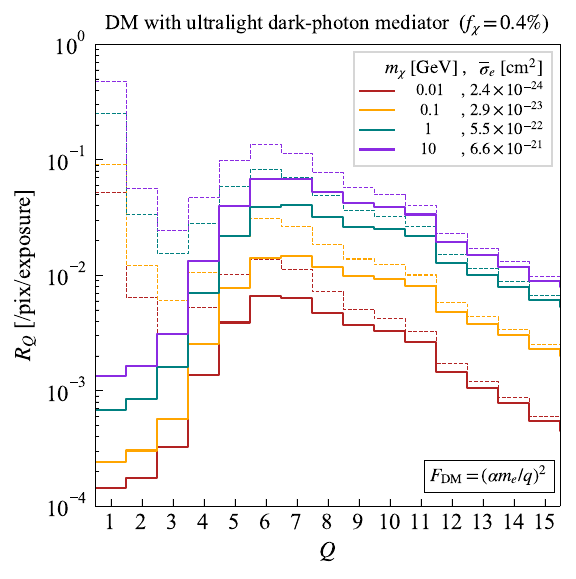}
    \caption{The spectrum of the interaction rate (per pixel per exposure of 3559.7~seconds), evaluated with different choices for $m_\chi$ and $\overline{\sigma}_e$, where each selected cross section maximizes the total signal rate at the detector for the corresponding mass. The solid lines include the screening effect in the HgCdTe on the rates. The screening is important, as can be seen by comparing the solid lines to the dashed lines, which neglect the screening effects in HgCdTe.}
    \label{fig: dm_binned_spectrum}
\end{figure}

\section{Dark matter spectrum in NIRSpec sensors}\label{sec:DMspectrum}

We use \texttt{QEDark}~\cite{Essig:2015cda} to calculate the crystal form factor $f_{\rm crys}$ for a HgTe and a CdTe target material, which is then used in Eq.~(\ref{eq: DM signal rate}) to calculate the DM-electron scattering rate. We account for the screening effects~\cite{Hochberg:2021pkt,Knapen:2021run} of the HgCdTe medium with the $1/|\epsilon(E_e,q)|^2$ factor, which is evaluated using the Lindhard formula~\cite{Dressel_Gruner_2002},
\begin{equation}
    \epsilon(E_e,q) = 1 + \frac{3\omega_p^2}{q^2v_F^2}\left(f(u_1)+f(u_2)\right)\,,
\end{equation}
with
\begin{equation}
\begin{split}
    k_F &= (3\pi^2 n_e)^{1/3}\,,\quad v_F=k_F/m_e\,,\\
    f(u)&=\frac{1}{2} + \frac{k_F}{4q}(1-u^2)\ln\frac{u+1}{u-1}\,,\\
    u_1 &=\frac{q}{2k_F} + \frac{E_e+i\Gamma}{q v_F}\,,\\
    u_2 &=\frac{q}{2k_F} - \frac{E_e+i\Gamma}{q v_F}\,,
\end{split}
\end{equation}
where $n_e$ is the number density of valence electrons, $k_F$ and $v_F$ are the Fermi momentum and velocity, and $\Gamma$ is the width of the plasmon peak, which we assume to be $0.1E_e$. We show the spectrum of the interaction rate $R_Q$ in Fig.~\ref{fig: dm_binned_spectrum} for four masses with reference cross sections that maximize the total signal rate for each mass (see Fig.~\ref{fig:DM-rates}). The solid lines are the spectrum with the $1/|\epsilon(E_e,q)|^2$ Lindhard screening factor, while the dashed lines neglect it. The screening effects are most significant for suppressing the rate for low charge yields $Q$ (soft scatterings with low energy transfer, $E_e$), but are also an $O(1)$ effect for larger $Q$.

\section{Probability density functions and statistical analysis}\label{sec:statistical}

To calculate the DM constraints, the probability distribution functions (PDF) of the pixel charge distribution needs to be understood for both the backgrounds and the DM signals. To obtain the PDF for the DM signals for a given mass $m_\chi$, reference cross section $\overline\sigma_e$, and DM fraction $f_\chi$,  we simulate DM events for an exposure of 3559.7~seconds in the pixels that survive after applying our all masks discussed in~\S\ref{sec:Pipeline-and-masks}. 
The total charge generated from DM in each pixel is given by:
\begin{equation}\label{eq: N_eDM}
    N_{e,\text{DM}} = \sum_{Q\geq1}Q\times N_Q\ ,
\end{equation}
where $N_Q$ denotes the number of events containing $Q$ electrons. For each $Q$, $N_Q$ follows a Poisson distribution
with mean $\lambda_Q=\text{exposure}\times R_Q(\overline\sigma_e)$, where $R_Q$ denotes the the rate of DM interactions that generate $Q$ electrons. We simulate the total charge in each pixel from DM by the weighted sum of Poisson samples of $N_Q$ according to Eq.~\eqref{eq: N_eDM}, and obtain the PDF of the pixel charge distribution, denoted as $f_{\text{DM}}(N_e;\overline\sigma_e)$. We note that $f_{\text{DM}}$ is not a Poisson distribution because it is a superposition of a set of Poisson samples with different weights.

We assume the backgrounds after all masks are applied  consist of a dark current that follows a single Poisson distribution:
\begin{equation}
    f_{\text{DC}}(N_e;\,\lambda_\text{DC}) = \text{Poisson}(N_{e};\,\lambda_\text{DC})\ ,
\end{equation}
where $\lambda_{\rm DC}$ denotes the mean of the Poisson distribution. We can then obtain the full PDF of both the background and the DM signal, $f(N_e;\lambda_\text{DC},\overline\sigma_e)$, by the 
convolution of $f_{\text{DM}}$ with $f_{\text{DC}}$:
\begin{equation}
    f(N_e;\,\overline\sigma_e,\lambda_\text{DC}) =\sum_{N=0}^{N_e}f_{\text{DM}}(N;\overline\sigma_e)f_{\text{DC}}(N_e-N,\lambda_\text{DC})\ .
\end{equation}
One can check that for $\overline\sigma_e=0$ (i.e., in the absence of a DM signal), $f(N_e;0,\lambda_\text{DC})=f_{\rm DC}(N_e;\lambda_\text{DC})$ as expected.

Before comparing this distribution with the data, the readout noise must be included.
%We model the readout noise with two Gaussian distributions, which mimics two sources of readout noise (see \S\ref{sec:DM-Constraints}). 
The pixel charge distribution is given by 
\begin{eqnarray}\label{eq:F_model_app}
    F(N_e)&=&N_{\rm tot}\,h\sum_n f(n;\overline\sigma_e,\lambda_{\rm DC})g_{\rm read}(N_e,n)\ ,\\
   g_{\rm read}(N_e,n)&\equiv&  r\, \textrm{Gauss}(N_e;n,\sigma_{1})+(1-r)\textrm{Gauss}(N_e;n,\sigma_{2})\ , \nonumber 
\end{eqnarray}
where $N_{\rm tot}$ is the total number of surviving pixels and $h$ is the additional normalization factor. We model the readout noise $g_{\rm read}$ using two Gaussian distributions, with $\textrm{Gauss}(N_e;n,\sigma)$ denoting the Gaussian distribution with mean $n$ and variance $\sigma$. The coefficients $r$ and $(1-r)$ ensure the probability density function $g_{\rm read}$ integrates to unity. 

\begin{figure*}[t]
    \centering
    \hspace{-20pt}
        \includegraphics[width=0.34\textwidth]{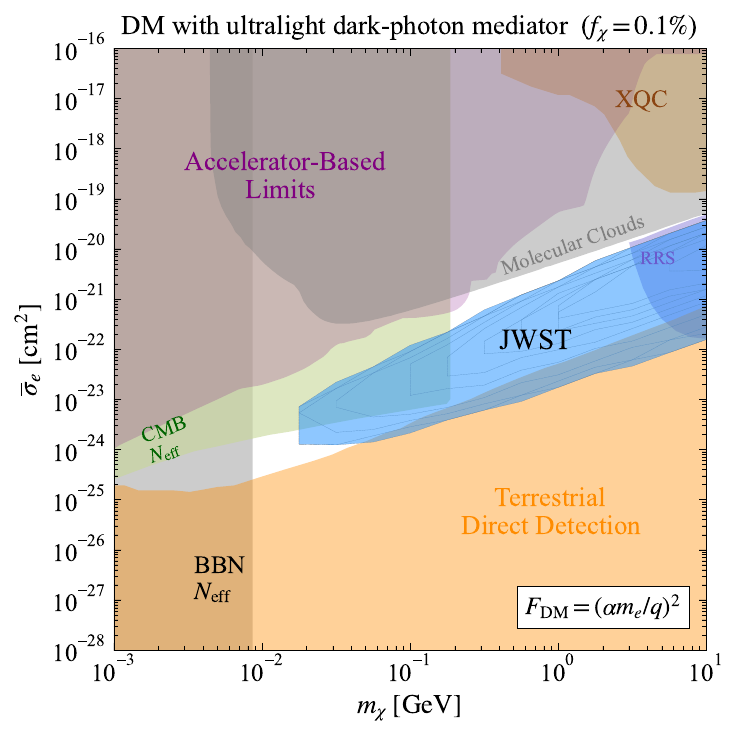}
    \hspace{-10pt}
        \includegraphics[width=0.34\textwidth]{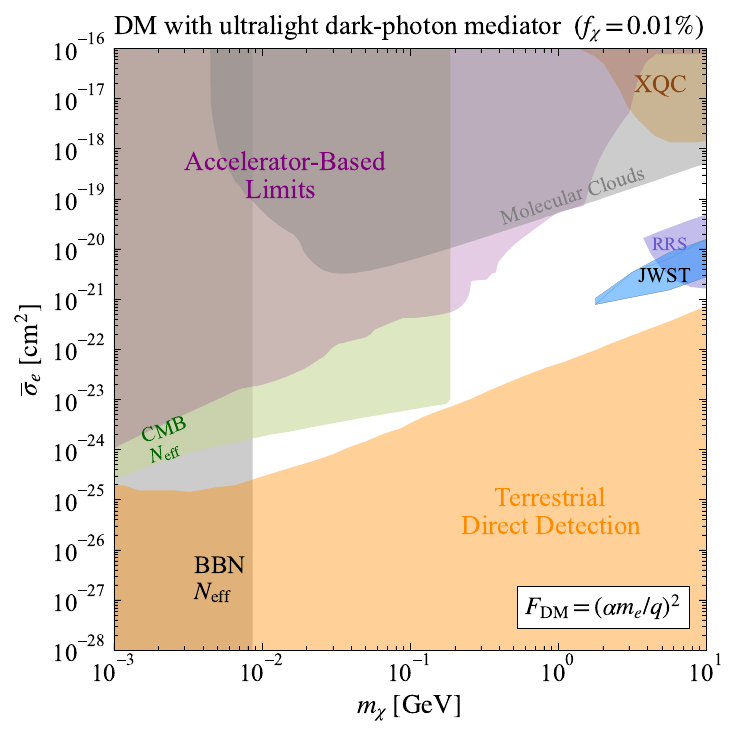}
    \hspace{-10pt}
    \includegraphics[width=0.34\textwidth]{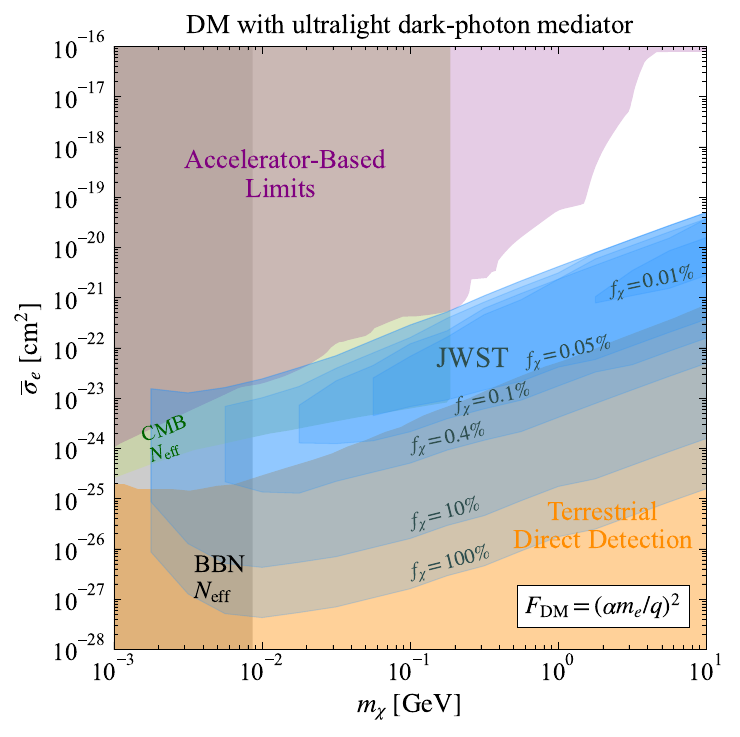}
    \hspace{-20pt}
    \caption{\textbf{Blue regions} show the 95\% CL constraints from 14 JWST NIRSpec dark images on the DM-electron scattering cross section $\overline{\sigma}_e$ versus the DM mass $m_\chi$ for DM interacting with ultralight dark-photon mediator for a sub-component of DM with fractional abundance $f_\chi=0.1\%$ (\textbf{left}) and $f_\chi=0.01\%$ (\textbf{middle}) (not all datasets provide a constraint at 95\%~CL).  The \textbf{right} plot shows how the JWST bound varies with DM subcomponent fractions $f_\chi$.  The other constraints are as in~Fig.~\ref{fig:JWST-constraint}.}
    \label{fig:JWST-constraint-other}
\end{figure*}

We derive a constraint on $\overline\sigma_e$ using the profile likelihood test analysis based on~\cite{Cowan:2010js} for each DM mass $m_\chi$ and fraction $f_\chi$. The likelihood function for the observed pixel charge distribution based on our model $F(N_e)$ is 
\begin{equation}
    L(\overline\sigma_e,\boldsymbol{\theta}) = \prod_{N_e}\text{Poisson}(F_\text{obs}(N_e);\,F(N_e))\,,
\end{equation}
where $\boldsymbol{\theta}\equiv(h,\lambda_\text{DC},\sigma_1,\sigma_2,r)$ denotes the nuisance parameters in our analysis.
The profile likelihood ratio is 
\begin{equation}
\lambda(\overline\sigma_e)=\frac{L(\overline\sigma_e,\hat{\hat{\boldsymbol{\theta}}})}{L(\hat{\overline\sigma}_e,\hat{\boldsymbol{\theta}})}\,,
\end{equation}
where $\hat{\hat{\boldsymbol{\theta}}}$ maximizes $L$ for the given $\overline\sigma_e$, and $(\hat{\overline\sigma}_e,\hat{\boldsymbol{\theta}})$ maximizes $L$ globally. We use the one-sided test statistics $q_\mu$~\cite{Cowan:2010js} to set the upper limit 
\begin{equation}
    q(\overline\sigma_e)=\left\{
    \begin{aligned}
    &-2\ln\lambda(\overline\sigma_e)\,,&s(\overline\sigma_e)>s(\hat{\overline\sigma}_e)\\
    &0\,,&s(\overline\sigma_e)\leq s(\hat{\overline\sigma}_e)
    \end{aligned}
    \right.\,,
\end{equation}
where $s(\overline\sigma_e)\equiv\sum Q\times R_Q(\overline\sigma_e)$ is the rate of total charges generated by DM. This test statistic ensures that the upper limit is set for $\overline\sigma_e$ such that $s(\overline\sigma_e)>s(\hat{\overline\sigma}_e)$, while the cases of $s(\overline\sigma_e)<s(\hat{\overline\sigma}_e)$ might indicate imperfect modeling of backgrounds and thus should not contribute to setting a limit on the signal strength. 

To derive the 95\%-CL limit on $\overline\sigma_e$, we need to find the value of $\overline\sigma_e$ for which the corresponding $p$-value equals 0.05. The data (pixel charge distribution) has many bins with a large number of entries, and we have checked that the resulting PDF of $q(\overline\sigma_e)$ agrees well with its asymptotic expression, which is given by the half-chi-square distribution. Following this asymptotic expression, we determine the 95\%-CL limit on $\overline\sigma_e$ when $q(\overline\sigma_e)=2.69$.

\section{JWST bounds for different dark matter abundances}\label{sec:other-fchi}

Fig.~\ref{fig:JWST-constraint-other} shows the 95\%~CL constraints on DM-electron scattering from the JWST NIRSpec data for fractional DM abundances $f_\chi = 0.1\%$ (left) and $f_\chi = 0.01\%$ (middle).  In Fig.~\ref{fig:JWST-constraint-other} (right), we compare how the JWST bound varies with $f_\chi$; the regions at high $\overline{\sigma}_e$ are excluded by bounds from the CMB~\cite{Boddy:2018wzy} for $f_\chi \gtrsim 0.4\%$ (not shown).

\section{Sample image after masking}\label{sec:sample-image}
\begin{figure*}[htbp]
    \centering
    \subfigure[]{
        \includegraphics[width=0.49\textwidth]{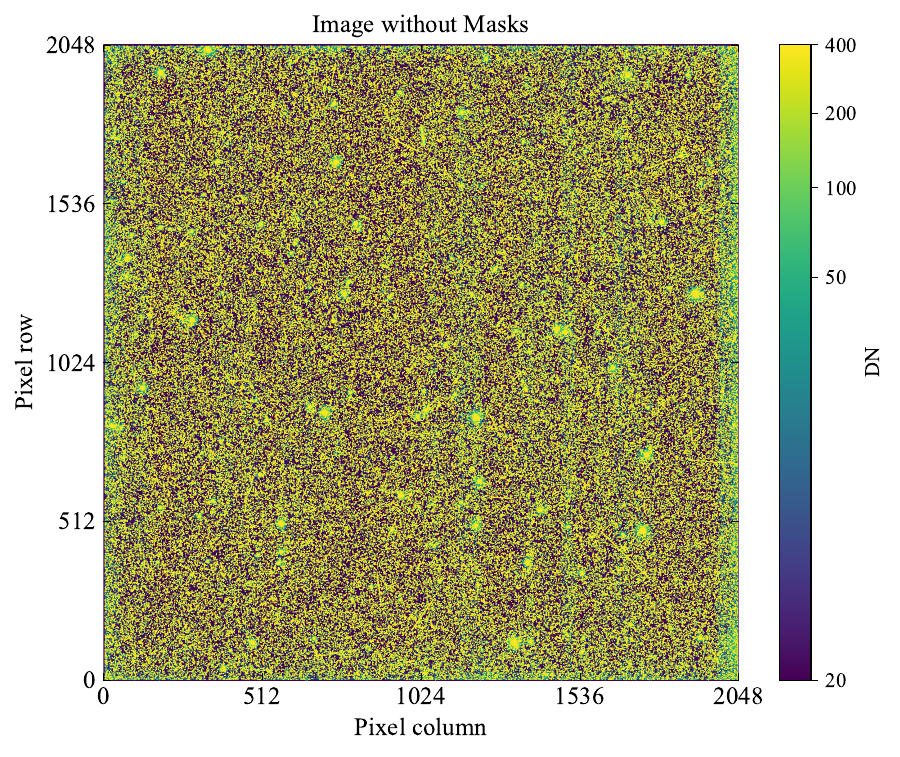}
    }
    \hspace{-10pt}
    \vspace{-20pt}
    \subfigure[]{
        \includegraphics[width=0.49\textwidth]{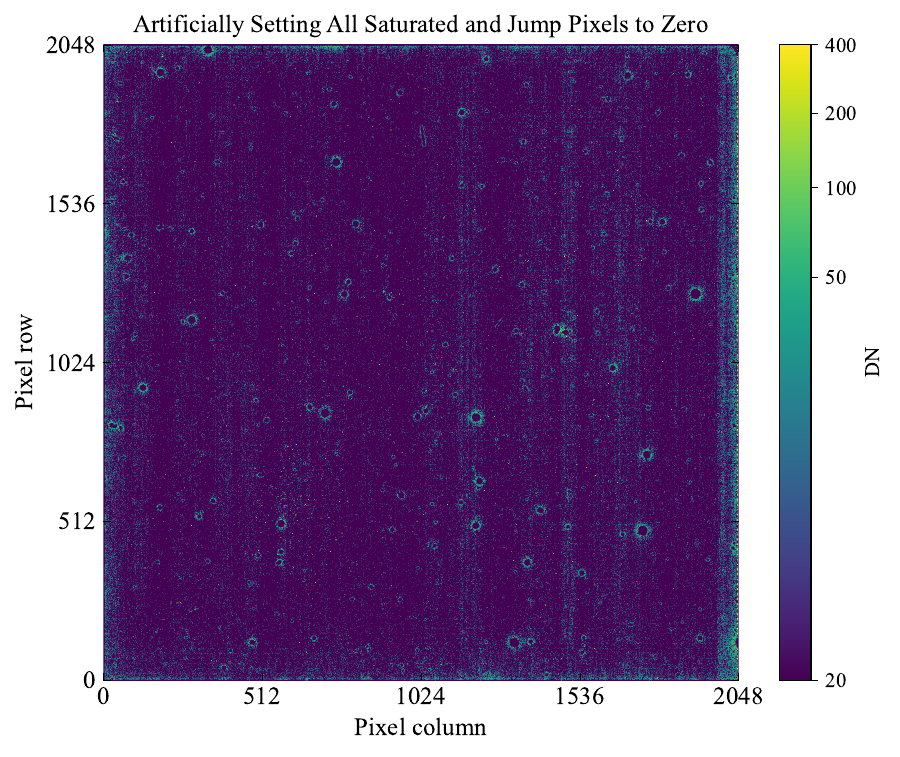}
    }
    \subfigure[]{
        \includegraphics[width=0.49\textwidth]{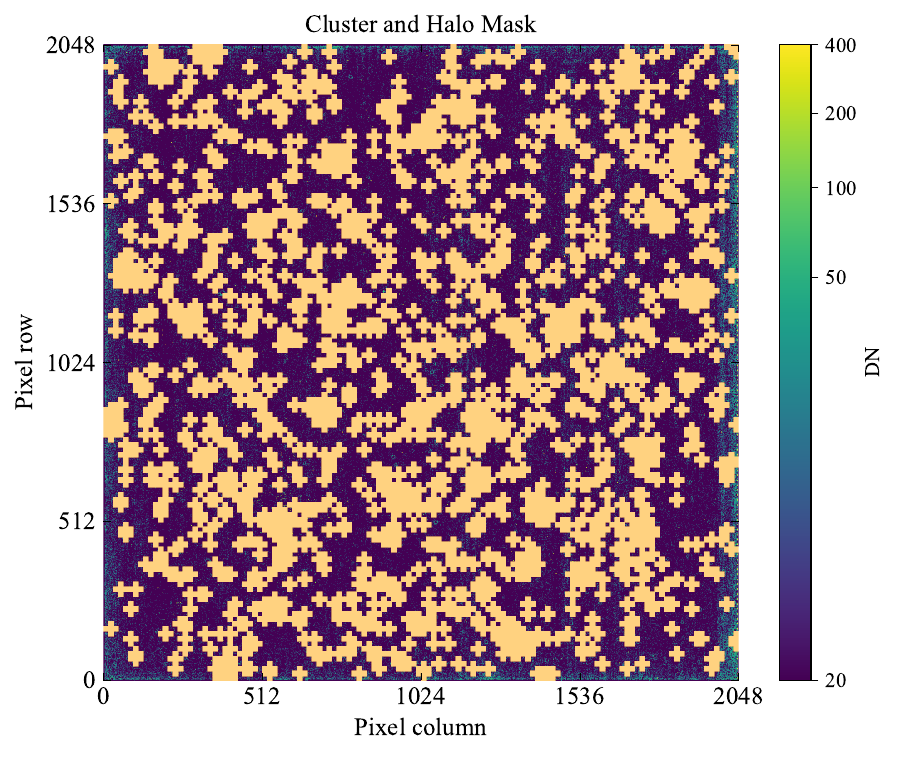}
    }
    \hspace{-10pt}
    \vspace{-20pt}
    \subfigure[]{
        \includegraphics[width=0.49\textwidth]{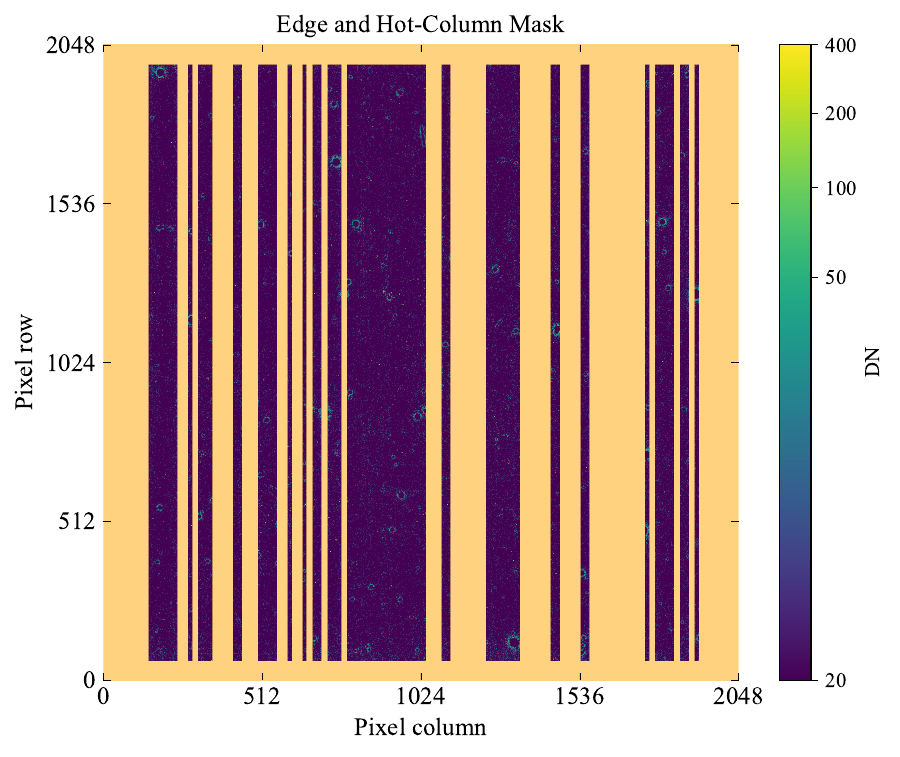}
    }
    \subfigure[]{
        \includegraphics[width=0.49\textwidth]{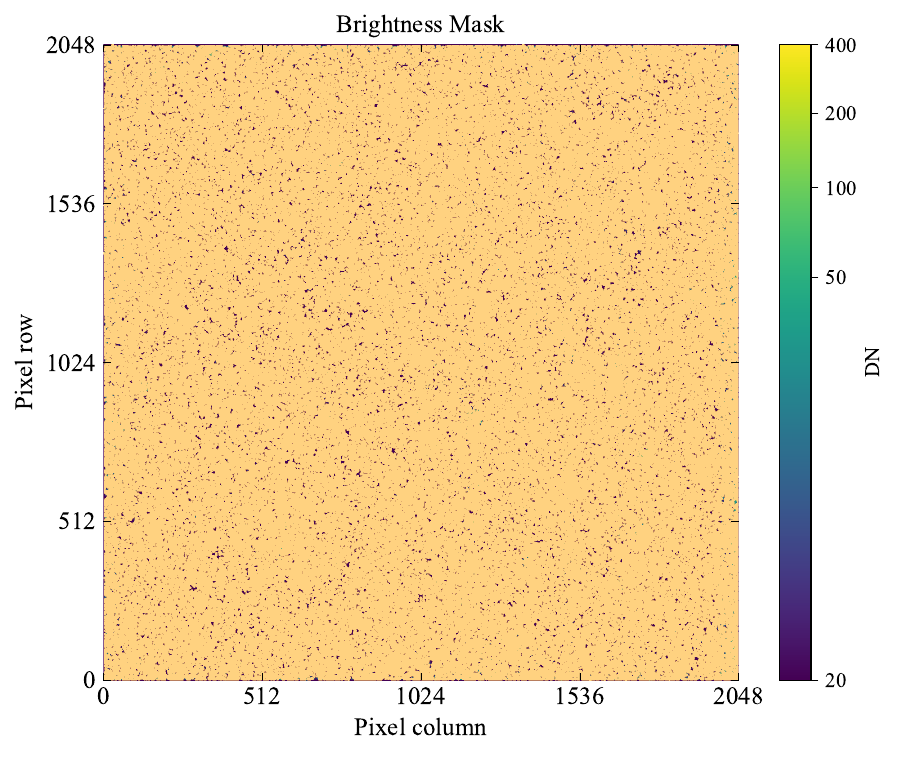}
    }
    \hspace{-10pt}
    \vspace{-15pt}
    \subfigure[]{
        \includegraphics[width=0.49\textwidth]{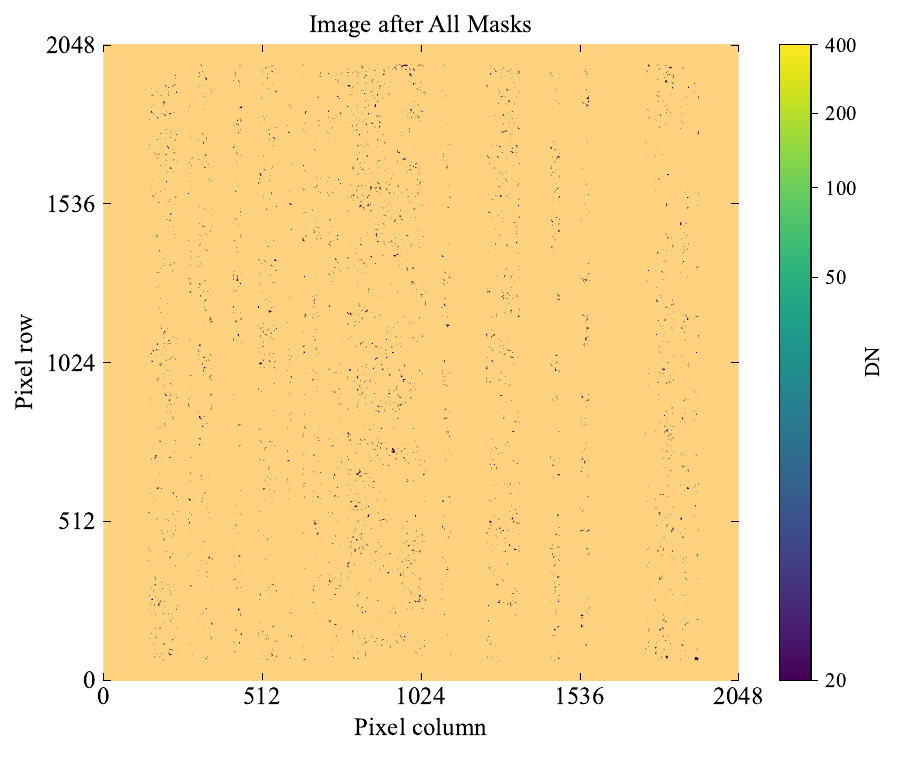}
    }
    \caption{\textbf{Top left} plot shows the image before masks. \textbf{Top right} plot shows the low-DN halos that surround the bright spots and the ``hot'' columns (for visualization purposes, this image has all pixels marked by the JWST pipeline \texttt{saturation} step or our \texttt{jump mask} set to 0~DN).  \textbf{Middle and bottom left} plots show the image after selectively applying the mask indicated  at the top of each image.  \textbf{Bottom right} plot shows the image after all masks, including the masking of pixels flagged by the JWST pipeline \texttt{saturation} or our \texttt{jump mask}.
    }
    \label{fig: masks}
\end{figure*}

Taking as an example the JWST dataset \texttt{obs\_id: jw01121008001\_02102\_00001\_nrs2}, we show in Fig.~\ref{fig: masks} the images of the total integrated DN value $y_\text{int}$ before and after applying various masks.
The first image (top left) is after applying the selected JWST pipeline steps (\texttt{superbias subtraction}, \texttt{reference pixel correction}, and \texttt{linearity correction}), without masking pixels flagged by the JWST pipeline \texttt{saturation} step, and without the custom masks. It shows many hot spots due to cosmic rays and other background events. To better demonstrate the low-DN events caused by these hot spots, we set the DN value of all pixels that are flagged by the JWST pipeline \texttt{saturation} step or our custom \texttt{jump mask} to zero in the second plot (top right), the two middle plots, and the bottom left plot. (We emphasize that this is done only here for better visualization, and not done in the data analysis.) The top right plot clearly shows the low-DN halos that surround the bright spots in the pre-mask image, which are possibly from Cherenkov radiation.  In addition, the image shows columns that contain pixels with larger values of DN than in other areas of the image, likely due to correlated noise. The pixels near the four edges also contain larger DN values than other areas. In the three following plots (middle row and bottom left), we show the effect of the custom masks individually, and in the bottom right plot, we show the image after applying all masks, including the JWST pipeline \texttt{saturation} and custom \texttt{jump mask}.

\section{List of Datasets}\label{sec:data-list}
The JWST datasets used in our analysis can be found at this link: \href{http://dx.doi.org/10.17909/q7x8-g105}{http://dx.doi.org/10.17909/q7x8-g105}. This includes the 30 datasets used for developing the masks, the fitting algorithm, and for generating constraints. 

\section{Pixel charge distributions and background-model fits}\label{sec: all background fittings}

\begin{figure*}
    \centering
    \subfigure[]{
        \includegraphics[width=0.33\textwidth]{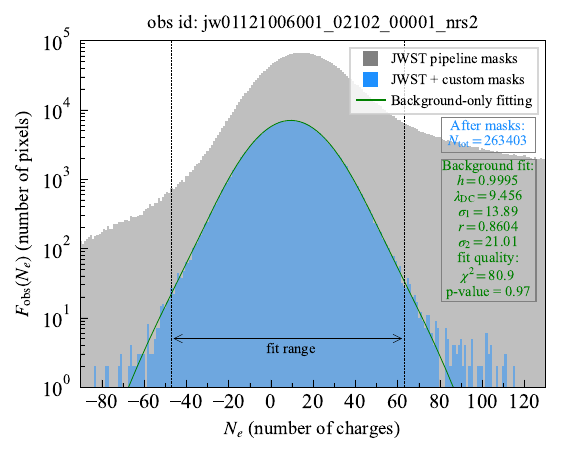}
    }
    \hspace{-16pt}
    \vspace{-26pt}
    \subfigure[]{
        \includegraphics[width=0.33\textwidth]{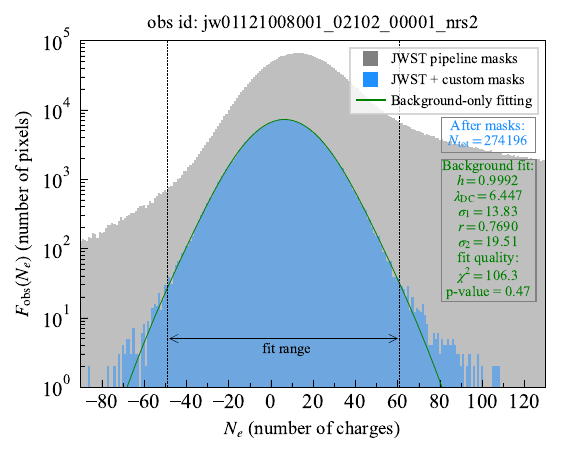}
    }
    \hspace{-16pt}
    \subfigure[]{
        \includegraphics[width=0.33\textwidth]{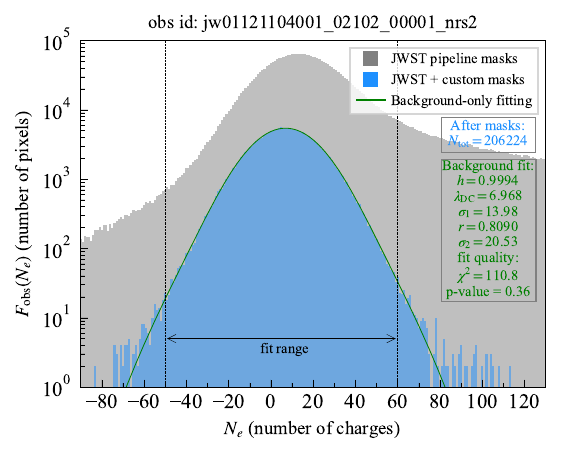}
    }
    \subfigure[]{
        \includegraphics[width=0.33\textwidth]{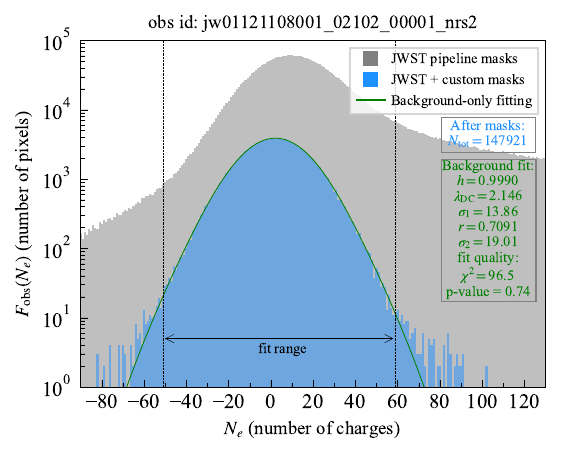}
    }
    \hspace{-16pt}
    \vspace{-26pt}
    \subfigure[]{
        \includegraphics[width=0.33\textwidth]{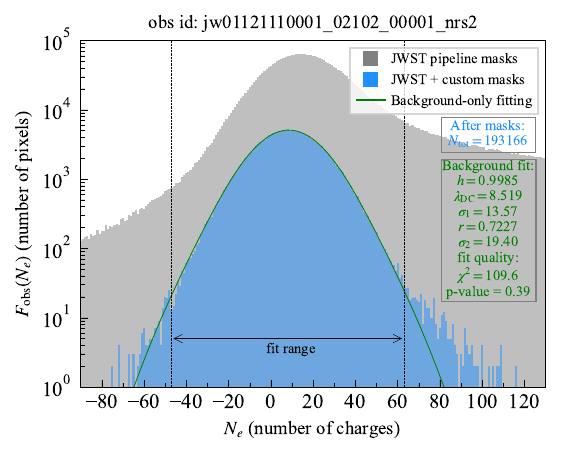}
    }
    \hspace{-16pt}
    \subfigure[]{
        \includegraphics[width=0.33\textwidth]{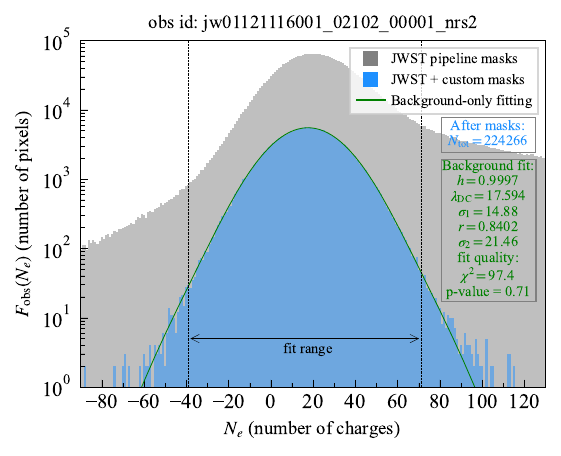}
    }
    \subfigure[]{
        \includegraphics[width=0.33\textwidth]{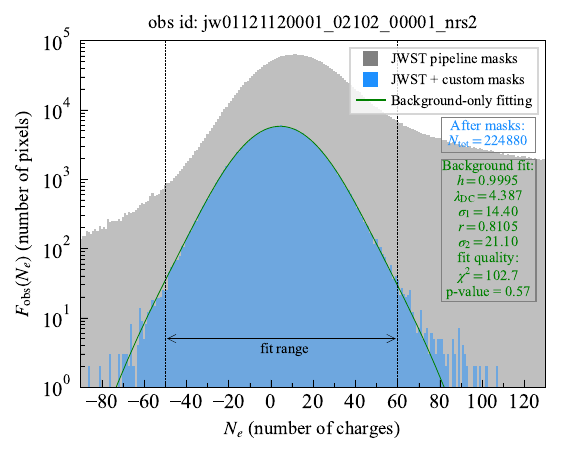}
    }
    \hspace{-16pt}
    \vspace{-26pt}
    \subfigure[]{
        \includegraphics[width=0.33\textwidth]{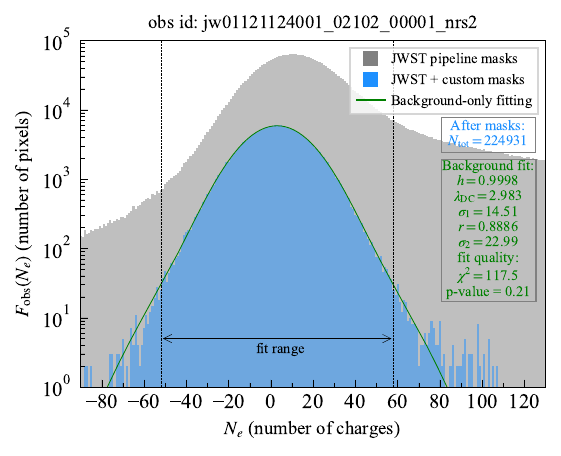}
    }
    \hspace{-16pt}
    \subfigure[]{
        \includegraphics[width=0.33\textwidth]{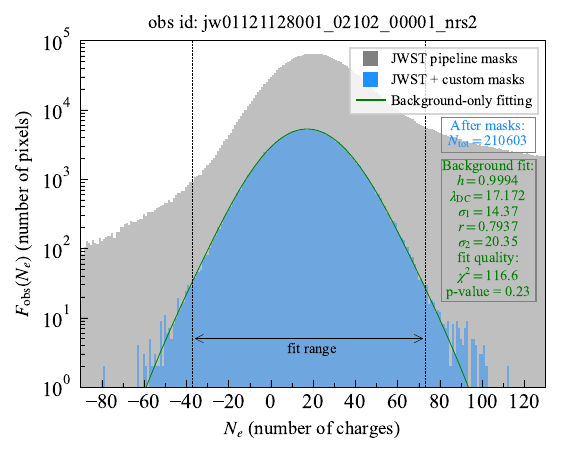}
    }
    \subfigure[]{
        \includegraphics[width=0.33\textwidth]{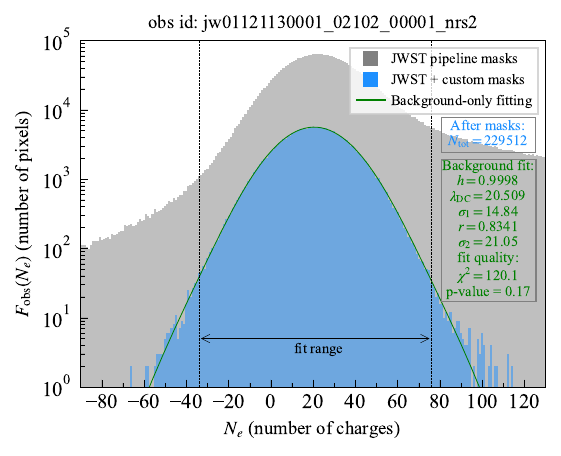}
    }
    \hspace{-16pt}
    \vspace{-26pt}
    \subfigure[]{
        \includegraphics[width=0.33\textwidth]{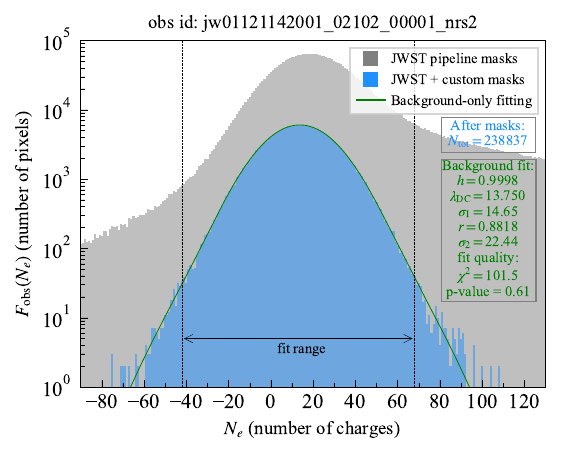}
    }
    \hspace{-16pt}
    \subfigure[]{
        \includegraphics[width=0.33\textwidth]{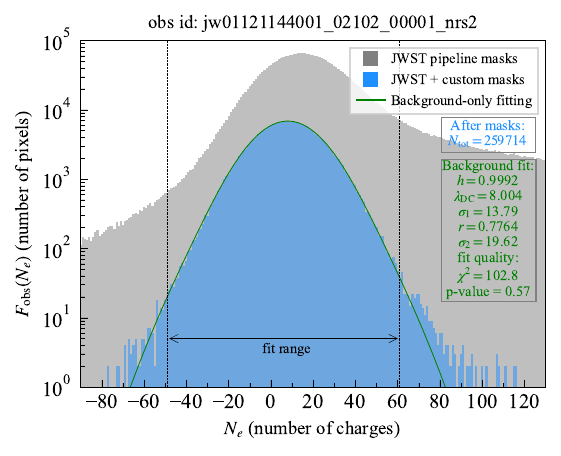}
    }
    \subfigure[]{
        \includegraphics[width=0.33\textwidth]{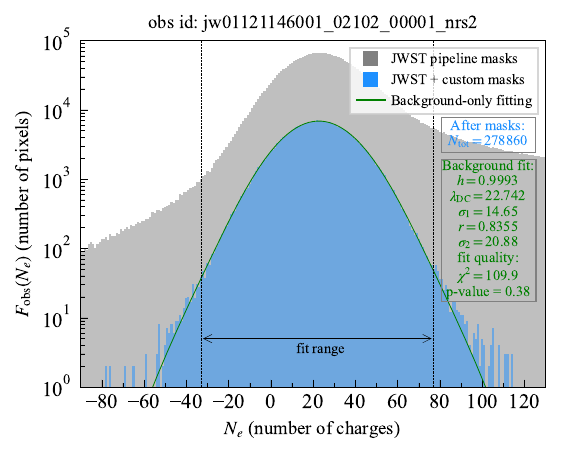}
    }
    \hspace{-16pt}
    \vspace{-26pt}
    \subfigure[]{
        \includegraphics[width=0.33\textwidth]{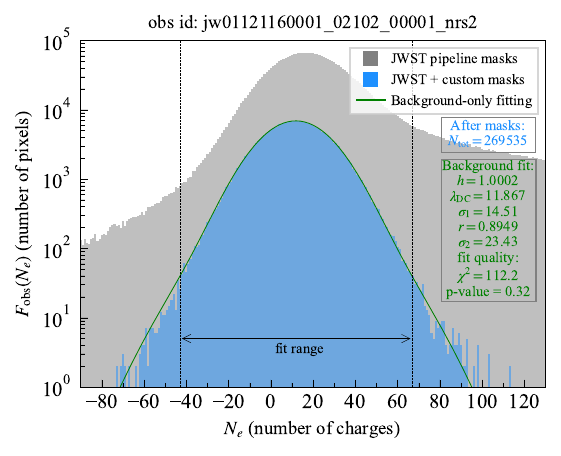}
    }
    \hspace{-8pt}
    \hspace{0.33\textwidth}
    \caption{Pixel charge distributions and background fits for the 14 datasets used in our analysis to derive constraints.  Each of these 14 datasets have a background fit that has a chi-square value of $\chi^2_\text{fit}\leq121$ ($p$-value$>0.15$). The best-fit background model for each image is shown in the legend. For more details see text and Fig.~\ref{fig: pvd pipeline mask}.}
    \label{fig: all 14 background fits}
\end{figure*}

Among the 30 datasets that we download, 14 of them are fit well by our background model with a chi-square value of $\chi_\text{fit}^2\leq121$ ($p$-value$>0.15$). These 14 datasets are used to generate the constraints. The pixel charge distributions and background fits for the 14 datasets are shown in Fig.~\ref{fig: all 14 background fits}.

\newpage

\bibliography{main.bbl}

\begin{thebibliography}{101}%
\makeatletter
\providecommand \@ifxundefined [1]{%
 \@ifx{#1\undefined}
}%
\providecommand \@ifnum [1]{%
 \ifnum #1\expandafter \@firstoftwo
 \else \expandafter \@secondoftwo
 \fi
}%
\providecommand \@ifx [1]{%
 \ifx #1\expandafter \@firstoftwo
 \else \expandafter \@secondoftwo
 \fi
}%
\providecommand \natexlab [1]{#1}%
\providecommand \enquote  [1]{``#1''}%
\providecommand \bibnamefont  [1]{#1}%
\providecommand \bibfnamefont [1]{#1}%
\providecommand \citenamefont [1]{#1}%
\providecommand \href@noop [0]{\@secondoftwo}%
\providecommand \href [0]{\begingroup \@sanitize@url \@href}%
\providecommand \@href[1]{\@@startlink{#1}\@@href}%
\providecommand \@@href[1]{\endgroup#1\@@endlink}%
\providecommand \@sanitize@url [0]{\catcode `\\12\catcode `\$12\catcode
  `\&12\catcode `\#12\catcode `\^12\catcode `\_12\catcode `\%12\relax}%
\providecommand \@@startlink[1]{}%
\providecommand \@@endlink[0]{}%
\providecommand \url  [0]{\begingroup\@sanitize@url \@url }%
\providecommand \@url [1]{\endgroup\@href {#1}{\urlprefix }}%
\providecommand \urlprefix  [0]{URL }%
\providecommand \Eprint [0]{\href }%
\providecommand \doibase [0]{http://dx.doi.org/}%
\providecommand \selectlanguage [0]{\@gobble}%
\providecommand \bibinfo  [0]{\@secondoftwo}%
\providecommand \bibfield  [0]{\@secondoftwo}%
\providecommand \translation [1]{[#1]}%
\providecommand \BibitemOpen [0]{}%
\providecommand \bibitemStop [0]{}%
\providecommand \bibitemNoStop [0]{.\EOS\space}%
\providecommand \EOS [0]{\spacefactor3000\relax}%
\providecommand \BibitemShut  [1]{\csname bibitem#1\endcsname}%
\let\auto@bib@innerbib\@empty
%</preamble>
\bibitem [{\citenamefont {Essig}\ \emph {et~al.}(2012)\citenamefont {Essig},
  \citenamefont {Mardon},\ and\ \citenamefont {Volansky}}]{Essig:2011nj}%
  \BibitemOpen
  \bibfield  {author} {\bibinfo {author} {\bibfnamefont {R.}~\bibnamefont
  {Essig}}, \bibinfo {author} {\bibfnamefont {J.}~\bibnamefont {Mardon}}, \
  and\ \bibinfo {author} {\bibfnamefont {T.}~\bibnamefont {Volansky}},\ }\href
  {\doibase 10.1103/PhysRevD.85.076007} {\bibfield  {journal} {\bibinfo
  {journal} {Phys. Rev. D}\ }\textbf {\bibinfo {volume} {85}},\ \bibinfo
  {pages} {076007} (\bibinfo {year} {2012})},\ \Eprint
  {http://arxiv.org/abs/1108.5383} {arXiv:1108.5383 [hep-ph]} \BibitemShut
  {NoStop}%
\bibitem [{\citenamefont {Tiffenberg}\ \emph {et~al.}(2017)\citenamefont
  {Tiffenberg}, \citenamefont {Sofo-Haro}, \citenamefont {Drlica-Wagner},
  \citenamefont {Essig}, \citenamefont {Guardincerri}, \citenamefont {Holland},
  \citenamefont {Volansky},\ and\ \citenamefont {Yu}}]{Tiffenberg:2017aac}%
  \BibitemOpen
  \bibfield  {author} {\bibinfo {author} {\bibfnamefont {J.}~\bibnamefont
  {Tiffenberg}}, \bibinfo {author} {\bibfnamefont {M.}~\bibnamefont
  {Sofo-Haro}}, \bibinfo {author} {\bibfnamefont {A.}~\bibnamefont
  {Drlica-Wagner}}, \bibinfo {author} {\bibfnamefont {R.}~\bibnamefont
  {Essig}}, \bibinfo {author} {\bibfnamefont {Y.}~\bibnamefont {Guardincerri}},
  \bibinfo {author} {\bibfnamefont {S.}~\bibnamefont {Holland}}, \bibinfo
  {author} {\bibfnamefont {T.}~\bibnamefont {Volansky}}, \ and\ \bibinfo
  {author} {\bibfnamefont {T.-T.}\ \bibnamefont {Yu}} (\bibinfo {collaboration}
  {SENSEI}),\ }\href {\doibase 10.1103/PhysRevLett.119.131802} {\bibfield
  {journal} {\bibinfo  {journal} {Phys. Rev. Lett.}\ }\textbf {\bibinfo
  {volume} {119}},\ \bibinfo {pages} {131802} (\bibinfo {year} {2017})},\
  \Eprint {http://arxiv.org/abs/1706.00028} {arXiv:1706.00028
  [physics.ins-det]} \BibitemShut {NoStop}%
\bibitem [{\citenamefont {Crisler}\ \emph {et~al.}(2018)\citenamefont
  {Crisler}, \citenamefont {Essig}, \citenamefont {Estrada}, \citenamefont
  {Fernandez}, \citenamefont {Tiffenberg}, \citenamefont {Sofo~haro},
  \citenamefont {Volansky},\ and\ \citenamefont {Yu}}]{Crisler:2018gci}%
  \BibitemOpen
  \bibfield  {author} {\bibinfo {author} {\bibfnamefont {M.}~\bibnamefont
  {Crisler}}, \bibinfo {author} {\bibfnamefont {R.}~\bibnamefont {Essig}},
  \bibinfo {author} {\bibfnamefont {J.}~\bibnamefont {Estrada}}, \bibinfo
  {author} {\bibfnamefont {G.}~\bibnamefont {Fernandez}}, \bibinfo {author}
  {\bibfnamefont {J.}~\bibnamefont {Tiffenberg}}, \bibinfo {author}
  {\bibfnamefont {M.}~\bibnamefont {Sofo~haro}}, \bibinfo {author}
  {\bibfnamefont {T.}~\bibnamefont {Volansky}}, \ and\ \bibinfo {author}
  {\bibfnamefont {T.-T.}\ \bibnamefont {Yu}} (\bibinfo {collaboration}
  {SENSEI}),\ }\href {\doibase 10.1103/PhysRevLett.121.061803} {\bibfield
  {journal} {\bibinfo  {journal} {Phys. Rev. Lett.}\ }\textbf {\bibinfo
  {volume} {121}},\ \bibinfo {pages} {061803} (\bibinfo {year} {2018})},\
  \Eprint {http://arxiv.org/abs/1804.00088} {arXiv:1804.00088 [hep-ex]}
  \BibitemShut {NoStop}%
\bibitem [{\citenamefont {Agnese}\ \emph {et~al.}(2018)\citenamefont {Agnese}
  \emph {et~al.}}]{Agnese:2018col}%
  \BibitemOpen
  \bibfield  {author} {\bibinfo {author} {\bibfnamefont {R.}~\bibnamefont
  {Agnese}} \emph {et~al.} (\bibinfo {collaboration} {SuperCDMS}),\ }\href
  {\doibase 10.1103/PhysRevLett.122.069901, 10.1103/PhysRevLett.121.051301}
  {\bibfield  {journal} {\bibinfo  {journal} {Phys. Rev. Lett.}\ }\textbf
  {\bibinfo {volume} {121}},\ \bibinfo {pages} {051301} (\bibinfo {year}
  {2018})},\ \bibinfo {note} {[erratum: Phys. Rev.
  Lett.122,no.6,069901(2019)]},\ \Eprint {http://arxiv.org/abs/1804.10697}
  {arXiv:1804.10697 [hep-ex]} \BibitemShut {NoStop}%
%%CITATION = ARXIV:1804.10697;%%
\bibitem [{\citenamefont {Abramoff}\ \emph {et~al.}(2019)\citenamefont
  {Abramoff} \emph {et~al.}}]{Abramoff:2019dfb}%
  \BibitemOpen
  \bibfield  {author} {\bibinfo {author} {\bibfnamefont {O.}~\bibnamefont
  {Abramoff}} \emph {et~al.} (\bibinfo {collaboration} {SENSEI}),\ }\href
  {\doibase 10.1103/PhysRevLett.122.161801} {\bibfield  {journal} {\bibinfo
  {journal} {Phys. Rev. Lett.}\ }\textbf {\bibinfo {volume} {122}},\ \bibinfo
  {pages} {161801} (\bibinfo {year} {2019})},\ \Eprint
  {http://arxiv.org/abs/1901.10478} {arXiv:1901.10478 [hep-ex]} \BibitemShut
  {NoStop}%
\bibitem [{\citenamefont {Aguilar-Arevalo}\ \emph {et~al.}(2019)\citenamefont
  {Aguilar-Arevalo} \emph {et~al.}}]{Aguilar-Arevalo:2019wdi}%
  \BibitemOpen
  \bibfield  {author} {\bibinfo {author} {\bibfnamefont {A.}~\bibnamefont
  {Aguilar-Arevalo}} \emph {et~al.} (\bibinfo {collaboration} {DAMIC}),\ }\href
  {\doibase 10.1103/PhysRevLett.123.181802} {\bibfield  {journal} {\bibinfo
  {journal} {Phys. Rev. Lett.}\ }\textbf {\bibinfo {volume} {123}},\ \bibinfo
  {pages} {181802} (\bibinfo {year} {2019})},\ \Eprint
  {http://arxiv.org/abs/1907.12628} {arXiv:1907.12628 [astro-ph.CO]}
  \BibitemShut {NoStop}%
%%CITATION = ARXIV:1907.12628;%%
\bibitem [{\citenamefont {Barak}\ \emph
  {et~al.}(2020{\natexlab{a}})\citenamefont {Barak} \emph
  {et~al.}}]{SENSEI:2020dpa}%
  \BibitemOpen
  \bibfield  {author} {\bibinfo {author} {\bibfnamefont {L.}~\bibnamefont
  {Barak}} \emph {et~al.} (\bibinfo {collaboration} {SENSEI}),\ }\href
  {\doibase 10.1103/PhysRevLett.125.171802} {\bibfield  {journal} {\bibinfo
  {journal} {Phys. Rev. Lett.}\ }\textbf {\bibinfo {volume} {125}},\ \bibinfo
  {pages} {171802} (\bibinfo {year} {2020}{\natexlab{a}})},\ \Eprint
  {http://arxiv.org/abs/2004.11378} {arXiv:2004.11378 [astro-ph.CO]}
  \BibitemShut {NoStop}%
\bibitem [{\citenamefont {Arnaud}\ \emph {et~al.}(2020)\citenamefont {Arnaud}
  \emph {et~al.}}]{Arnaud:2020svb}%
  \BibitemOpen
  \bibfield  {author} {\bibinfo {author} {\bibfnamefont {Q.}~\bibnamefont
  {Arnaud}} \emph {et~al.} (\bibinfo {collaboration} {EDELWEISS}),\ }\href
  {\doibase 10.1103/PhysRevLett.125.141301} {\bibfield  {journal} {\bibinfo
  {journal} {Phys. Rev. Lett.}\ }\textbf {\bibinfo {volume} {125}},\ \bibinfo
  {pages} {141301} (\bibinfo {year} {2020})},\ \Eprint
  {http://arxiv.org/abs/2003.01046} {arXiv:2003.01046 [astro-ph.GA]}
  \BibitemShut {NoStop}%
\bibitem [{\citenamefont {Amaral}\ \emph {et~al.}(2020)\citenamefont {Amaral}
  \emph {et~al.}}]{Amaral:2020ryn}%
  \BibitemOpen
  \bibfield  {author} {\bibinfo {author} {\bibfnamefont {D.~W.}\ \bibnamefont
  {Amaral}} \emph {et~al.} (\bibinfo {collaboration} {SuperCDMS}),\ }\href
  {\doibase 10.1103/PhysRevD.102.091101} {\bibfield  {journal} {\bibinfo
  {journal} {Phys. Rev. D}\ }\textbf {\bibinfo {volume} {102}},\ \bibinfo
  {pages} {091101} (\bibinfo {year} {2020})},\ \Eprint
  {http://arxiv.org/abs/2005.14067} {arXiv:2005.14067 [hep-ex]} \BibitemShut
  {NoStop}%
\bibitem [{\citenamefont {Arnquist}\ \emph
  {et~al.}(2023{\natexlab{a}})\citenamefont {Arnquist} \emph
  {et~al.}}]{DAMIC_2022}%
  \BibitemOpen
  \bibfield  {author} {\bibinfo {author} {\bibfnamefont {I.}~\bibnamefont
  {Arnquist}} \emph {et~al.} (\bibinfo {collaboration} {DAMIC-M}),\ }\href
  {\doibase 10.21468/SciPostPhysProc.12.014} {\bibfield  {journal} {\bibinfo
  {journal} {SciPost Phys. Proc.}\ ,\ \bibinfo {pages} {014}} (\bibinfo {year}
  {2023}{\natexlab{a}})}\BibitemShut {NoStop}%
\bibitem [{\citenamefont {Zhang}\ \emph {et~al.}(2022)\citenamefont {Zhang}
  \emph {et~al.}}]{CDEX:2022kcd}%
  \BibitemOpen
  \bibfield  {author} {\bibinfo {author} {\bibfnamefont {Z.~Y.}\ \bibnamefont
  {Zhang}} \emph {et~al.} (\bibinfo {collaboration} {CDEX}),\ }\href {\doibase
  10.1103/PhysRevLett.129.221301} {\bibfield  {journal} {\bibinfo  {journal}
  {Phys. Rev. Lett.}\ }\textbf {\bibinfo {volume} {129}},\ \bibinfo {pages}
  {221301} (\bibinfo {year} {2022})},\ \Eprint
  {http://arxiv.org/abs/2206.04128} {arXiv:2206.04128 [hep-ex]} \BibitemShut
  {NoStop}%
\bibitem [{\citenamefont {Essig}\ \emph {et~al.}(2022)\citenamefont {Essig}
  \emph {et~al.}}]{Essig:2022dfa}%
  \BibitemOpen
  \bibfield  {author} {\bibinfo {author} {\bibfnamefont {R.}~\bibnamefont
  {Essig}} \emph {et~al.},\ }in\ \href@noop {} {\emph {\bibinfo {booktitle}
  {{Snowmass 2021}}}}\ (\bibinfo {year} {2022})\ \Eprint
  {http://arxiv.org/abs/2203.08297} {arXiv:2203.08297 [hep-ph]} \BibitemShut
  {NoStop}%
\bibitem [{\citenamefont {Arnquist}\ \emph
  {et~al.}(2023{\natexlab{b}})\citenamefont {Arnquist} \emph
  {et~al.}}]{DAMIC-M:2023gxo}%
  \BibitemOpen
  \bibfield  {author} {\bibinfo {author} {\bibfnamefont {I.}~\bibnamefont
  {Arnquist}} \emph {et~al.} (\bibinfo {collaboration} {DAMIC-M}),\ }\href
  {\doibase 10.1103/PhysRevLett.130.171003} {\bibfield  {journal} {\bibinfo
  {journal} {Phys. Rev. Lett.}\ }\textbf {\bibinfo {volume} {130}},\ \bibinfo
  {pages} {171003} (\bibinfo {year} {2023}{\natexlab{b}})},\ \Eprint
  {http://arxiv.org/abs/2302.02372} {arXiv:2302.02372 [hep-ex]} \BibitemShut
  {NoStop}%
\bibitem [{\citenamefont {Arnquist}\ \emph {et~al.}(2024)\citenamefont
  {Arnquist} \emph {et~al.}}]{DAMIC-M:2023hgj}%
  \BibitemOpen
  \bibfield  {author} {\bibinfo {author} {\bibfnamefont {I.}~\bibnamefont
  {Arnquist}} \emph {et~al.} (\bibinfo {collaboration} {DAMIC-M}),\ }\href
  {\doibase 10.1103/PhysRevLett.132.101006} {\bibfield  {journal} {\bibinfo
  {journal} {Phys. Rev. Lett.}\ }\textbf {\bibinfo {volume} {132}},\ \bibinfo
  {pages} {101006} (\bibinfo {year} {2024})},\ \Eprint
  {http://arxiv.org/abs/2307.07251} {arXiv:2307.07251 [hep-ex]} \BibitemShut
  {NoStop}%
\bibitem [{\citenamefont {Adari}\ \emph {et~al.}(2023)\citenamefont {Adari}
  \emph {et~al.}}]{SENSEI:2023zdf}%
  \BibitemOpen
  \bibfield  {author} {\bibinfo {author} {\bibfnamefont {P.}~\bibnamefont
  {Adari}} \emph {et~al.} (\bibinfo {collaboration} {SENSEI}),\ }\href@noop {}
  {\  (\bibinfo {year} {2023})},\ \Eprint {http://arxiv.org/abs/2312.13342}
  {arXiv:2312.13342 [astro-ph.CO]} \BibitemShut {NoStop}%
\bibitem [{\citenamefont {Albakry}\ \emph {et~al.}(2024)\citenamefont {Albakry}
  \emph {et~al.}}]{SuperCDMS:2024yiv}%
  \BibitemOpen
  \bibfield  {author} {\bibinfo {author} {\bibfnamefont {M.~F.}\ \bibnamefont
  {Albakry}} \emph {et~al.} (\bibinfo {collaboration} {SuperCDMS}),\
  }\href@noop {} {\  (\bibinfo {year} {2024})},\ \Eprint
  {http://arxiv.org/abs/2407.08085} {arXiv:2407.08085 [hep-ex]} \BibitemShut
  {NoStop}%
\bibitem [{\citenamefont {Starkman}\ \emph {et~al.}(1990)\citenamefont
  {Starkman}, \citenamefont {Gould}, \citenamefont {Esmailzadeh},\ and\
  \citenamefont {Dimopoulos}}]{Starkman:1990nj}%
  \BibitemOpen
  \bibfield  {author} {\bibinfo {author} {\bibfnamefont {G.~D.}\ \bibnamefont
  {Starkman}}, \bibinfo {author} {\bibfnamefont {A.}~\bibnamefont {Gould}},
  \bibinfo {author} {\bibfnamefont {R.}~\bibnamefont {Esmailzadeh}}, \ and\
  \bibinfo {author} {\bibfnamefont {S.}~\bibnamefont {Dimopoulos}},\ }\href
  {\doibase 10.1103/PhysRevD.41.3594} {\bibfield  {journal} {\bibinfo
  {journal} {Phys. Rev.}\ }\textbf {\bibinfo {volume} {D41}},\ \bibinfo {pages}
  {3594} (\bibinfo {year} {1990})}\BibitemShut {NoStop}%
%%CITATION = PHRVA,D41,3594;%%
\bibitem [{\citenamefont {Zaharijas}\ and\ \citenamefont
  {Farrar}(2005)}]{Zaharijas:2004jv}%
  \BibitemOpen
  \bibfield  {author} {\bibinfo {author} {\bibfnamefont {G.}~\bibnamefont
  {Zaharijas}}\ and\ \bibinfo {author} {\bibfnamefont {G.~R.}\ \bibnamefont
  {Farrar}},\ }\href {\doibase 10.1103/PhysRevD.72.083502} {\bibfield
  {journal} {\bibinfo  {journal} {Phys. Rev.}\ }\textbf {\bibinfo {volume}
  {D72}},\ \bibinfo {pages} {083502} (\bibinfo {year} {2005})},\ \Eprint
  {http://arxiv.org/abs/astro-ph/0406531} {arXiv:astro-ph/0406531 [astro-ph]}
  \BibitemShut {NoStop}%
%%CITATION = ASTRO-PH/0406531;%%
\bibitem [{\citenamefont {Emken}\ \emph {et~al.}(2019)\citenamefont {Emken},
  \citenamefont {Essig}, \citenamefont {Kouvaris},\ and\ \citenamefont
  {Sholapurkar}}]{Emken:2019tni}%
  \BibitemOpen
  \bibfield  {author} {\bibinfo {author} {\bibfnamefont {T.}~\bibnamefont
  {Emken}}, \bibinfo {author} {\bibfnamefont {R.}~\bibnamefont {Essig}},
  \bibinfo {author} {\bibfnamefont {C.}~\bibnamefont {Kouvaris}}, \ and\
  \bibinfo {author} {\bibfnamefont {M.}~\bibnamefont {Sholapurkar}},\ }\href
  {\doibase 10.1088/1475-7516/2019/09/070} {\bibfield  {journal} {\bibinfo
  {journal} {JCAP}\ }\textbf {\bibinfo {volume} {09}},\ \bibinfo {pages} {070}
  (\bibinfo {year} {2019})},\ \Eprint {http://arxiv.org/abs/1905.06348}
  {arXiv:1905.06348 [hep-ph]} \BibitemShut {NoStop}%
\bibitem [{\citenamefont {Chen}\ \emph {et~al.}(2002)\citenamefont {Chen},
  \citenamefont {Hannestad},\ and\ \citenamefont {Scherrer}}]{Chen:2002yh}%
  \BibitemOpen
  \bibfield  {author} {\bibinfo {author} {\bibfnamefont {X.-l.}\ \bibnamefont
  {Chen}}, \bibinfo {author} {\bibfnamefont {S.}~\bibnamefont {Hannestad}}, \
  and\ \bibinfo {author} {\bibfnamefont {R.~J.}\ \bibnamefont {Scherrer}},\
  }\href {\doibase 10.1103/PhysRevD.65.123515} {\bibfield  {journal} {\bibinfo
  {journal} {Phys. Rev.}\ }\textbf {\bibinfo {volume} {D65}},\ \bibinfo {pages}
  {123515} (\bibinfo {year} {2002})},\ \Eprint
  {http://arxiv.org/abs/astro-ph/0202496} {arXiv:astro-ph/0202496 [astro-ph]}
  \BibitemShut {NoStop}%
%%CITATION = ASTRO-PH/0202496;%%
\bibitem [{\citenamefont {Dvorkin}\ \emph {et~al.}(2014)\citenamefont
  {Dvorkin}, \citenamefont {Blum},\ and\ \citenamefont
  {Kamionkowski}}]{Dvorkin:2013cea}%
  \BibitemOpen
  \bibfield  {author} {\bibinfo {author} {\bibfnamefont {C.}~\bibnamefont
  {Dvorkin}}, \bibinfo {author} {\bibfnamefont {K.}~\bibnamefont {Blum}}, \
  and\ \bibinfo {author} {\bibfnamefont {M.}~\bibnamefont {Kamionkowski}},\
  }\href {\doibase 10.1103/PhysRevD.89.023519} {\bibfield  {journal} {\bibinfo
  {journal} {Phys. Rev.}\ }\textbf {\bibinfo {volume} {D89}},\ \bibinfo {pages}
  {023519} (\bibinfo {year} {2014})},\ \Eprint {http://arxiv.org/abs/1311.2937}
  {arXiv:1311.2937 [astro-ph.CO]} \BibitemShut {NoStop}%
%%CITATION = ARXIV:1311.2937;%%
\bibitem [{\citenamefont {Gluscevic}\ and\ \citenamefont
  {Boddy}(2018)}]{Gluscevic:2017ywp}%
  \BibitemOpen
  \bibfield  {author} {\bibinfo {author} {\bibfnamefont {V.}~\bibnamefont
  {Gluscevic}}\ and\ \bibinfo {author} {\bibfnamefont {K.~K.}\ \bibnamefont
  {Boddy}},\ }\href {\doibase 10.1103/PhysRevLett.121.081301} {\bibfield
  {journal} {\bibinfo  {journal} {Phys. Rev. Lett.}\ }\textbf {\bibinfo
  {volume} {121}},\ \bibinfo {pages} {081301} (\bibinfo {year} {2018})},\
  \Eprint {http://arxiv.org/abs/1712.07133} {arXiv:1712.07133 [astro-ph.CO]}
  \BibitemShut {NoStop}%
%%CITATION = ARXIV:1712.07133;%%
\bibitem [{\citenamefont {McDermott}\ \emph {et~al.}(2011)\citenamefont
  {McDermott}, \citenamefont {Yu},\ and\ \citenamefont
  {Zurek}}]{McDermott:2010pa}%
  \BibitemOpen
  \bibfield  {author} {\bibinfo {author} {\bibfnamefont {S.~D.}\ \bibnamefont
  {McDermott}}, \bibinfo {author} {\bibfnamefont {H.-B.}\ \bibnamefont {Yu}}, \
  and\ \bibinfo {author} {\bibfnamefont {K.~M.}\ \bibnamefont {Zurek}},\ }\href
  {\doibase 10.1103/PhysRevD.83.063509} {\bibfield  {journal} {\bibinfo
  {journal} {Phys. Rev.}\ }\textbf {\bibinfo {volume} {D83}},\ \bibinfo {pages}
  {063509} (\bibinfo {year} {2011})},\ \Eprint {http://arxiv.org/abs/1011.2907}
  {arXiv:1011.2907 [hep-ph]} \BibitemShut {NoStop}%
%%CITATION = ARXIV:1011.2907;%%
\bibitem [{\citenamefont {Slatyer}(2016)}]{Slatyer:2015jla}%
  \BibitemOpen
  \bibfield  {author} {\bibinfo {author} {\bibfnamefont {T.~R.}\ \bibnamefont
  {Slatyer}},\ }\href {\doibase 10.1103/PhysRevD.93.023527} {\bibfield
  {journal} {\bibinfo  {journal} {Phys. Rev.}\ }\textbf {\bibinfo {volume}
  {D93}},\ \bibinfo {pages} {023527} (\bibinfo {year} {2016})},\ \Eprint
  {http://arxiv.org/abs/1506.03811} {arXiv:1506.03811 [hep-ph]} \BibitemShut
  {NoStop}%
%%CITATION = ARXIV:1506.03811;%%
\bibitem [{\citenamefont {Dolgov}\ \emph {et~al.}(2013)\citenamefont {Dolgov},
  \citenamefont {Dubovsky}, \citenamefont {Rubtsov},\ and\ \citenamefont
  {Tkachev}}]{Dolgov:2013una}%
  \BibitemOpen
  \bibfield  {author} {\bibinfo {author} {\bibfnamefont {A.~D.}\ \bibnamefont
  {Dolgov}}, \bibinfo {author} {\bibfnamefont {S.~L.}\ \bibnamefont
  {Dubovsky}}, \bibinfo {author} {\bibfnamefont {G.~I.}\ \bibnamefont
  {Rubtsov}}, \ and\ \bibinfo {author} {\bibfnamefont {I.~I.}\ \bibnamefont
  {Tkachev}},\ }\href {\doibase 10.1103/PhysRevD.88.117701} {\bibfield
  {journal} {\bibinfo  {journal} {Phys. Rev.}\ }\textbf {\bibinfo {volume}
  {D88}},\ \bibinfo {pages} {117701} (\bibinfo {year} {2013})},\ \Eprint
  {http://arxiv.org/abs/1310.2376} {arXiv:1310.2376 [hep-ph]} \BibitemShut
  {NoStop}%
%%CITATION = ARXIV:1310.2376;%%
\bibitem [{\citenamefont {Dubovsky}\ \emph {et~al.}(2004)\citenamefont
  {Dubovsky}, \citenamefont {Gorbunov},\ and\ \citenamefont
  {Rubtsov}}]{Dubovsky:2003yn}%
  \BibitemOpen
  \bibfield  {author} {\bibinfo {author} {\bibfnamefont {S.~L.}\ \bibnamefont
  {Dubovsky}}, \bibinfo {author} {\bibfnamefont {D.~S.}\ \bibnamefont
  {Gorbunov}}, \ and\ \bibinfo {author} {\bibfnamefont {G.~I.}\ \bibnamefont
  {Rubtsov}},\ }\href {\doibase 10.1134/1.1675909} {\bibfield  {journal}
  {\bibinfo  {journal} {JETP Lett.}\ }\textbf {\bibinfo {volume} {79}},\
  \bibinfo {pages} {1} (\bibinfo {year} {2004})},\ \bibinfo {note} {[Pisma Zh.
  Eksp. Teor. Fiz.79,3(2004)]},\ \Eprint {http://arxiv.org/abs/hep-ph/0311189}
  {arXiv:hep-ph/0311189 [hep-ph]} \BibitemShut {NoStop}%
%%CITATION = HEP-PH/0311189;%%
\bibitem [{\citenamefont {Xu}\ \emph {et~al.}(2018)\citenamefont {Xu},
  \citenamefont {Dvorkin},\ and\ \citenamefont {Chael}}]{Xu:2018efh}%
  \BibitemOpen
  \bibfield  {author} {\bibinfo {author} {\bibfnamefont {W.~L.}\ \bibnamefont
  {Xu}}, \bibinfo {author} {\bibfnamefont {C.}~\bibnamefont {Dvorkin}}, \ and\
  \bibinfo {author} {\bibfnamefont {A.}~\bibnamefont {Chael}},\ }\href
  {\doibase 10.1103/PhysRevD.97.103530} {\bibfield  {journal} {\bibinfo
  {journal} {Phys. Rev.}\ }\textbf {\bibinfo {volume} {D97}},\ \bibinfo {pages}
  {103530} (\bibinfo {year} {2018})},\ \Eprint
  {http://arxiv.org/abs/1802.06788} {arXiv:1802.06788 [astro-ph.CO]}
  \BibitemShut {NoStop}%
%%CITATION = ARXIV:1802.06788;%%
\bibitem [{\citenamefont {Ali-Haïmoud}\ \emph {et~al.}(2015)\citenamefont
  {Ali-Haïmoud}, \citenamefont {Chluba},\ and\ \citenamefont
  {Kamionkowski}}]{Ali-Haimoud:2015pwa}%
  \BibitemOpen
  \bibfield  {author} {\bibinfo {author} {\bibfnamefont {Y.}~\bibnamefont
  {Ali-Haïmoud}}, \bibinfo {author} {\bibfnamefont {J.}~\bibnamefont
  {Chluba}}, \ and\ \bibinfo {author} {\bibfnamefont {M.}~\bibnamefont
  {Kamionkowski}},\ }\href {\doibase 10.1103/PhysRevLett.115.071304} {\bibfield
   {journal} {\bibinfo  {journal} {Phys. Rev. Lett.}\ }\textbf {\bibinfo
  {volume} {115}},\ \bibinfo {pages} {071304} (\bibinfo {year} {2015})},\
  \Eprint {http://arxiv.org/abs/1506.04745} {arXiv:1506.04745 [astro-ph.CO]}
  \BibitemShut {NoStop}%
%%CITATION = ARXIV:1506.04745;%%
\bibitem [{\citenamefont {Boddy}\ \emph {et~al.}(2018)\citenamefont {Boddy},
  \citenamefont {Gluscevic}, \citenamefont {Poulin}, \citenamefont {Kovetz},
  \citenamefont {Kamionkowski},\ and\ \citenamefont {Barkana}}]{Boddy:2018wzy}%
  \BibitemOpen
  \bibfield  {author} {\bibinfo {author} {\bibfnamefont {K.~K.}\ \bibnamefont
  {Boddy}}, \bibinfo {author} {\bibfnamefont {V.}~\bibnamefont {Gluscevic}},
  \bibinfo {author} {\bibfnamefont {V.}~\bibnamefont {Poulin}}, \bibinfo
  {author} {\bibfnamefont {E.~D.}\ \bibnamefont {Kovetz}}, \bibinfo {author}
  {\bibfnamefont {M.}~\bibnamefont {Kamionkowski}}, \ and\ \bibinfo {author}
  {\bibfnamefont {R.}~\bibnamefont {Barkana}},\ }\href {\doibase
  10.1103/PhysRevD.98.123506} {\bibfield  {journal} {\bibinfo  {journal} {Phys.
  Rev. D}\ }\textbf {\bibinfo {volume} {98}},\ \bibinfo {pages} {123506}
  (\bibinfo {year} {2018})},\ \Eprint {http://arxiv.org/abs/1808.00001}
  {arXiv:1808.00001 [astro-ph.CO]} \BibitemShut {NoStop}%
\bibitem [{\citenamefont {Vogel}\ and\ \citenamefont
  {Redondo}(2014)}]{Vogel:2013raa}%
  \BibitemOpen
  \bibfield  {author} {\bibinfo {author} {\bibfnamefont {H.}~\bibnamefont
  {Vogel}}\ and\ \bibinfo {author} {\bibfnamefont {J.}~\bibnamefont
  {Redondo}},\ }\href {\doibase 10.1088/1475-7516/2014/02/029} {\bibfield
  {journal} {\bibinfo  {journal} {JCAP}\ }\textbf {\bibinfo {volume} {1402}},\
  \bibinfo {pages} {029} (\bibinfo {year} {2014})},\ \Eprint
  {http://arxiv.org/abs/1311.2600} {arXiv:1311.2600 [hep-ph]} \BibitemShut
  {NoStop}%
%%CITATION = ARXIV:1311.2600;%%
\bibitem [{\citenamefont {Davidson}\ \emph {et~al.}(2000)\citenamefont
  {Davidson}, \citenamefont {Hannestad},\ and\ \citenamefont
  {Raffelt}}]{Davidson:2000hf}%
  \BibitemOpen
  \bibfield  {author} {\bibinfo {author} {\bibfnamefont {S.}~\bibnamefont
  {Davidson}}, \bibinfo {author} {\bibfnamefont {S.}~\bibnamefont {Hannestad}},
  \ and\ \bibinfo {author} {\bibfnamefont {G.}~\bibnamefont {Raffelt}},\ }\href
  {\doibase 10.1088/1126-6708/2000/05/003} {\bibfield  {journal} {\bibinfo
  {journal} {JHEP}\ }\textbf {\bibinfo {volume} {05}},\ \bibinfo {pages} {003}
  (\bibinfo {year} {2000})},\ \Eprint {http://arxiv.org/abs/hep-ph/0001179}
  {arXiv:hep-ph/0001179 [hep-ph]} \BibitemShut {NoStop}%
%%CITATION = HEP-PH/0001179;%%
\bibitem [{\citenamefont {Creque-Sarbinowski}\ \emph
  {et~al.}(2019)\citenamefont {Creque-Sarbinowski}, \citenamefont {Ji},
  \citenamefont {Kovetz},\ and\ \citenamefont
  {Kamionkowski}}]{Creque-Sarbinowski:2019mcm}%
  \BibitemOpen
  \bibfield  {author} {\bibinfo {author} {\bibfnamefont {C.}~\bibnamefont
  {Creque-Sarbinowski}}, \bibinfo {author} {\bibfnamefont {L.}~\bibnamefont
  {Ji}}, \bibinfo {author} {\bibfnamefont {E.~D.}\ \bibnamefont {Kovetz}}, \
  and\ \bibinfo {author} {\bibfnamefont {M.}~\bibnamefont {Kamionkowski}},\
  }\href {\doibase 10.1103/PhysRevD.100.023528} {\bibfield  {journal} {\bibinfo
   {journal} {Phys. Rev. D}\ }\textbf {\bibinfo {volume} {100}},\ \bibinfo
  {pages} {023528} (\bibinfo {year} {2019})},\ \Eprint
  {http://arxiv.org/abs/1903.09154} {arXiv:1903.09154 [astro-ph.CO]}
  \BibitemShut {NoStop}%
\bibitem [{\citenamefont {Boehm}\ \emph {et~al.}(2013)\citenamefont {Boehm},
  \citenamefont {Dolan},\ and\ \citenamefont {McCabe}}]{Boehm:2013jpa}%
  \BibitemOpen
  \bibfield  {author} {\bibinfo {author} {\bibfnamefont {C.}~\bibnamefont
  {Boehm}}, \bibinfo {author} {\bibfnamefont {M.~J.}\ \bibnamefont {Dolan}}, \
  and\ \bibinfo {author} {\bibfnamefont {C.}~\bibnamefont {McCabe}},\ }\href
  {\doibase 10.1088/1475-7516/2013/08/041} {\bibfield  {journal} {\bibinfo
  {journal} {JCAP}\ }\textbf {\bibinfo {volume} {1308}},\ \bibinfo {pages}
  {041} (\bibinfo {year} {2013})},\ \Eprint {http://arxiv.org/abs/1303.6270}
  {arXiv:1303.6270 [hep-ph]} \BibitemShut {NoStop}%
%%CITATION = ARXIV:1303.6270;%%
\bibitem [{\citenamefont {Bhoonah}\ \emph {et~al.}(2019)\citenamefont
  {Bhoonah}, \citenamefont {Bramante}, \citenamefont {Elahi},\ and\
  \citenamefont {Schon}}]{Bhoonah:2018gjb}%
  \BibitemOpen
  \bibfield  {author} {\bibinfo {author} {\bibfnamefont {A.}~\bibnamefont
  {Bhoonah}}, \bibinfo {author} {\bibfnamefont {J.}~\bibnamefont {Bramante}},
  \bibinfo {author} {\bibfnamefont {F.}~\bibnamefont {Elahi}}, \ and\ \bibinfo
  {author} {\bibfnamefont {S.}~\bibnamefont {Schon}},\ }\href {\doibase
  10.1103/PhysRevD.100.023001} {\bibfield  {journal} {\bibinfo  {journal}
  {Phys. Rev.}\ }\textbf {\bibinfo {volume} {D100}},\ \bibinfo {pages} {023001}
  (\bibinfo {year} {2019})},\ \Eprint {http://arxiv.org/abs/1812.10919}
  {arXiv:1812.10919 [hep-ph]} \BibitemShut {NoStop}%
%%CITATION = ARXIV:1812.10919;%%
\bibitem [{\citenamefont {Chang}\ \emph {et~al.}(2018)\citenamefont {Chang},
  \citenamefont {Essig},\ and\ \citenamefont {McDermott}}]{Chang:2018rso}%
  \BibitemOpen
  \bibfield  {author} {\bibinfo {author} {\bibfnamefont {J.~H.}\ \bibnamefont
  {Chang}}, \bibinfo {author} {\bibfnamefont {R.}~\bibnamefont {Essig}}, \ and\
  \bibinfo {author} {\bibfnamefont {S.~D.}\ \bibnamefont {McDermott}},\ }\href
  {\doibase 10.1007/JHEP09(2018)051} {\bibfield  {journal} {\bibinfo  {journal}
  {JHEP}\ }\textbf {\bibinfo {volume} {09}},\ \bibinfo {pages} {051} (\bibinfo
  {year} {2018})},\ \Eprint {http://arxiv.org/abs/1803.00993} {arXiv:1803.00993
  [hep-ph]} \BibitemShut {NoStop}%
%%CITATION = ARXIV:1803.00993;%%
\bibitem [{\citenamefont {Prinz}\ \emph {et~al.}(1998)\citenamefont {Prinz}
  \emph {et~al.}}]{Prinz:1998ua}%
  \BibitemOpen
  \bibfield  {author} {\bibinfo {author} {\bibfnamefont {A.~A.}\ \bibnamefont
  {Prinz}} \emph {et~al.},\ }\href {\doibase 10.1103/PhysRevLett.81.1175}
  {\bibfield  {journal} {\bibinfo  {journal} {Phys. Rev. Lett.}\ }\textbf
  {\bibinfo {volume} {81}},\ \bibinfo {pages} {1175} (\bibinfo {year}
  {1998})},\ \Eprint {http://arxiv.org/abs/hep-ex/9804008}
  {arXiv:hep-ex/9804008 [hep-ex]} \BibitemShut {NoStop}%
%%CITATION = HEP-EX/9804008;%%
\bibitem [{\citenamefont {Magill}\ \emph {et~al.}(2019)\citenamefont {Magill},
  \citenamefont {Plestid}, \citenamefont {Pospelov},\ and\ \citenamefont
  {Tsai}}]{Magill:2018tbb}%
  \BibitemOpen
  \bibfield  {author} {\bibinfo {author} {\bibfnamefont {G.}~\bibnamefont
  {Magill}}, \bibinfo {author} {\bibfnamefont {R.}~\bibnamefont {Plestid}},
  \bibinfo {author} {\bibfnamefont {M.}~\bibnamefont {Pospelov}}, \ and\
  \bibinfo {author} {\bibfnamefont {Y.-D.}\ \bibnamefont {Tsai}},\ }\href
  {\doibase 10.1103/PhysRevLett.122.071801} {\bibfield  {journal} {\bibinfo
  {journal} {Phys. Rev. Lett.}\ }\textbf {\bibinfo {volume} {122}},\ \bibinfo
  {pages} {071801} (\bibinfo {year} {2019})},\ \Eprint
  {http://arxiv.org/abs/1806.03310} {arXiv:1806.03310 [hep-ph]} \BibitemShut
  {NoStop}%
%%CITATION = ARXIV:1806.03310;%%
\bibitem [{\citenamefont {Acciarri}\ \emph {et~al.}(2020)\citenamefont
  {Acciarri} \emph {et~al.}}]{ArgoNeuT:2019ckq}%
  \BibitemOpen
  \bibfield  {author} {\bibinfo {author} {\bibfnamefont {R.}~\bibnamefont
  {Acciarri}} \emph {et~al.} (\bibinfo {collaboration} {ArgoNeuT}),\ }\href
  {\doibase 10.1103/PhysRevLett.124.131801} {\bibfield  {journal} {\bibinfo
  {journal} {Phys. Rev. Lett.}\ }\textbf {\bibinfo {volume} {124}},\ \bibinfo
  {pages} {131801} (\bibinfo {year} {2020})},\ \Eprint
  {http://arxiv.org/abs/1911.07996} {arXiv:1911.07996 [hep-ex]} \BibitemShut
  {NoStop}%
\bibitem [{\citenamefont {Harnik}\ \emph {et~al.}(2021)\citenamefont {Harnik},
  \citenamefont {Plestid}, \citenamefont {Pospelov},\ and\ \citenamefont
  {Ramani}}]{Harnik:2020ugb}%
  \BibitemOpen
  \bibfield  {author} {\bibinfo {author} {\bibfnamefont {R.}~\bibnamefont
  {Harnik}}, \bibinfo {author} {\bibfnamefont {R.}~\bibnamefont {Plestid}},
  \bibinfo {author} {\bibfnamefont {M.}~\bibnamefont {Pospelov}}, \ and\
  \bibinfo {author} {\bibfnamefont {H.}~\bibnamefont {Ramani}},\ }\href
  {\doibase 10.1103/PhysRevD.103.075029} {\bibfield  {journal} {\bibinfo
  {journal} {Phys. Rev. D}\ }\textbf {\bibinfo {volume} {103}},\ \bibinfo
  {pages} {075029} (\bibinfo {year} {2021})},\ \Eprint
  {http://arxiv.org/abs/2010.11190} {arXiv:2010.11190 [hep-ph]} \BibitemShut
  {NoStop}%
\bibitem [{\citenamefont {Plestid}\ \emph {et~al.}(2020)\citenamefont
  {Plestid}, \citenamefont {Takhistov}, \citenamefont {Tsai}, \citenamefont
  {Bringmann}, \citenamefont {Kusenko},\ and\ \citenamefont
  {Pospelov}}]{Plestid:2020kdm}%
  \BibitemOpen
  \bibfield  {author} {\bibinfo {author} {\bibfnamefont {R.}~\bibnamefont
  {Plestid}}, \bibinfo {author} {\bibfnamefont {V.}~\bibnamefont {Takhistov}},
  \bibinfo {author} {\bibfnamefont {Y.-D.}\ \bibnamefont {Tsai}}, \bibinfo
  {author} {\bibfnamefont {T.}~\bibnamefont {Bringmann}}, \bibinfo {author}
  {\bibfnamefont {A.}~\bibnamefont {Kusenko}}, \ and\ \bibinfo {author}
  {\bibfnamefont {M.}~\bibnamefont {Pospelov}},\ }\href {\doibase
  10.1103/PhysRevD.102.115032} {\bibfield  {journal} {\bibinfo  {journal}
  {Phys. Rev. D}\ }\textbf {\bibinfo {volume} {102}},\ \bibinfo {pages}
  {115032} (\bibinfo {year} {2020})},\ \Eprint
  {http://arxiv.org/abs/2002.11732} {arXiv:2002.11732 [hep-ph]} \BibitemShut
  {NoStop}%
\bibitem [{\citenamefont {Ball}\ \emph
  {et~al.}(2020{\natexlab{a}})\citenamefont {Ball}, \citenamefont {Beauregard},
  \citenamefont {Brooke}, \citenamefont {Campagnari}, \citenamefont {Carrigan},
  \citenamefont {Citron}, \citenamefont {De~La~Haye}, \citenamefont {De~Roeck},
  \citenamefont {Elskens}, \citenamefont {Franco}, \citenamefont {Ezeldine},
  \citenamefont {Francis}, \citenamefont {Gastal}, \citenamefont {Ghimire},
  \citenamefont {Goldstein}, \citenamefont {Golf}, \citenamefont {Guiang},
  \citenamefont {Haas}, \citenamefont {Heller}, \citenamefont {Hill},
  \citenamefont {Lavezzo}, \citenamefont {Loos}, \citenamefont {Lowette},
  \citenamefont {Magill}, \citenamefont {Manley}, \citenamefont {Marsh},
  \citenamefont {Miller}, \citenamefont {Odegard}, \citenamefont {Saab},
  \citenamefont {Sahili}, \citenamefont {Schmitz}, \citenamefont {Setti},
  \citenamefont {Shakeshaft}, \citenamefont {Stuart}, \citenamefont
  {Swiatlowski}, \citenamefont {Yoo}, \citenamefont {Zaraket},\ and\
  \citenamefont {Zheng}}]{PhysRevD.102.032002}%
  \BibitemOpen
  \bibfield  {author} {\bibinfo {author} {\bibfnamefont {A.}~\bibnamefont
  {Ball}}, \bibinfo {author} {\bibfnamefont {G.}~\bibnamefont {Beauregard}},
  \bibinfo {author} {\bibfnamefont {J.}~\bibnamefont {Brooke}}, \bibinfo
  {author} {\bibfnamefont {C.}~\bibnamefont {Campagnari}}, \bibinfo {author}
  {\bibfnamefont {M.}~\bibnamefont {Carrigan}}, \bibinfo {author}
  {\bibfnamefont {M.}~\bibnamefont {Citron}}, \bibinfo {author} {\bibfnamefont
  {J.}~\bibnamefont {De~La~Haye}}, \bibinfo {author} {\bibfnamefont
  {A.}~\bibnamefont {De~Roeck}}, \bibinfo {author} {\bibfnamefont
  {Y.}~\bibnamefont {Elskens}}, \bibinfo {author} {\bibfnamefont {R.~E.}\
  \bibnamefont {Franco}}, \bibinfo {author} {\bibfnamefont {M.}~\bibnamefont
  {Ezeldine}}, \bibinfo {author} {\bibfnamefont {B.}~\bibnamefont {Francis}},
  \bibinfo {author} {\bibfnamefont {M.}~\bibnamefont {Gastal}}, \bibinfo
  {author} {\bibfnamefont {M.}~\bibnamefont {Ghimire}}, \bibinfo {author}
  {\bibfnamefont {J.}~\bibnamefont {Goldstein}}, \bibinfo {author}
  {\bibfnamefont {F.}~\bibnamefont {Golf}}, \bibinfo {author} {\bibfnamefont
  {J.}~\bibnamefont {Guiang}}, \bibinfo {author} {\bibfnamefont
  {A.}~\bibnamefont {Haas}}, \bibinfo {author} {\bibfnamefont {R.}~\bibnamefont
  {Heller}}, \bibinfo {author} {\bibfnamefont {C.~S.}\ \bibnamefont {Hill}},
  \bibinfo {author} {\bibfnamefont {L.}~\bibnamefont {Lavezzo}}, \bibinfo
  {author} {\bibfnamefont {R.}~\bibnamefont {Loos}}, \bibinfo {author}
  {\bibfnamefont {S.}~\bibnamefont {Lowette}}, \bibinfo {author} {\bibfnamefont
  {G.}~\bibnamefont {Magill}}, \bibinfo {author} {\bibfnamefont
  {B.}~\bibnamefont {Manley}}, \bibinfo {author} {\bibfnamefont
  {B.}~\bibnamefont {Marsh}}, \bibinfo {author} {\bibfnamefont {D.~W.}\
  \bibnamefont {Miller}}, \bibinfo {author} {\bibfnamefont {B.}~\bibnamefont
  {Odegard}}, \bibinfo {author} {\bibfnamefont {F.~R.}\ \bibnamefont {Saab}},
  \bibinfo {author} {\bibfnamefont {J.}~\bibnamefont {Sahili}}, \bibinfo
  {author} {\bibfnamefont {R.}~\bibnamefont {Schmitz}}, \bibinfo {author}
  {\bibfnamefont {F.}~\bibnamefont {Setti}}, \bibinfo {author} {\bibfnamefont
  {H.}~\bibnamefont {Shakeshaft}}, \bibinfo {author} {\bibfnamefont
  {D.}~\bibnamefont {Stuart}}, \bibinfo {author} {\bibfnamefont
  {M.}~\bibnamefont {Swiatlowski}}, \bibinfo {author} {\bibfnamefont
  {J.}~\bibnamefont {Yoo}}, \bibinfo {author} {\bibfnamefont {H.}~\bibnamefont
  {Zaraket}}, \ and\ \bibinfo {author} {\bibfnamefont {H.}~\bibnamefont
  {Zheng}},\ }\href {\doibase 10.1103/PhysRevD.102.032002} {\bibfield
  {journal} {\bibinfo  {journal} {Phys. Rev. D}\ }\textbf {\bibinfo {volume}
  {102}},\ \bibinfo {pages} {032002} (\bibinfo {year}
  {2020}{\natexlab{a}})}\BibitemShut {NoStop}%
\bibitem [{\citenamefont {Barak}\ \emph {et~al.}(2024)\citenamefont {Barak}
  \emph {et~al.}}]{SENSEI:2023gie}%
  \BibitemOpen
  \bibfield  {author} {\bibinfo {author} {\bibfnamefont {L.}~\bibnamefont
  {Barak}} \emph {et~al.} (\bibinfo {collaboration} {SENSEI}),\ }\href
  {\doibase 10.1103/PhysRevLett.133.071801} {\bibfield  {journal} {\bibinfo
  {journal} {Phys. Rev. Lett.}\ }\textbf {\bibinfo {volume} {133}},\ \bibinfo
  {pages} {071801} (\bibinfo {year} {2024})},\ \Eprint
  {http://arxiv.org/abs/2305.04964} {arXiv:2305.04964 [hep-ex]} \BibitemShut
  {NoStop}%
\bibitem [{\citenamefont {Muñoz}\ and\ \citenamefont
  {Loeb}(2018)}]{Munoz:2018pzp}%
  \BibitemOpen
  \bibfield  {author} {\bibinfo {author} {\bibfnamefont {J.~B.}\ \bibnamefont
  {Muñoz}}\ and\ \bibinfo {author} {\bibfnamefont {A.}~\bibnamefont {Loeb}},\
  }\href {\doibase 10.1038/s41586-018-0151-x} {\bibfield  {journal} {\bibinfo
  {journal} {Nature}\ }\textbf {\bibinfo {volume} {557}},\ \bibinfo {pages}
  {684} (\bibinfo {year} {2018})},\ \Eprint {http://arxiv.org/abs/1802.10094}
  {arXiv:1802.10094 [astro-ph.CO]} \BibitemShut {NoStop}%
%%CITATION = ARXIV:1802.10094;%%
\bibitem [{\citenamefont {Liu}\ \emph {et~al.}(2019)\citenamefont {Liu},
  \citenamefont {Outmezguine}, \citenamefont {Redigolo},\ and\ \citenamefont
  {Volansky}}]{Liu:2019knx}%
  \BibitemOpen
  \bibfield  {author} {\bibinfo {author} {\bibfnamefont {H.}~\bibnamefont
  {Liu}}, \bibinfo {author} {\bibfnamefont {N.~J.}\ \bibnamefont
  {Outmezguine}}, \bibinfo {author} {\bibfnamefont {D.}~\bibnamefont
  {Redigolo}}, \ and\ \bibinfo {author} {\bibfnamefont {T.}~\bibnamefont
  {Volansky}},\ }\href {\doibase 10.1103/PhysRevD.100.123011} {\bibfield
  {journal} {\bibinfo  {journal} {Phys. Rev. D}\ }\textbf {\bibinfo {volume}
  {100}},\ \bibinfo {pages} {123011} (\bibinfo {year} {2019})},\ \Eprint
  {http://arxiv.org/abs/1908.06986} {arXiv:1908.06986 [hep-ph]} \BibitemShut
  {NoStop}%
\bibitem [{\citenamefont {Rich}\ \emph {et~al.}(1987)\citenamefont {Rich},
  \citenamefont {Rocchia},\ and\ \citenamefont {Spiro}}]{Rich:1987st}%
  \BibitemOpen
  \bibfield  {author} {\bibinfo {author} {\bibfnamefont {J.}~\bibnamefont
  {Rich}}, \bibinfo {author} {\bibfnamefont {R.}~\bibnamefont {Rocchia}}, \
  and\ \bibinfo {author} {\bibfnamefont {M.}~\bibnamefont {Spiro}},\ }\bibfield
   {booktitle} {\emph {\bibinfo {booktitle} {{New and Exotic Phenomena.
  Proceedings, 7th Moriond Workshop, Les Arcs, France, January 24-31, 1987}}},\
  }\href {\doibase 10.1016/0370-2693(87)90788-X} {\bibfield  {journal}
  {\bibinfo  {journal} {Phys. Lett.}\ }\textbf {\bibinfo {volume} {B194}},\
  \bibinfo {pages} {173} (\bibinfo {year} {1987})},\ \bibinfo {note}
  {[,221(1987)]}\BibitemShut {NoStop}%
%%CITATION = PHLTA,B194,173;%%
\bibitem [{\citenamefont {Mahdawi}\ and\ \citenamefont
  {Farrar}(2018)}]{Mahdawi:2018euy}%
  \BibitemOpen
  \bibfield  {author} {\bibinfo {author} {\bibfnamefont {M.~S.}\ \bibnamefont
  {Mahdawi}}\ and\ \bibinfo {author} {\bibfnamefont {G.~R.}\ \bibnamefont
  {Farrar}},\ }\href {\doibase 10.1088/1475-7516/2018/10/007} {\bibfield
  {journal} {\bibinfo  {journal} {JCAP}\ }\textbf {\bibinfo {volume} {1810}},\
  \bibinfo {pages} {007} (\bibinfo {year} {2018})},\ \Eprint
  {http://arxiv.org/abs/1804.03073} {arXiv:1804.03073 [hep-ph]} \BibitemShut
  {NoStop}%
%%CITATION = ARXIV:1804.03073;%%
\bibitem [{\citenamefont {Prabhu}\ and\ \citenamefont
  {Blanco}(2023)}]{Prabhu:2022dtm}%
  \BibitemOpen
  \bibfield  {author} {\bibinfo {author} {\bibfnamefont {A.}~\bibnamefont
  {Prabhu}}\ and\ \bibinfo {author} {\bibfnamefont {C.}~\bibnamefont
  {Blanco}},\ }\href {\doibase 10.1103/PhysRevD.108.035035} {\bibfield
  {journal} {\bibinfo  {journal} {Phys. Rev. D}\ }\textbf {\bibinfo {volume}
  {108}},\ \bibinfo {pages} {035035} (\bibinfo {year} {2023})},\ \Eprint
  {http://arxiv.org/abs/2211.05787} {arXiv:2211.05787 [hep-ph]} \BibitemShut
  {NoStop}%
\bibitem [{\citenamefont {Buen-Abad}\ \emph {et~al.}(2022)\citenamefont
  {Buen-Abad}, \citenamefont {Essig}, \citenamefont {McKeen},\ and\
  \citenamefont {Zhong}}]{Buen-Abad:2021mvc}%
  \BibitemOpen
  \bibfield  {author} {\bibinfo {author} {\bibfnamefont {M.~A.}\ \bibnamefont
  {Buen-Abad}}, \bibinfo {author} {\bibfnamefont {R.}~\bibnamefont {Essig}},
  \bibinfo {author} {\bibfnamefont {D.}~\bibnamefont {McKeen}}, \ and\ \bibinfo
  {author} {\bibfnamefont {Y.-M.}\ \bibnamefont {Zhong}},\ }\href {\doibase
  10.1016/j.physrep.2022.02.006} {\bibfield  {journal} {\bibinfo  {journal}
  {Phys. Rept.}\ }\textbf {\bibinfo {volume} {961}},\ \bibinfo {pages} {1}
  (\bibinfo {year} {2022})},\ \Eprint {http://arxiv.org/abs/2107.12377}
  {arXiv:2107.12377 [astro-ph.CO]} \BibitemShut {NoStop}%
\bibitem [{\citenamefont {Nguyen}\ \emph {et~al.}(2021)\citenamefont {Nguyen},
  \citenamefont {Sarnaaik}, \citenamefont {Boddy}, \citenamefont {Nadler},\
  and\ \citenamefont {Gluscevic}}]{Nguyen:2021cnb}%
  \BibitemOpen
  \bibfield  {author} {\bibinfo {author} {\bibfnamefont {D.~V.}\ \bibnamefont
  {Nguyen}}, \bibinfo {author} {\bibfnamefont {D.}~\bibnamefont {Sarnaaik}},
  \bibinfo {author} {\bibfnamefont {K.~K.}\ \bibnamefont {Boddy}}, \bibinfo
  {author} {\bibfnamefont {E.~O.}\ \bibnamefont {Nadler}}, \ and\ \bibinfo
  {author} {\bibfnamefont {V.}~\bibnamefont {Gluscevic}},\ }\href {\doibase
  10.1103/PhysRevD.104.103521} {\bibfield  {journal} {\bibinfo  {journal}
  {Phys. Rev. D}\ }\textbf {\bibinfo {volume} {104}},\ \bibinfo {pages}
  {103521} (\bibinfo {year} {2021})},\ \Eprint
  {http://arxiv.org/abs/2107.12380} {arXiv:2107.12380 [astro-ph.CO]}
  \BibitemShut {NoStop}%
\bibitem [{\citenamefont {Kelly}\ and\ \citenamefont
  {Tsai}(2019)}]{Kelly:2018brz}%
  \BibitemOpen
  \bibfield  {author} {\bibinfo {author} {\bibfnamefont {K.~J.}\ \bibnamefont
  {Kelly}}\ and\ \bibinfo {author} {\bibfnamefont {Y.-D.}\ \bibnamefont
  {Tsai}},\ }\href {\doibase 10.1103/PhysRevD.100.015043} {\bibfield  {journal}
  {\bibinfo  {journal} {Phys. Rev. D}\ }\textbf {\bibinfo {volume} {100}},\
  \bibinfo {pages} {015043} (\bibinfo {year} {2019})},\ \Eprint
  {http://arxiv.org/abs/1812.03998} {arXiv:1812.03998 [hep-ph]} \BibitemShut
  {NoStop}%
\bibitem [{\citenamefont {Gninenko}\ \emph {et~al.}(2019)\citenamefont
  {Gninenko}, \citenamefont {Kirpichnikov},\ and\ \citenamefont
  {Krasnikov}}]{Gninenko:2018ter}%
  \BibitemOpen
  \bibfield  {author} {\bibinfo {author} {\bibfnamefont {S.~N.}\ \bibnamefont
  {Gninenko}}, \bibinfo {author} {\bibfnamefont {D.~V.}\ \bibnamefont
  {Kirpichnikov}}, \ and\ \bibinfo {author} {\bibfnamefont {N.~V.}\
  \bibnamefont {Krasnikov}},\ }\href {\doibase 10.1103/PhysRevD.100.035003}
  {\bibfield  {journal} {\bibinfo  {journal} {Phys. Rev. D}\ }\textbf {\bibinfo
  {volume} {100}},\ \bibinfo {pages} {035003} (\bibinfo {year} {2019})},\
  \Eprint {http://arxiv.org/abs/1810.06856} {arXiv:1810.06856 [hep-ph]}
  \BibitemShut {NoStop}%
\bibitem [{\citenamefont {Foroughi-Abari}\ \emph {et~al.}(2021)\citenamefont
  {Foroughi-Abari}, \citenamefont {Kling},\ and\ \citenamefont
  {Tsai}}]{Foroughi-Abari:2020qar}%
  \BibitemOpen
  \bibfield  {author} {\bibinfo {author} {\bibfnamefont {S.}~\bibnamefont
  {Foroughi-Abari}}, \bibinfo {author} {\bibfnamefont {F.}~\bibnamefont
  {Kling}}, \ and\ \bibinfo {author} {\bibfnamefont {Y.-D.}\ \bibnamefont
  {Tsai}},\ }\href {\doibase 10.1103/PhysRevD.104.035014} {\bibfield  {journal}
  {\bibinfo  {journal} {Phys. Rev. D}\ }\textbf {\bibinfo {volume} {104}},\
  \bibinfo {pages} {035014} (\bibinfo {year} {2021})},\ \Eprint
  {http://arxiv.org/abs/2010.07941} {arXiv:2010.07941 [hep-ph]} \BibitemShut
  {NoStop}%
\bibitem [{\citenamefont {Perez}\ \emph {et~al.}(2024)\citenamefont {Perez}
  \emph {et~al.}}]{Oscura:2023qch}%
  \BibitemOpen
  \bibfield  {author} {\bibinfo {author} {\bibfnamefont {S.}~\bibnamefont
  {Perez}} \emph {et~al.} (\bibinfo {collaboration} {Oscura}),\ }\href
  {\doibase 10.1007/JHEP02(2024)072} {\bibfield  {journal} {\bibinfo  {journal}
  {JHEP}\ }\textbf {\bibinfo {volume} {02}},\ \bibinfo {pages} {072} (\bibinfo
  {year} {2024})},\ \Eprint {http://arxiv.org/abs/2304.08625} {arXiv:2304.08625
  [hep-ex]} \BibitemShut {NoStop}%
\bibitem [{\citenamefont {de~Montigny}\ \emph {et~al.}(2023)\citenamefont
  {de~Montigny}, \citenamefont {Ouimet}, \citenamefont {Pinfold}, \citenamefont
  {Shaa},\ and\ \citenamefont {Staelens}}]{deMontigny:2023qft}%
  \BibitemOpen
  \bibfield  {author} {\bibinfo {author} {\bibfnamefont {M.}~\bibnamefont
  {de~Montigny}}, \bibinfo {author} {\bibfnamefont {P.-P.~A.}\ \bibnamefont
  {Ouimet}}, \bibinfo {author} {\bibfnamefont {J.}~\bibnamefont {Pinfold}},
  \bibinfo {author} {\bibfnamefont {A.}~\bibnamefont {Shaa}}, \ and\ \bibinfo
  {author} {\bibfnamefont {M.}~\bibnamefont {Staelens}},\ }\href@noop {} {\
  (\bibinfo {year} {2023})},\ \Eprint {http://arxiv.org/abs/2307.07855}
  {arXiv:2307.07855 [hep-ph]} \BibitemShut {NoStop}%
\bibitem [{\citenamefont {Kalliokoski}\ \emph {et~al.}(2024)\citenamefont
  {Kalliokoski}, \citenamefont {Mitsou}, \citenamefont {de~Montigny},
  \citenamefont {Mukhopadhyay}, \citenamefont {Ouimet}, \citenamefont
  {Pinfold}, \citenamefont {Shaa},\ and\ \citenamefont
  {Staelens}}]{Kalliokoski:2023cgw}%
  \BibitemOpen
  \bibfield  {author} {\bibinfo {author} {\bibfnamefont {M.}~\bibnamefont
  {Kalliokoski}}, \bibinfo {author} {\bibfnamefont {V.~A.}\ \bibnamefont
  {Mitsou}}, \bibinfo {author} {\bibfnamefont {M.}~\bibnamefont {de~Montigny}},
  \bibinfo {author} {\bibfnamefont {A.}~\bibnamefont {Mukhopadhyay}}, \bibinfo
  {author} {\bibfnamefont {P.-P.~A.}\ \bibnamefont {Ouimet}}, \bibinfo {author}
  {\bibfnamefont {J.}~\bibnamefont {Pinfold}}, \bibinfo {author} {\bibfnamefont
  {A.}~\bibnamefont {Shaa}}, \ and\ \bibinfo {author} {\bibfnamefont
  {M.}~\bibnamefont {Staelens}},\ }\href {\doibase 10.1007/JHEP04(2024)137}
  {\bibfield  {journal} {\bibinfo  {journal} {JHEP}\ }\textbf {\bibinfo
  {volume} {04}},\ \bibinfo {pages} {137} (\bibinfo {year} {2024})},\ \Eprint
  {http://arxiv.org/abs/2311.02185} {arXiv:2311.02185 [hep-ph]} \BibitemShut
  {NoStop}%
\bibitem [{\citenamefont {Tsai}\ \emph {et~al.}(2024)\citenamefont {Tsai},
  \citenamefont {Hwang}, \citenamefont {Schmitz}, \citenamefont {Citron},
  \citenamefont {Gunthoti}, \citenamefont {Steenis}, \citenamefont {Jeong},
  \citenamefont {Moon}, \citenamefont {Yoo},\ and\ \citenamefont
  {Liu}}]{Tsai:2024wdh}%
  \BibitemOpen
  \bibfield  {author} {\bibinfo {author} {\bibfnamefont {Y.-D.}\ \bibnamefont
  {Tsai}}, \bibinfo {author} {\bibfnamefont {I.}~\bibnamefont {Hwang}},
  \bibinfo {author} {\bibfnamefont {R.}~\bibnamefont {Schmitz}}, \bibinfo
  {author} {\bibfnamefont {M.}~\bibnamefont {Citron}}, \bibinfo {author}
  {\bibfnamefont {K.}~\bibnamefont {Gunthoti}}, \bibinfo {author}
  {\bibfnamefont {J.}~\bibnamefont {Steenis}}, \bibinfo {author} {\bibfnamefont
  {H.}~\bibnamefont {Jeong}}, \bibinfo {author} {\bibfnamefont
  {H.}~\bibnamefont {Moon}}, \bibinfo {author} {\bibfnamefont {J.~H.}\
  \bibnamefont {Yoo}}, \ and\ \bibinfo {author} {\bibfnamefont {M.~X.}\
  \bibnamefont {Liu}},\ }\href@noop {} {\  (\bibinfo {year} {2024})},\ \Eprint
  {http://arxiv.org/abs/2407.07142} {arXiv:2407.07142 [hep-ph]} \BibitemShut
  {NoStop}%
\bibitem [{\citenamefont {Pospelov}\ and\ \citenamefont
  {Ramani}(2021)}]{Pospelov:2020ktu}%
  \BibitemOpen
  \bibfield  {author} {\bibinfo {author} {\bibfnamefont {M.}~\bibnamefont
  {Pospelov}}\ and\ \bibinfo {author} {\bibfnamefont {H.}~\bibnamefont
  {Ramani}},\ }\href {\doibase 10.1103/PhysRevD.103.115031} {\bibfield
  {journal} {\bibinfo  {journal} {Phys. Rev. D}\ }\textbf {\bibinfo {volume}
  {103}},\ \bibinfo {pages} {115031} (\bibinfo {year} {2021})},\ \Eprint
  {http://arxiv.org/abs/2012.03957} {arXiv:2012.03957 [hep-ph]} \BibitemShut
  {NoStop}%
\bibitem [{\citenamefont {Budker}\ \emph {et~al.}(2022)\citenamefont {Budker},
  \citenamefont {Graham}, \citenamefont {Ramani}, \citenamefont
  {Schmidt-Kaler}, \citenamefont {Smorra},\ and\ \citenamefont
  {Ulmer}}]{Budker:2021quh}%
  \BibitemOpen
  \bibfield  {author} {\bibinfo {author} {\bibfnamefont {D.}~\bibnamefont
  {Budker}}, \bibinfo {author} {\bibfnamefont {P.~W.}\ \bibnamefont {Graham}},
  \bibinfo {author} {\bibfnamefont {H.}~\bibnamefont {Ramani}}, \bibinfo
  {author} {\bibfnamefont {F.}~\bibnamefont {Schmidt-Kaler}}, \bibinfo {author}
  {\bibfnamefont {C.}~\bibnamefont {Smorra}}, \ and\ \bibinfo {author}
  {\bibfnamefont {S.}~\bibnamefont {Ulmer}},\ }\href {\doibase
  10.1103/PRXQuantum.3.010330} {\bibfield  {journal} {\bibinfo  {journal} {PRX
  Quantum}\ }\textbf {\bibinfo {volume} {3}},\ \bibinfo {pages} {010330}
  (\bibinfo {year} {2022})},\ \Eprint {http://arxiv.org/abs/2108.05283}
  {arXiv:2108.05283 [hep-ph]} \BibitemShut {NoStop}%
\bibitem [{\citenamefont {Berlin}\ \emph {et~al.}(2022)\citenamefont {Berlin},
  \citenamefont {Liu}, \citenamefont {Pospelov},\ and\ \citenamefont
  {Ramani}}]{Berlin:2021zbv}%
  \BibitemOpen
  \bibfield  {author} {\bibinfo {author} {\bibfnamefont {A.}~\bibnamefont
  {Berlin}}, \bibinfo {author} {\bibfnamefont {H.}~\bibnamefont {Liu}},
  \bibinfo {author} {\bibfnamefont {M.}~\bibnamefont {Pospelov}}, \ and\
  \bibinfo {author} {\bibfnamefont {H.}~\bibnamefont {Ramani}},\ }\href
  {\doibase 10.1103/PhysRevD.105.095028} {\bibfield  {journal} {\bibinfo
  {journal} {Phys. Rev. D}\ }\textbf {\bibinfo {volume} {105}},\ \bibinfo
  {pages} {095028} (\bibinfo {year} {2022})},\ \Eprint
  {http://arxiv.org/abs/2110.06217} {arXiv:2110.06217 [hep-ph]} \BibitemShut
  {NoStop}%
\bibitem [{\citenamefont {Kim}\ \emph {et~al.}(2021)\citenamefont {Kim},
  \citenamefont {Hwang},\ and\ \citenamefont {Yoo}}]{Kim:2021eix}%
  \BibitemOpen
  \bibfield  {author} {\bibinfo {author} {\bibfnamefont {J.~H.}\ \bibnamefont
  {Kim}}, \bibinfo {author} {\bibfnamefont {I.~S.}\ \bibnamefont {Hwang}}, \
  and\ \bibinfo {author} {\bibfnamefont {J.~H.}\ \bibnamefont {Yoo}},\ }\href
  {\doibase 10.1007/JHEP05(2021)031} {\bibfield  {journal} {\bibinfo  {journal}
  {JHEP}\ }\textbf {\bibinfo {volume} {05}},\ \bibinfo {pages} {031} (\bibinfo
  {year} {2021})},\ \Eprint {http://arxiv.org/abs/2102.11493} {arXiv:2102.11493
  [hep-ex]} \BibitemShut {NoStop}%
\bibitem [{\citenamefont {Kling}\ \emph {et~al.}(2023)\citenamefont {Kling},
  \citenamefont {Kuo}, \citenamefont {Trojanowski},\ and\ \citenamefont
  {Tsai}}]{Kling:2022ykt}%
  \BibitemOpen
  \bibfield  {author} {\bibinfo {author} {\bibfnamefont {F.}~\bibnamefont
  {Kling}}, \bibinfo {author} {\bibfnamefont {J.-L.}\ \bibnamefont {Kuo}},
  \bibinfo {author} {\bibfnamefont {S.}~\bibnamefont {Trojanowski}}, \ and\
  \bibinfo {author} {\bibfnamefont {Y.-D.}\ \bibnamefont {Tsai}},\ }\href
  {\doibase 10.1016/j.nuclphysb.2023.116103} {\bibfield  {journal} {\bibinfo
  {journal} {Nucl. Phys. B}\ }\textbf {\bibinfo {volume} {987}},\ \bibinfo
  {pages} {116103} (\bibinfo {year} {2023})},\ \Eprint
  {http://arxiv.org/abs/2205.09137} {arXiv:2205.09137 [hep-ph]} \BibitemShut
  {NoStop}%
\bibitem [{\citenamefont {McKeen}\ \emph {et~al.}(2022)\citenamefont {McKeen},
  \citenamefont {Moore}, \citenamefont {Morrissey}, \citenamefont {Pospelov},\
  and\ \citenamefont {Ramani}}]{McKeen:2022poo}%
  \BibitemOpen
  \bibfield  {author} {\bibinfo {author} {\bibfnamefont {D.}~\bibnamefont
  {McKeen}}, \bibinfo {author} {\bibfnamefont {M.}~\bibnamefont {Moore}},
  \bibinfo {author} {\bibfnamefont {D.~E.}\ \bibnamefont {Morrissey}}, \bibinfo
  {author} {\bibfnamefont {M.}~\bibnamefont {Pospelov}}, \ and\ \bibinfo
  {author} {\bibfnamefont {H.}~\bibnamefont {Ramani}},\ }\href {\doibase
  10.1103/PhysRevD.106.035011} {\bibfield  {journal} {\bibinfo  {journal}
  {Phys. Rev. D}\ }\textbf {\bibinfo {volume} {106}},\ \bibinfo {pages}
  {035011} (\bibinfo {year} {2022})},\ \Eprint
  {http://arxiv.org/abs/2202.08840} {arXiv:2202.08840 [hep-ph]} \BibitemShut
  {NoStop}%
\bibitem [{\citenamefont {Berlin}\ \emph {et~al.}(2024)\citenamefont {Berlin},
  \citenamefont {Liu}, \citenamefont {Pospelov},\ and\ \citenamefont
  {Ramani}}]{Berlin:2023zpn}%
  \BibitemOpen
  \bibfield  {author} {\bibinfo {author} {\bibfnamefont {A.}~\bibnamefont
  {Berlin}}, \bibinfo {author} {\bibfnamefont {H.}~\bibnamefont {Liu}},
  \bibinfo {author} {\bibfnamefont {M.}~\bibnamefont {Pospelov}}, \ and\
  \bibinfo {author} {\bibfnamefont {H.}~\bibnamefont {Ramani}},\ }\href
  {\doibase 10.1103/PhysRevD.109.075027} {\bibfield  {journal} {\bibinfo
  {journal} {Phys. Rev. D}\ }\textbf {\bibinfo {volume} {109}},\ \bibinfo
  {pages} {075027} (\bibinfo {year} {2024})},\ \Eprint
  {http://arxiv.org/abs/2302.06619} {arXiv:2302.06619 [hep-ph]} \BibitemShut
  {NoStop}%
\bibitem [{\citenamefont {McKeen}\ \emph {et~al.}(2023)\citenamefont {McKeen},
  \citenamefont {Morrissey}, \citenamefont {Pospelov}, \citenamefont {Ramani},\
  and\ \citenamefont {Ray}}]{McKeen:2023ztq}%
  \BibitemOpen
  \bibfield  {author} {\bibinfo {author} {\bibfnamefont {D.}~\bibnamefont
  {McKeen}}, \bibinfo {author} {\bibfnamefont {D.~E.}\ \bibnamefont
  {Morrissey}}, \bibinfo {author} {\bibfnamefont {M.}~\bibnamefont {Pospelov}},
  \bibinfo {author} {\bibfnamefont {H.}~\bibnamefont {Ramani}}, \ and\ \bibinfo
  {author} {\bibfnamefont {A.}~\bibnamefont {Ray}},\ }\href {\doibase
  10.1103/PhysRevLett.131.011005} {\bibfield  {journal} {\bibinfo  {journal}
  {Phys. Rev. Lett.}\ }\textbf {\bibinfo {volume} {131}},\ \bibinfo {pages}
  {011005} (\bibinfo {year} {2023})},\ \Eprint
  {http://arxiv.org/abs/2303.03416} {arXiv:2303.03416 [hep-ph]} \BibitemShut
  {NoStop}%
\bibitem [{\citenamefont {Pospelov}\ and\ \citenamefont
  {Ray}(2024)}]{Pospelov:2023mlz}%
  \BibitemOpen
  \bibfield  {author} {\bibinfo {author} {\bibfnamefont {M.}~\bibnamefont
  {Pospelov}}\ and\ \bibinfo {author} {\bibfnamefont {A.}~\bibnamefont {Ray}},\
  }\href {\doibase 10.1088/1475-7516/2024/01/029} {\bibfield  {journal}
  {\bibinfo  {journal} {JCAP}\ }\textbf {\bibinfo {volume} {01}},\ \bibinfo
  {pages} {029} (\bibinfo {year} {2024})},\ \Eprint
  {http://arxiv.org/abs/2309.10032} {arXiv:2309.10032 [hep-ph]} \BibitemShut
  {NoStop}%
\bibitem [{\citenamefont {Ema}\ \emph {et~al.}(2024)\citenamefont {Ema},
  \citenamefont {Pospelov},\ and\ \citenamefont {Ray}}]{Ema:2024oce}%
  \BibitemOpen
  \bibfield  {author} {\bibinfo {author} {\bibfnamefont {Y.}~\bibnamefont
  {Ema}}, \bibinfo {author} {\bibfnamefont {M.}~\bibnamefont {Pospelov}}, \
  and\ \bibinfo {author} {\bibfnamefont {A.}~\bibnamefont {Ray}},\ }\href
  {\doibase 10.1007/JHEP07(2024)094} {\bibfield  {journal} {\bibinfo  {journal}
  {JHEP}\ }\textbf {\bibinfo {volume} {07}},\ \bibinfo {pages} {094} (\bibinfo
  {year} {2024})},\ \Eprint {http://arxiv.org/abs/2402.03431} {arXiv:2402.03431
  [hep-ph]} \BibitemShut {NoStop}%
\bibitem [{\citenamefont {Snowden-Ifft}\ \emph {et~al.}(1990)\citenamefont
  {Snowden-Ifft}, \citenamefont {Barwick},\ and\ \citenamefont
  {Price}}]{snowden1990search}%
  \BibitemOpen
  \bibfield  {author} {\bibinfo {author} {\bibfnamefont {D.}~\bibnamefont
  {Snowden-Ifft}}, \bibinfo {author} {\bibfnamefont {S.}~\bibnamefont
  {Barwick}}, \ and\ \bibinfo {author} {\bibfnamefont {P.}~\bibnamefont
  {Price}},\ }\href@noop {} {\bibfield  {journal} {\bibinfo  {journal} {The
  Astrophysical Journal}\ }\textbf {\bibinfo {volume} {364}},\ \bibinfo {pages}
  {L25} (\bibinfo {year} {1990})}\BibitemShut {NoStop}%
\bibitem [{\citenamefont {{Snowden-Ifft}}\ \emph {et~al.}(1990)\citenamefont
  {{Snowden-Ifft}}, \citenamefont {{Barwick}},\ and\ \citenamefont
  {{Price}}}]{1990ApJ...364L..25S}%
  \BibitemOpen
  \bibfield  {author} {\bibinfo {author} {\bibfnamefont {D.~P.}\ \bibnamefont
  {{Snowden-Ifft}}}, \bibinfo {author} {\bibfnamefont {S.~W.}\ \bibnamefont
  {{Barwick}}}, \ and\ \bibinfo {author} {\bibfnamefont {P.~B.}\ \bibnamefont
  {{Price}}},\ }\href {\doibase 10.1086/185866} {\bibfield  {journal} {\bibinfo
   {journal} {Astrophysical Journal Letters}\ }\textbf {\bibinfo {volume}
  {364}},\ \bibinfo {pages} {L25} (\bibinfo {year} {1990})}\BibitemShut
  {NoStop}%
\bibitem [{\citenamefont {Wandelt}\ \emph {et~al.}(2000)\citenamefont
  {Wandelt}, \citenamefont {Dave}, \citenamefont {Farrar}, \citenamefont
  {McGuire}, \citenamefont {Spergel},\ and\ \citenamefont
  {Steinhardt}}]{Wandelt:2000ad}%
  \BibitemOpen
  \bibfield  {author} {\bibinfo {author} {\bibfnamefont {B.~D.}\ \bibnamefont
  {Wandelt}}, \bibinfo {author} {\bibfnamefont {R.}~\bibnamefont {Dave}},
  \bibinfo {author} {\bibfnamefont {G.~R.}\ \bibnamefont {Farrar}}, \bibinfo
  {author} {\bibfnamefont {P.~C.}\ \bibnamefont {McGuire}}, \bibinfo {author}
  {\bibfnamefont {D.~N.}\ \bibnamefont {Spergel}}, \ and\ \bibinfo {author}
  {\bibfnamefont {P.~J.}\ \bibnamefont {Steinhardt}},\ }in\ \href@noop {}
  {\emph {\bibinfo {booktitle} {{4th International Symposium on Sources and
  Detection of Dark Matter in the Universe (DM 2000)}}}}\ (\bibinfo {year}
  {2000})\ pp.\ \bibinfo {pages} {263--274},\ \Eprint
  {http://arxiv.org/abs/astro-ph/0006344} {arXiv:astro-ph/0006344} \BibitemShut
  {NoStop}%
\bibitem [{\citenamefont {Erickcek}\ \emph {et~al.}(2007)\citenamefont
  {Erickcek}, \citenamefont {Steinhardt}, \citenamefont {McCammon},\ and\
  \citenamefont {McGuire}}]{Erickcek:2007jv}%
  \BibitemOpen
  \bibfield  {author} {\bibinfo {author} {\bibfnamefont {A.~L.}\ \bibnamefont
  {Erickcek}}, \bibinfo {author} {\bibfnamefont {P.~J.}\ \bibnamefont
  {Steinhardt}}, \bibinfo {author} {\bibfnamefont {D.}~\bibnamefont
  {McCammon}}, \ and\ \bibinfo {author} {\bibfnamefont {P.~C.}\ \bibnamefont
  {McGuire}},\ }\href {\doibase 10.1103/PhysRevD.76.042007} {\bibfield
  {journal} {\bibinfo  {journal} {Phys. Rev.}\ }\textbf {\bibinfo {volume}
  {D76}},\ \bibinfo {pages} {042007} (\bibinfo {year} {2007})},\ \Eprint
  {http://arxiv.org/abs/0704.0794} {arXiv:0704.0794 [astro-ph]} \BibitemShut
  {NoStop}%
%%CITATION = ARXIV:0704.0794;%%
\bibitem [{\citenamefont {Adams}\ \emph {et~al.}(2020)\citenamefont {Adams}
  \emph {et~al.}}]{Adams:2019nbz}%
  \BibitemOpen
  \bibfield  {author} {\bibinfo {author} {\bibfnamefont {J.~S.}\ \bibnamefont
  {Adams}} \emph {et~al.},\ }\href {\doibase 10.1007/s10909-019-02307-2}
  {\bibfield  {journal} {\bibinfo  {journal} {J. Low Temp. Phys.}\ }\textbf
  {\bibinfo {volume} {199}},\ \bibinfo {pages} {1072} (\bibinfo {year}
  {2020})},\ \Eprint {http://arxiv.org/abs/1908.09010} {arXiv:1908.09010
  [astro-ph.IM]} \BibitemShut {NoStop}%
\bibitem [{\citenamefont {Saffold}\ \emph {et~al.}(2024)\citenamefont
  {Saffold}, \citenamefont {Alpine}, \citenamefont {Essig}, \citenamefont
  {Estrada}, \citenamefont {Kim}, \citenamefont {Kubik},\ and\ \citenamefont
  {Lembeck}}]{Saffold:2024lsj}%
  \BibitemOpen
  \bibfield  {author} {\bibinfo {author} {\bibfnamefont {N.}~\bibnamefont
  {Saffold}}, \bibinfo {author} {\bibfnamefont {P.~M.}\ \bibnamefont {Alpine}},
  \bibinfo {author} {\bibfnamefont {R.}~\bibnamefont {Essig}}, \bibinfo
  {author} {\bibfnamefont {J.}~\bibnamefont {Estrada}}, \bibinfo {author}
  {\bibfnamefont {T.}~\bibnamefont {Kim}}, \bibinfo {author} {\bibfnamefont
  {D.}~\bibnamefont {Kubik}}, \ and\ \bibinfo {author} {\bibfnamefont {M.~F.}\
  \bibnamefont {Lembeck}},\ }\href {\doibase 10.22323/1.441.0062} {\bibfield
  {journal} {\bibinfo  {journal} {PoS}\ }\textbf {\bibinfo {volume}
  {TAUP2023}},\ \bibinfo {pages} {062} (\bibinfo {year} {2024})}\BibitemShut
  {NoStop}%
\bibitem [{\citenamefont {Holdom}(1986)}]{Holdom:1985ag}%
  \BibitemOpen
  \bibfield  {author} {\bibinfo {author} {\bibfnamefont {B.}~\bibnamefont
  {Holdom}},\ }\href {\doibase 10.1016/0370-2693(86)91377-8} {\bibfield
  {journal} {\bibinfo  {journal} {Phys. Lett.}\ }\textbf {\bibinfo {volume}
  {166B}},\ \bibinfo {pages} {196} (\bibinfo {year} {1986})}\BibitemShut
  {NoStop}%
%%CITATION = PHLTA,166B,196;%%
\bibitem [{\citenamefont {Galison}\ and\ \citenamefont
  {Manohar}(1984)}]{Galison:1983pa}%
  \BibitemOpen
  \bibfield  {author} {\bibinfo {author} {\bibfnamefont {P.}~\bibnamefont
  {Galison}}\ and\ \bibinfo {author} {\bibfnamefont {A.}~\bibnamefont
  {Manohar}},\ }\href {\doibase 10.1016/0370-2693(84)91161-4} {\bibfield
  {journal} {\bibinfo  {journal} {Phys. Lett.}\ }\textbf {\bibinfo {volume}
  {136B}},\ \bibinfo {pages} {279} (\bibinfo {year} {1984})}\BibitemShut
  {NoStop}%
%%CITATION = PHLTA,136B,279;%%
\bibitem [{\citenamefont {Essig}\ \emph {et~al.}(2016)\citenamefont {Essig},
  \citenamefont {Fernandez-Serra}, \citenamefont {Mardon}, \citenamefont
  {Soto}, \citenamefont {Volansky},\ and\ \citenamefont {Yu}}]{Essig:2015cda}%
  \BibitemOpen
  \bibfield  {author} {\bibinfo {author} {\bibfnamefont {R.}~\bibnamefont
  {Essig}}, \bibinfo {author} {\bibfnamefont {M.}~\bibnamefont
  {Fernandez-Serra}}, \bibinfo {author} {\bibfnamefont {J.}~\bibnamefont
  {Mardon}}, \bibinfo {author} {\bibfnamefont {A.}~\bibnamefont {Soto}},
  \bibinfo {author} {\bibfnamefont {T.}~\bibnamefont {Volansky}}, \ and\
  \bibinfo {author} {\bibfnamefont {T.-T.}\ \bibnamefont {Yu}},\ }\href
  {\doibase 10.1007/JHEP05(2016)046} {\bibfield  {journal} {\bibinfo  {journal}
  {JHEP}\ }\textbf {\bibinfo {volume} {05}},\ \bibinfo {pages} {046} (\bibinfo
  {year} {2016})},\ \Eprint {http://arxiv.org/abs/1509.01598} {arXiv:1509.01598
  [hep-ph]} \BibitemShut {NoStop}%
\bibitem [{\citenamefont {Hochberg}\ \emph {et~al.}(2021)\citenamefont
  {Hochberg}, \citenamefont {Kahn}, \citenamefont {Kurinsky}, \citenamefont
  {Lehmann}, \citenamefont {Yu},\ and\ \citenamefont
  {Berggren}}]{Hochberg:2021pkt}%
  \BibitemOpen
  \bibfield  {author} {\bibinfo {author} {\bibfnamefont {Y.}~\bibnamefont
  {Hochberg}}, \bibinfo {author} {\bibfnamefont {Y.}~\bibnamefont {Kahn}},
  \bibinfo {author} {\bibfnamefont {N.}~\bibnamefont {Kurinsky}}, \bibinfo
  {author} {\bibfnamefont {B.~V.}\ \bibnamefont {Lehmann}}, \bibinfo {author}
  {\bibfnamefont {T.~C.}\ \bibnamefont {Yu}}, \ and\ \bibinfo {author}
  {\bibfnamefont {K.~K.}\ \bibnamefont {Berggren}},\ }\href {\doibase
  10.1103/PhysRevLett.127.151802} {\bibfield  {journal} {\bibinfo  {journal}
  {Phys. Rev. Lett.}\ }\textbf {\bibinfo {volume} {127}},\ \bibinfo {pages}
  {151802} (\bibinfo {year} {2021})},\ \Eprint
  {http://arxiv.org/abs/2101.08263} {arXiv:2101.08263 [hep-ph]} \BibitemShut
  {NoStop}%
\bibitem [{\citenamefont {Knapen}\ \emph {et~al.}(2021)\citenamefont {Knapen},
  \citenamefont {Kozaczuk},\ and\ \citenamefont {Lin}}]{Knapen:2021run}%
  \BibitemOpen
  \bibfield  {author} {\bibinfo {author} {\bibfnamefont {S.}~\bibnamefont
  {Knapen}}, \bibinfo {author} {\bibfnamefont {J.}~\bibnamefont {Kozaczuk}}, \
  and\ \bibinfo {author} {\bibfnamefont {T.}~\bibnamefont {Lin}},\ }\href
  {\doibase 10.1103/PhysRevD.104.015031} {\bibfield  {journal} {\bibinfo
  {journal} {Phys. Rev. D}\ }\textbf {\bibinfo {volume} {104}},\ \bibinfo
  {pages} {015031} (\bibinfo {year} {2021})},\ \Eprint
  {http://arxiv.org/abs/2101.08275} {arXiv:2101.08275 [hep-ph]} \BibitemShut
  {NoStop}%
\bibitem [{\citenamefont {Knapen}\ \emph {et~al.}(2022)\citenamefont {Knapen},
  \citenamefont {Kozaczuk},\ and\ \citenamefont {Lin}}]{Knapen:2021bwg}%
  \BibitemOpen
  \bibfield  {author} {\bibinfo {author} {\bibfnamefont {S.}~\bibnamefont
  {Knapen}}, \bibinfo {author} {\bibfnamefont {J.}~\bibnamefont {Kozaczuk}}, \
  and\ \bibinfo {author} {\bibfnamefont {T.}~\bibnamefont {Lin}},\ }\href
  {\doibase 10.1103/PhysRevD.105.015014} {\bibfield  {journal} {\bibinfo
  {journal} {Phys. Rev. D}\ }\textbf {\bibinfo {volume} {105}},\ \bibinfo
  {pages} {015014} (\bibinfo {year} {2022})},\ \Eprint
  {http://arxiv.org/abs/2104.12786} {arXiv:2104.12786 [hep-ph]} \BibitemShut
  {NoStop}%
\bibitem [{\citenamefont {Griffin}\ \emph {et~al.}(2021)\citenamefont
  {Griffin}, \citenamefont {Inzani}, \citenamefont {Trickle}, \citenamefont
  {Zhang},\ and\ \citenamefont {Zurek}}]{Griffin:2021znd}%
  \BibitemOpen
  \bibfield  {author} {\bibinfo {author} {\bibfnamefont {S.~M.}\ \bibnamefont
  {Griffin}}, \bibinfo {author} {\bibfnamefont {K.}~\bibnamefont {Inzani}},
  \bibinfo {author} {\bibfnamefont {T.}~\bibnamefont {Trickle}}, \bibinfo
  {author} {\bibfnamefont {Z.}~\bibnamefont {Zhang}}, \ and\ \bibinfo {author}
  {\bibfnamefont {K.~M.}\ \bibnamefont {Zurek}},\ }\href {\doibase
  10.1103/PhysRevD.104.095015} {\bibfield  {journal} {\bibinfo  {journal}
  {Phys. Rev. D}\ }\textbf {\bibinfo {volume} {104}},\ \bibinfo {pages}
  {095015} (\bibinfo {year} {2021})},\ \Eprint
  {http://arxiv.org/abs/2105.05253} {arXiv:2105.05253 [hep-ph]} \BibitemShut
  {NoStop}%
\bibitem [{\citenamefont {Kahn}\ and\ \citenamefont
  {Lin}(2022)}]{Kahn:2021ttr}%
  \BibitemOpen
  \bibfield  {author} {\bibinfo {author} {\bibfnamefont {Y.}~\bibnamefont
  {Kahn}}\ and\ \bibinfo {author} {\bibfnamefont {T.}~\bibnamefont {Lin}},\
  }\href {\doibase 10.1088/1361-6633/ac5f63} {\bibfield  {journal} {\bibinfo
  {journal} {Rept. Prog. Phys.}\ }\textbf {\bibinfo {volume} {85}},\ \bibinfo
  {pages} {066901} (\bibinfo {year} {2022})},\ \Eprint
  {http://arxiv.org/abs/2108.03239} {arXiv:2108.03239 [hep-ph]} \BibitemShut
  {NoStop}%
\bibitem [{\citenamefont {Trickle}(2023)}]{Trickle:2022fwt}%
  \BibitemOpen
  \bibfield  {author} {\bibinfo {author} {\bibfnamefont {T.}~\bibnamefont
  {Trickle}},\ }\href {\doibase 10.1103/PhysRevD.107.035035} {\bibfield
  {journal} {\bibinfo  {journal} {Phys. Rev. D}\ }\textbf {\bibinfo {volume}
  {107}},\ \bibinfo {pages} {035035} (\bibinfo {year} {2023})},\ \Eprint
  {http://arxiv.org/abs/2210.14917} {arXiv:2210.14917 [hep-ph]} \BibitemShut
  {NoStop}%
\bibitem [{\citenamefont {Dreyer}\ \emph {et~al.}(2024)\citenamefont {Dreyer},
  \citenamefont {Essig}, \citenamefont {Fernandez-Serra}, \citenamefont
  {Singal},\ and\ \citenamefont {Zhen}}]{Dreyer:2023ovn}%
  \BibitemOpen
  \bibfield  {author} {\bibinfo {author} {\bibfnamefont {C.~E.}\ \bibnamefont
  {Dreyer}}, \bibinfo {author} {\bibfnamefont {R.}~\bibnamefont {Essig}},
  \bibinfo {author} {\bibfnamefont {M.}~\bibnamefont {Fernandez-Serra}},
  \bibinfo {author} {\bibfnamefont {A.}~\bibnamefont {Singal}}, \ and\ \bibinfo
  {author} {\bibfnamefont {C.}~\bibnamefont {Zhen}},\ }\href {\doibase
  10.1103/PhysRevD.109.115008} {\bibfield  {journal} {\bibinfo  {journal}
  {Phys. Rev. D}\ }\textbf {\bibinfo {volume} {109}},\ \bibinfo {pages}
  {115008} (\bibinfo {year} {2024})},\ \Eprint
  {http://arxiv.org/abs/2306.14944} {arXiv:2306.14944 [hep-ph]} \BibitemShut
  {NoStop}%
\bibitem [{QEd()}]{QEdark}%
  \BibitemOpen
  \href@noop {} {\enquote {\bibinfo {title} {{\tt QEDark}},}\ }\bibinfo
  {howpublished} {\url{https://github.com/tientienyu/QEdark}}\BibitemShut
  {NoStop}%
\bibitem [{\citenamefont {Dreyer}\ \emph {et~al.}(pear)\citenamefont {Dreyer},
  \citenamefont {Essig}, \citenamefont {Fernandez-Serra}, \citenamefont
  {Hott},\ and\ \citenamefont {Singal}}]{Dreyer:in-progress}%
  \BibitemOpen
  \bibfield  {author} {\bibinfo {author} {\bibfnamefont {C.~E.}\ \bibnamefont
  {Dreyer}}, \bibinfo {author} {\bibfnamefont {R.}~\bibnamefont {Essig}},
  \bibinfo {author} {\bibfnamefont {M.}~\bibnamefont {Fernandez-Serra}},
  \bibinfo {author} {\bibfnamefont {M.}~\bibnamefont {Hott}}, \ and\ \bibinfo
  {author} {\bibfnamefont {A.}~\bibnamefont {Singal}},\ }\href@noop {} {\
  (\bibinfo {year} {to~appear})}\BibitemShut {NoStop}%
\bibitem [{\citenamefont {Dressel}\ and\ \citenamefont
  {Grüner}(2002)}]{Dressel_Gruner_2002}%
  \BibitemOpen
  \bibfield  {author} {\bibinfo {author} {\bibfnamefont {M.}~\bibnamefont
  {Dressel}}\ and\ \bibinfo {author} {\bibfnamefont {G.}~\bibnamefont
  {Grüner}},\ }\href@noop {} {\emph {\bibinfo {title} {Electrodynamics of
  Solids: Optical Properties of Electrons in Matter}}}\ (\bibinfo  {publisher}
  {Cambridge University Press},\ \bibinfo {year} {2002})\BibitemShut {NoStop}%
\bibitem [{\citenamefont {Baxter}\ \emph {et~al.}(2021)\citenamefont {Baxter}
  \emph {et~al.}}]{Baxter:2021pqo}%
  \BibitemOpen
  \bibfield  {author} {\bibinfo {author} {\bibfnamefont {D.}~\bibnamefont
  {Baxter}} \emph {et~al.},\ }\href {\doibase 10.1140/epjc/s10052-021-09655-y}
  {\bibfield  {journal} {\bibinfo  {journal} {Eur. Phys. J. C}\ }\textbf
  {\bibinfo {volume} {81}},\ \bibinfo {pages} {907} (\bibinfo {year} {2021})},\
  \Eprint {http://arxiv.org/abs/2105.00599} {arXiv:2105.00599 [hep-ex]}
  \BibitemShut {NoStop}%
\bibitem [{\citenamefont {Jakobsen}\ \emph {et~al.}(2022)\citenamefont
  {Jakobsen}, \citenamefont {Ferruit}, \citenamefont {{Alves de Oliveira}},
  \citenamefont {Arribas}, \citenamefont {Bagnasco}, \citenamefont {Barho},
  \citenamefont {Beck}, \citenamefont {Birkmann}, \citenamefont {B{\"{o}}ker},
  \citenamefont {Bunker}, \citenamefont {Charlot}, \citenamefont {de~Jong},
  \citenamefont {de~Marchi}, \citenamefont {Ehrenwinkler}, \citenamefont
  {Falcolini}, \citenamefont {Fels}, \citenamefont {Franx}, \citenamefont
  {Franz}, \citenamefont {Funke}, \citenamefont {Giardino}, \citenamefont
  {Gnata}, \citenamefont {Holota}, \citenamefont {Honnen}, \citenamefont
  {Jensen}, \citenamefont {Jentsch}, \citenamefont {Johnson}, \citenamefont
  {Jollet}, \citenamefont {Karl}, \citenamefont {Kling}, \citenamefont
  {K{\"{o}}hler}, \citenamefont {Kolm}, \citenamefont {Kumari}, \citenamefont
  {Lander}, \citenamefont {Lemke}, \citenamefont {L{\'{o}}pez-Caniego},
  \citenamefont {L{\"{u}}tzgendorf}, \citenamefont {Maiolino}, \citenamefont
  {Manjavacas}, \citenamefont {Marston}, \citenamefont {Maschmann},
  \citenamefont {Maurer}, \citenamefont {Messerschmidt}, \citenamefont
  {Moseley}, \citenamefont {Mosner}, \citenamefont {Mott}, \citenamefont
  {Muzerolle}, \citenamefont {Pirzkal}, \citenamefont {Pittet}, \citenamefont
  {Plitzke}, \citenamefont {Posselt}, \citenamefont {Rapp}, \citenamefont
  {Rauscher}, \citenamefont {Rawle}, \citenamefont {Rix}, \citenamefont
  {R{\"{o}}del}, \citenamefont {Rumler}, \citenamefont {Sabbi}, \citenamefont
  {Salvignol}, \citenamefont {Schmid}, \citenamefont {Sirianni}, \citenamefont
  {Smith}, \citenamefont {Strada}, \citenamefont {te~Plate}, \citenamefont
  {Valenti}, \citenamefont {Wettemann}, \citenamefont {Wiehe}, \citenamefont
  {Wiesmayer}, \citenamefont {Willott}, \citenamefont {Wright}, \citenamefont
  {Zeidler},\ and\ \citenamefont {Zincke}}]{Jakobsen_2022}%
  \BibitemOpen
  \bibfield  {author} {\bibinfo {author} {\bibfnamefont {P.}~\bibnamefont
  {Jakobsen}}, \bibinfo {author} {\bibfnamefont {P.}~\bibnamefont {Ferruit}},
  \bibinfo {author} {\bibfnamefont {C.}~\bibnamefont {{Alves de Oliveira}}},
  \bibinfo {author} {\bibfnamefont {S.}~\bibnamefont {Arribas}}, \bibinfo
  {author} {\bibfnamefont {G.}~\bibnamefont {Bagnasco}}, \bibinfo {author}
  {\bibfnamefont {R.}~\bibnamefont {Barho}}, \bibinfo {author} {\bibfnamefont
  {T.}~\bibnamefont {Beck}}, \bibinfo {author} {\bibfnamefont {S.}~\bibnamefont
  {Birkmann}}, \bibinfo {author} {\bibfnamefont {T.}~\bibnamefont
  {B{\"{o}}ker}}, \bibinfo {author} {\bibfnamefont {A.}~\bibnamefont {Bunker}},
  \bibinfo {author} {\bibfnamefont {S.}~\bibnamefont {Charlot}}, \bibinfo
  {author} {\bibfnamefont {P.}~\bibnamefont {de~Jong}}, \bibinfo {author}
  {\bibfnamefont {G.}~\bibnamefont {de~Marchi}}, \bibinfo {author}
  {\bibfnamefont {R.}~\bibnamefont {Ehrenwinkler}}, \bibinfo {author}
  {\bibfnamefont {M.}~\bibnamefont {Falcolini}}, \bibinfo {author}
  {\bibfnamefont {R.}~\bibnamefont {Fels}}, \bibinfo {author} {\bibfnamefont
  {M.}~\bibnamefont {Franx}}, \bibinfo {author} {\bibfnamefont
  {D.}~\bibnamefont {Franz}}, \bibinfo {author} {\bibfnamefont
  {M.}~\bibnamefont {Funke}}, \bibinfo {author} {\bibfnamefont
  {G.}~\bibnamefont {Giardino}}, \bibinfo {author} {\bibfnamefont
  {X.}~\bibnamefont {Gnata}}, \bibinfo {author} {\bibfnamefont
  {W.}~\bibnamefont {Holota}}, \bibinfo {author} {\bibfnamefont
  {K.}~\bibnamefont {Honnen}}, \bibinfo {author} {\bibfnamefont
  {P.}~\bibnamefont {Jensen}}, \bibinfo {author} {\bibfnamefont
  {M.}~\bibnamefont {Jentsch}}, \bibinfo {author} {\bibfnamefont
  {T.}~\bibnamefont {Johnson}}, \bibinfo {author} {\bibfnamefont
  {D.}~\bibnamefont {Jollet}}, \bibinfo {author} {\bibfnamefont
  {H.}~\bibnamefont {Karl}}, \bibinfo {author} {\bibfnamefont {G.}~\bibnamefont
  {Kling}}, \bibinfo {author} {\bibfnamefont {J.}~\bibnamefont {K{\"{o}}hler}},
  \bibinfo {author} {\bibfnamefont {M.~G.}\ \bibnamefont {Kolm}}, \bibinfo
  {author} {\bibfnamefont {N.}~\bibnamefont {Kumari}}, \bibinfo {author}
  {\bibfnamefont {M.}~\bibnamefont {Lander}}, \bibinfo {author} {\bibfnamefont
  {R.}~\bibnamefont {Lemke}}, \bibinfo {author} {\bibfnamefont
  {M.}~\bibnamefont {L{\'{o}}pez-Caniego}}, \bibinfo {author} {\bibfnamefont
  {N.}~\bibnamefont {L{\"{u}}tzgendorf}}, \bibinfo {author} {\bibfnamefont
  {R.}~\bibnamefont {Maiolino}}, \bibinfo {author} {\bibfnamefont
  {E.}~\bibnamefont {Manjavacas}}, \bibinfo {author} {\bibfnamefont
  {A.}~\bibnamefont {Marston}}, \bibinfo {author} {\bibfnamefont
  {M.}~\bibnamefont {Maschmann}}, \bibinfo {author} {\bibfnamefont
  {R.}~\bibnamefont {Maurer}}, \bibinfo {author} {\bibfnamefont
  {B.}~\bibnamefont {Messerschmidt}}, \bibinfo {author} {\bibfnamefont
  {S.}~\bibnamefont {Moseley}}, \bibinfo {author} {\bibfnamefont
  {P.}~\bibnamefont {Mosner}}, \bibinfo {author} {\bibfnamefont
  {D.}~\bibnamefont {Mott}}, \bibinfo {author} {\bibfnamefont {J.}~\bibnamefont
  {Muzerolle}}, \bibinfo {author} {\bibfnamefont {N.}~\bibnamefont {Pirzkal}},
  \bibinfo {author} {\bibfnamefont {J.~F.}\ \bibnamefont {Pittet}}, \bibinfo
  {author} {\bibfnamefont {A.}~\bibnamefont {Plitzke}}, \bibinfo {author}
  {\bibfnamefont {W.}~\bibnamefont {Posselt}}, \bibinfo {author} {\bibfnamefont
  {B.}~\bibnamefont {Rapp}}, \bibinfo {author} {\bibfnamefont {B.}~\bibnamefont
  {Rauscher}}, \bibinfo {author} {\bibfnamefont {T.}~\bibnamefont {Rawle}},
  \bibinfo {author} {\bibfnamefont {H.~W.}\ \bibnamefont {Rix}}, \bibinfo
  {author} {\bibfnamefont {A.}~\bibnamefont {R{\"{o}}del}}, \bibinfo {author}
  {\bibfnamefont {P.}~\bibnamefont {Rumler}}, \bibinfo {author} {\bibfnamefont
  {E.}~\bibnamefont {Sabbi}}, \bibinfo {author} {\bibfnamefont {J.~C.}\
  \bibnamefont {Salvignol}}, \bibinfo {author} {\bibfnamefont {T.}~\bibnamefont
  {Schmid}}, \bibinfo {author} {\bibfnamefont {M.}~\bibnamefont {Sirianni}},
  \bibinfo {author} {\bibfnamefont {C.}~\bibnamefont {Smith}}, \bibinfo
  {author} {\bibfnamefont {P.}~\bibnamefont {Strada}}, \bibinfo {author}
  {\bibfnamefont {M.}~\bibnamefont {te~Plate}}, \bibinfo {author}
  {\bibfnamefont {J.}~\bibnamefont {Valenti}}, \bibinfo {author} {\bibfnamefont
  {T.}~\bibnamefont {Wettemann}}, \bibinfo {author} {\bibfnamefont
  {T.}~\bibnamefont {Wiehe}}, \bibinfo {author} {\bibfnamefont
  {M.}~\bibnamefont {Wiesmayer}}, \bibinfo {author} {\bibfnamefont
  {C.}~\bibnamefont {Willott}}, \bibinfo {author} {\bibfnamefont
  {R.}~\bibnamefont {Wright}}, \bibinfo {author} {\bibfnamefont
  {P.}~\bibnamefont {Zeidler}}, \ and\ \bibinfo {author} {\bibfnamefont
  {C.}~\bibnamefont {Zincke}},\ }\href {\doibase 10.1051/0004-6361/202142663}
  {\bibfield  {journal} {\bibinfo  {journal} {A{\&}A}\ }\textbf {\bibinfo
  {volume} {661}},\ \bibinfo {pages} {A80} (\bibinfo {year}
  {2022})}\BibitemShut {NoStop}%
\bibitem [{\citenamefont {Birkmann}\ \emph {et~al.}(2022)\citenamefont
  {Birkmann}, \citenamefont {Giardino}, \citenamefont {Sirianni}, \citenamefont
  {Ferruit}, \citenamefont {Rauscher}, \citenamefont {{Alves de Oliveira}},
  \citenamefont {B{\"{o}}ker}, \citenamefont {Kumari}, \citenamefont
  {L{\"{u}}tzgendorf}, \citenamefont {Manjavacas}, \citenamefont {Proffitt},
  \citenamefont {Rawle}, \citenamefont {te~Plate},\ and\ \citenamefont
  {Zeidler}}]{Birkmann2022}%
  \BibitemOpen
  \bibfield  {author} {\bibinfo {author} {\bibfnamefont {S.~M.}\ \bibnamefont
  {Birkmann}}, \bibinfo {author} {\bibfnamefont {G.}~\bibnamefont {Giardino}},
  \bibinfo {author} {\bibfnamefont {M.}~\bibnamefont {Sirianni}}, \bibinfo
  {author} {\bibfnamefont {P.}~\bibnamefont {Ferruit}}, \bibinfo {author}
  {\bibfnamefont {B.~J.}\ \bibnamefont {Rauscher}}, \bibinfo {author}
  {\bibfnamefont {C.}~\bibnamefont {{Alves de Oliveira}}}, \bibinfo {author}
  {\bibfnamefont {T.}~\bibnamefont {B{\"{o}}ker}}, \bibinfo {author}
  {\bibfnamefont {N.}~\bibnamefont {Kumari}}, \bibinfo {author} {\bibfnamefont
  {N.}~\bibnamefont {L{\"{u}}tzgendorf}}, \bibinfo {author} {\bibfnamefont
  {E.}~\bibnamefont {Manjavacas}}, \bibinfo {author} {\bibfnamefont
  {C.}~\bibnamefont {Proffitt}}, \bibinfo {author} {\bibfnamefont
  {T.}~\bibnamefont {Rawle}}, \bibinfo {author} {\bibfnamefont
  {M.}~\bibnamefont {te~Plate}}, \ and\ \bibinfo {author} {\bibfnamefont
  {P.}~\bibnamefont {Zeidler}},\ }\href {\doibase 10.1117/12.2629545}
  {\bibfield  {journal} {\bibinfo  {journal} {Proc SPIE}\ }\textbf {\bibinfo
  {volume} {12180}},\ \bibinfo {pages} {101} (\bibinfo {year}
  {2022})}\BibitemShut {NoStop}%
\bibitem [{\citenamefont {{{These estimates of materials surrounding the
  NIRSpec detector were informed, in part, by private conversations with the
  NIRSpec System Engineer, Maurice te Plate}}}()}]{JWST_communication}%
  \BibitemOpen
  \bibfield  {author} {\bibinfo {author} {\bibnamefont {{{These estimates of
  materials surrounding the NIRSpec detector were informed, in part, by private
  conversations with the NIRSpec System Engineer, Maurice te Plate}}}},\
  }\href@noop {} {}\BibitemShut {NoStop}%
\bibitem [{\citenamefont {Emken}\ and\ \citenamefont
  {Kouvaris}(2019)}]{Emken2018a}%
  \BibitemOpen
  \bibfield  {author} {\bibinfo {author} {\bibfnamefont {T.}~\bibnamefont
  {Emken}}\ and\ \bibinfo {author} {\bibfnamefont {C.}~\bibnamefont
  {Kouvaris}},\ }\href@noop {} {\enquote {\bibinfo {title}
  {\textnormal{DaMaSCUS-CRUST v1.1},
  \href{https://ascl.net/1803.001}{\textnormal{[ascl:1803.001]}}\textnormal{
  Available at
  \href{https://github.com/temken/damascus-crust}{https://github.com/temken/}
  and archived under [\href{https://doi.org/10.5281/zenodo.2846401}{DOI:
  10.5281/zenodo.2846401}]}},}\ } (\bibinfo {year} {2018-2019})\BibitemShut
  {NoStop}%
\bibitem [{\citenamefont {Emken}\ and\ \citenamefont
  {Kouvaris}(2018)}]{Emken:2018run}%
  \BibitemOpen
  \bibfield  {author} {\bibinfo {author} {\bibfnamefont {T.}~\bibnamefont
  {Emken}}\ and\ \bibinfo {author} {\bibfnamefont {C.}~\bibnamefont
  {Kouvaris}},\ }\href {\doibase 10.1103/PhysRevD.97.115047} {\bibfield
  {journal} {\bibinfo  {journal} {Phys. Rev.}\ }\textbf {\bibinfo {volume}
  {D97}},\ \bibinfo {pages} {115047} (\bibinfo {year} {2018})},\ \Eprint
  {http://arxiv.org/abs/1802.04764} {arXiv:1802.04764 [hep-ph]} \BibitemShut
  {NoStop}%
%%CITATION = ARXIV:1802.04764;%%
\bibitem [{MAS()}]{MAST-website}%
  \BibitemOpen
  \href@noop {} {\enquote {\bibinfo {title} {{Barbara A.~Mikulski Archive for
  Space Telescopes}},}\ }\bibinfo {howpublished}
  {\url{https://mast.stsci.edu/portal/Mashup/Clients/Mast/Portal.html}}\BibitemShut
  {NoStop}%
\bibitem [{\citenamefont {{JWST User Documentation:
  \url{https://jwst-docs.stsci.edu/jwst-near-infrared-spectrograph/nirspec-instrumentation/nirspec-detectors/nirspec-detector-performance}}}()}]{JWST_gain}%
  \BibitemOpen
  \bibfield  {author} {\bibinfo {author} {\bibnamefont {{JWST User
  Documentation:
  \url{https://jwst-docs.stsci.edu/jwst-near-infrared-spectrograph/nirspec-instrumentation/nirspec-detectors/nirspec-detector-performance}}}},\
  }\href@noop {} {}\BibitemShut {NoStop}%
\bibitem [{\citenamefont {{Bushouse}}\ \emph {et~al.}(2023)\citenamefont
  {{Bushouse}}, \citenamefont {{Eisenhamer}}, \citenamefont {{Dencheva}},
  \citenamefont {{Davies}}, \citenamefont {{Greenfield}}, \citenamefont
  {{Morrison}}, \citenamefont {{Hodge}}, \citenamefont {{Simon}}, \citenamefont
  {{Grumm}}, \citenamefont {{Droettboom}}, \citenamefont {{Slavich}},
  \citenamefont {{Sosey}}, \citenamefont {{Pauly}}, \citenamefont {{Miller}},
  \citenamefont {{Jedrzejewski}}, \citenamefont {{Hack}}, \citenamefont
  {{Davis}}, \citenamefont {{Crawford}}, \citenamefont {{Law}}, \citenamefont
  {{Gordon}}, \citenamefont {{Regan}}, \citenamefont {{Cara}}, \citenamefont
  {{MacDonald}}, \citenamefont {{Bradley}}, \citenamefont {{Shanahan}},
  \citenamefont {{Jamieson}}, \citenamefont {{Teodoro}},\ and\ \citenamefont
  {{Williams}}}]{Bushouse2023}%
  \BibitemOpen
  \bibfield  {author} {\bibinfo {author} {\bibfnamefont {H.}~\bibnamefont
  {{Bushouse}}}, \bibinfo {author} {\bibfnamefont {J.}~\bibnamefont
  {{Eisenhamer}}}, \bibinfo {author} {\bibfnamefont {N.}~\bibnamefont
  {{Dencheva}}}, \bibinfo {author} {\bibfnamefont {J.}~\bibnamefont
  {{Davies}}}, \bibinfo {author} {\bibfnamefont {P.}~\bibnamefont
  {{Greenfield}}}, \bibinfo {author} {\bibfnamefont {J.}~\bibnamefont
  {{Morrison}}}, \bibinfo {author} {\bibfnamefont {P.}~\bibnamefont {{Hodge}}},
  \bibinfo {author} {\bibfnamefont {B.}~\bibnamefont {{Simon}}}, \bibinfo
  {author} {\bibfnamefont {D.}~\bibnamefont {{Grumm}}}, \bibinfo {author}
  {\bibfnamefont {M.}~\bibnamefont {{Droettboom}}}, \bibinfo {author}
  {\bibfnamefont {E.}~\bibnamefont {{Slavich}}}, \bibinfo {author}
  {\bibfnamefont {M.}~\bibnamefont {{Sosey}}}, \bibinfo {author} {\bibfnamefont
  {T.}~\bibnamefont {{Pauly}}}, \bibinfo {author} {\bibfnamefont
  {T.}~\bibnamefont {{Miller}}}, \bibinfo {author} {\bibfnamefont
  {R.}~\bibnamefont {{Jedrzejewski}}}, \bibinfo {author} {\bibfnamefont
  {W.}~\bibnamefont {{Hack}}}, \bibinfo {author} {\bibfnamefont
  {D.}~\bibnamefont {{Davis}}}, \bibinfo {author} {\bibfnamefont
  {S.}~\bibnamefont {{Crawford}}}, \bibinfo {author} {\bibfnamefont
  {D.}~\bibnamefont {{Law}}}, \bibinfo {author} {\bibfnamefont
  {K.}~\bibnamefont {{Gordon}}}, \bibinfo {author} {\bibfnamefont
  {M.}~\bibnamefont {{Regan}}}, \bibinfo {author} {\bibfnamefont
  {M.}~\bibnamefont {{Cara}}}, \bibinfo {author} {\bibfnamefont
  {K.}~\bibnamefont {{MacDonald}}}, \bibinfo {author} {\bibfnamefont
  {L.}~\bibnamefont {{Bradley}}}, \bibinfo {author} {\bibfnamefont
  {C.}~\bibnamefont {{Shanahan}}}, \bibinfo {author} {\bibfnamefont
  {W.}~\bibnamefont {{Jamieson}}}, \bibinfo {author} {\bibfnamefont
  {M.}~\bibnamefont {{Teodoro}}}, \ and\ \bibinfo {author} {\bibfnamefont
  {T.}~\bibnamefont {{Williams}}},\ }\href {\doibase 10.5281/zenodo.6984365}
  {\enquote {\bibinfo {title} {{JWST Calibration Pipeline}},}\ }\bibinfo
  {howpublished} {Zenodo} (\bibinfo {year} {2023})\BibitemShut {NoStop}%
\bibitem [{\citenamefont {Rauscher}\ \emph {et~al.}(2017)\citenamefont
  {Rauscher}, \citenamefont {Arendt}, \citenamefont {Fixsen}, \citenamefont
  {Greenhouse}, \citenamefont {Lander}, \citenamefont {Lindler}, \citenamefont
  {Loose}, \citenamefont {Moseley}, \citenamefont {Mott}, \citenamefont {Wen},
  \citenamefont {Wilson},\ and\ \citenamefont {Xenophontos}}]{Rauscher_2017}%
  \BibitemOpen
  \bibfield  {author} {\bibinfo {author} {\bibfnamefont {B.~J.}\ \bibnamefont
  {Rauscher}}, \bibinfo {author} {\bibfnamefont {R.~G.}\ \bibnamefont
  {Arendt}}, \bibinfo {author} {\bibfnamefont {D.~J.}\ \bibnamefont {Fixsen}},
  \bibinfo {author} {\bibfnamefont {M.~A.}\ \bibnamefont {Greenhouse}},
  \bibinfo {author} {\bibfnamefont {M.}~\bibnamefont {Lander}}, \bibinfo
  {author} {\bibfnamefont {D.}~\bibnamefont {Lindler}}, \bibinfo {author}
  {\bibfnamefont {M.}~\bibnamefont {Loose}}, \bibinfo {author} {\bibfnamefont
  {S.~H.}\ \bibnamefont {Moseley}}, \bibinfo {author} {\bibfnamefont {D.~B.}\
  \bibnamefont {Mott}}, \bibinfo {author} {\bibfnamefont {Y.}~\bibnamefont
  {Wen}}, \bibinfo {author} {\bibfnamefont {D.~V.}\ \bibnamefont {Wilson}}, \
  and\ \bibinfo {author} {\bibfnamefont {C.}~\bibnamefont {Xenophontos}},\
  }\href {\doibase 10.1088/1538-3873/aa83fd} {\bibfield  {journal} {\bibinfo
  {journal} {Publications of the Astronomical Society of the Pacific}\ }\textbf
  {\bibinfo {volume} {129}},\ \bibinfo {pages} {105003} (\bibinfo {year}
  {2017})}\BibitemShut {NoStop}%
\bibitem [{\citenamefont {{JWST User Documentation:
  \url{https://jwst-docs.stsci.edu/depreciated-jdox-articles/data-artifacts-and-features/snowballs-and-shower-artifacts}}}()}]{JWST_snowball}%
  \BibitemOpen
  \bibfield  {author} {\bibinfo {author} {\bibnamefont {{JWST User
  Documentation:
  \url{https://jwst-docs.stsci.edu/depreciated-jdox-articles/data-artifacts-and-features/snowballs-and-shower-artifacts}}}},\
  }\href@noop {} {}\BibitemShut {NoStop}%
\bibitem [{\citenamefont {Barak}\ \emph
  {et~al.}(2020{\natexlab{b}})\citenamefont {Barak}, \citenamefont {Bloch},
  \citenamefont {Cababie}, \citenamefont {Cancelo}, \citenamefont {Chaplinsky},
  \citenamefont {Chierchie}, \citenamefont {Crisler}, \citenamefont
  {Drlica-Wagner}, \citenamefont {Essig}, \citenamefont {Estrada},
  \citenamefont {Etzion}, \citenamefont {Moroni}, \citenamefont {Gift},
  \citenamefont {Munagavalasa}, \citenamefont {Orly}, \citenamefont
  {Rodrigues}, \citenamefont {Singal}, \citenamefont {Haro}, \citenamefont
  {Stefanazzi}, \citenamefont {Tiffenberg}, \citenamefont {Uemura},
  \citenamefont {Volansky},\ and\ \citenamefont {Yu}}]{sensei2020}%
  \BibitemOpen
  \bibfield  {author} {\bibinfo {author} {\bibfnamefont {L.}~\bibnamefont
  {Barak}}, \bibinfo {author} {\bibfnamefont {I.~M.}\ \bibnamefont {Bloch}},
  \bibinfo {author} {\bibfnamefont {M.}~\bibnamefont {Cababie}}, \bibinfo
  {author} {\bibfnamefont {G.}~\bibnamefont {Cancelo}}, \bibinfo {author}
  {\bibfnamefont {L.}~\bibnamefont {Chaplinsky}}, \bibinfo {author}
  {\bibfnamefont {F.}~\bibnamefont {Chierchie}}, \bibinfo {author}
  {\bibfnamefont {M.}~\bibnamefont {Crisler}}, \bibinfo {author} {\bibfnamefont
  {A.}~\bibnamefont {Drlica-Wagner}}, \bibinfo {author} {\bibfnamefont
  {R.}~\bibnamefont {Essig}}, \bibinfo {author} {\bibfnamefont
  {J.}~\bibnamefont {Estrada}}, \bibinfo {author} {\bibfnamefont
  {E.}~\bibnamefont {Etzion}}, \bibinfo {author} {\bibfnamefont {G.~F.}\
  \bibnamefont {Moroni}}, \bibinfo {author} {\bibfnamefont {D.}~\bibnamefont
  {Gift}}, \bibinfo {author} {\bibfnamefont {S.}~\bibnamefont {Munagavalasa}},
  \bibinfo {author} {\bibfnamefont {A.}~\bibnamefont {Orly}}, \bibinfo {author}
  {\bibfnamefont {D.}~\bibnamefont {Rodrigues}}, \bibinfo {author}
  {\bibfnamefont {A.}~\bibnamefont {Singal}}, \bibinfo {author} {\bibfnamefont
  {M.~S.}\ \bibnamefont {Haro}}, \bibinfo {author} {\bibfnamefont
  {L.}~\bibnamefont {Stefanazzi}}, \bibinfo {author} {\bibfnamefont
  {J.}~\bibnamefont {Tiffenberg}}, \bibinfo {author} {\bibfnamefont
  {S.}~\bibnamefont {Uemura}}, \bibinfo {author} {\bibfnamefont
  {T.}~\bibnamefont {Volansky}}, \ and\ \bibinfo {author} {\bibfnamefont
  {T.-T.}\ \bibnamefont {Yu}} (\bibinfo {collaboration} {SENSEI
  Collaboration}),\ }\href {\doibase 10.1103/PhysRevLett.125.171802} {\bibfield
   {journal} {\bibinfo  {journal} {Phys. Rev. Lett.}\ }\textbf {\bibinfo
  {volume} {125}},\ \bibinfo {pages} {171802} (\bibinfo {year}
  {2020}{\natexlab{b}})}\BibitemShut {NoStop}%
\bibitem [{\citenamefont {Du}\ \emph {et~al.}(2022)\citenamefont {Du},
  \citenamefont {Egana-Ugrinovic}, \citenamefont {Essig},\ and\ \citenamefont
  {Sholapurkar}}]{Du_2022}%
  \BibitemOpen
  \bibfield  {author} {\bibinfo {author} {\bibfnamefont {P.}~\bibnamefont
  {Du}}, \bibinfo {author} {\bibfnamefont {D.}~\bibnamefont {Egana-Ugrinovic}},
  \bibinfo {author} {\bibfnamefont {R.}~\bibnamefont {Essig}}, \ and\ \bibinfo
  {author} {\bibfnamefont {M.}~\bibnamefont {Sholapurkar}},\ }\href {\doibase
  10.1103/PhysRevX.12.011009} {\bibfield  {journal} {\bibinfo  {journal} {Phys.
  Rev. X}\ }\textbf {\bibinfo {volume} {12}},\ \bibinfo {pages} {011009}
  (\bibinfo {year} {2022})}\BibitemShut {NoStop}%
\bibitem [{\citenamefont {Du}\ \emph {et~al.}(2024)\citenamefont {Du},
  \citenamefont {Ega\~na Ugrinovic}, \citenamefont {Essig},\ and\ \citenamefont
  {Sholapurkar}}]{Du:2023soy}%
  \BibitemOpen
  \bibfield  {author} {\bibinfo {author} {\bibfnamefont {P.}~\bibnamefont
  {Du}}, \bibinfo {author} {\bibfnamefont {D.}~\bibnamefont {Ega\~na
  Ugrinovic}}, \bibinfo {author} {\bibfnamefont {R.}~\bibnamefont {Essig}}, \
  and\ \bibinfo {author} {\bibfnamefont {M.}~\bibnamefont {Sholapurkar}},\
  }\href {\doibase 10.1007/JHEP01(2024)164} {\bibfield  {journal} {\bibinfo
  {journal} {JHEP}\ }\textbf {\bibinfo {volume} {01}},\ \bibinfo {pages} {164}
  (\bibinfo {year} {2024})},\ \Eprint {http://arxiv.org/abs/2310.03068}
  {arXiv:2310.03068 [hep-ph]} \BibitemShut {NoStop}%
\bibitem [{\citenamefont {Ball}\ \emph
  {et~al.}(2020{\natexlab{b}})\citenamefont {Ball} \emph
  {et~al.}}]{Ball:2020dnx}%
  \BibitemOpen
  \bibfield  {author} {\bibinfo {author} {\bibfnamefont {A.}~\bibnamefont
  {Ball}} \emph {et~al.},\ }\href {\doibase 10.1103/PhysRevD.102.032002}
  {\bibfield  {journal} {\bibinfo  {journal} {Phys. Rev. D}\ }\textbf {\bibinfo
  {volume} {102}},\ \bibinfo {pages} {032002} (\bibinfo {year}
  {2020}{\natexlab{b}})},\ \Eprint {http://arxiv.org/abs/2005.06518}
  {arXiv:2005.06518 [hep-ex]} \BibitemShut {NoStop}%
\bibitem [{\citenamefont {Cowan}\ \emph {et~al.}(2011)\citenamefont {Cowan},
  \citenamefont {Cranmer}, \citenamefont {Gross},\ and\ \citenamefont
  {Vitells}}]{Cowan:2010js}%
  \BibitemOpen
  \bibfield  {author} {\bibinfo {author} {\bibfnamefont {G.}~\bibnamefont
  {Cowan}}, \bibinfo {author} {\bibfnamefont {K.}~\bibnamefont {Cranmer}},
  \bibinfo {author} {\bibfnamefont {E.}~\bibnamefont {Gross}}, \ and\ \bibinfo
  {author} {\bibfnamefont {O.}~\bibnamefont {Vitells}},\ }\href {\doibase
  10.1140/epjc/s10052-011-1554-0} {\bibfield  {journal} {\bibinfo  {journal}
  {Eur. Phys. J. C}\ }\textbf {\bibinfo {volume} {71}},\ \bibinfo {pages}
  {1554} (\bibinfo {year} {2011})},\ \bibinfo {note} {[Erratum: Eur.Phys.J.C
  73, 2501 (2013)]},\ \Eprint {http://arxiv.org/abs/1007.1727} {arXiv:1007.1727
  [physics.data-an]} \BibitemShut {NoStop}%
\end{thebibliography}%

\end{document}